# Simulating Anisotropic Thermal Conduction in Supernova Remnants I : Numerics and the Evolution of Remnants

By

Dinshaw S. Balsara (dbalsara@nd.edu), David A. Tilley (dtilley@nd.edu) and J. Christopher Howk (jhowk@nd.edu)

Physics Department, University of Notre Dame

**Key Words :** Supernova Remnants, Thermal Conduction, X-ray emission, Magnetohydrodynamics, Numerical Schemes

**Running Head :** Anisotropic Thermal Conduction in SNRs I

**Mailing Address:**

Physics Department

College of Science

University of Notre Dame

225 Nieuwland Science Hall

Notre Dame, IN 46556

**Phone :** (574) 631-9639

**Fax :** (574) 631-5952

**Abstract**

Anisotropic thermal conduction plays an important role in various astrophysical systems. One of the most stringent tests of thermal conduction can be found in supernova remnants. In this paper we study anisotropic thermal conduction and examine the physical nature of the flux of thermal conduction in the classical and saturated limits. We also present a temporally second-order accurate implicit-explicit scheme for the time-update of thermal conduction terms within a numerical MHD scheme. Several useful tests are presented showing that the method works well.

Several simulations of supernova remnants are presented for a range of ISM parameters. The role of thermal conduction in such remnants has been studied. We find that thermal conduction produces cooler temperatures and higher densities in the hot gas bubbles that form in the remnants. The effect of thermal conduction in changing the thermal characteristics of the hot gas bubble increases as the remnant propagates through denser ISMs. Remnants evolving in denser ISMs are shown to make a faster transition to a centre-bright x-ray morphology, with the trend emerging earlier in hard x-rays than in the soft x-rays.

## I) Introduction

Thermal conduction (TC henceforth) plays an important role in determining the transport of energy in various astrophysical systems. The importance of TC in the evolution of supernova remnants (SNRs henceforth) was catalogued in Chevalier (1975), White & Long (1991, WL henceforth), Slavin & Cox (1992,1993, SC92 and SC93 henceforth), Cui & Cox (1992) and Cox et al (1999). Forslund (1970) showed that when classical TC causes the electrons to travel faster than the ion sound speed, free streaming of the electrons is strongly impeded by plasma instabilities. Cowie & McKee (1977, CM henceforth) realized that when the electron mean free path is larger than the temperature scale length the TC will be limited to its saturated value. CM showed the importance of saturated TC in the early evolution of SNRs. TC is also important in several other astrophysical systems. The role of TC in shock-cloud interactions was examined in Klein, McKee & Colella (1994) who found that it plays an important role in redistributing the energy in crushed clouds. The usefulness of TC in determining the survivability of high-velocity clouds as they impinge on the Galactic disk was studied in Maller & Bullock (2004). The relevance of TC in the formation and evolution of turbulent boundary layers that form between the hot and warm phases of the interstellar medium (ISM henceforth) was examined by Borkowski, Balbus, & Fristrom (1990) and Begelman & McKee (1990) and observations along those lines have been presented by Oegerle et al (2005), Bowen et al (2005), Savage & Lehner (2005) and Dixon, Sankrit & Otte (2006). TC also plays an important role in modulating the thermal instability in colliding streams of warm interstellar gas, Heitsch et al (2005). Field (1965) and Piontek & Ostriker (2004) showed the importance of TC in modulating the thermal instability and setting a lower bound to the length scales on which molecular clouds can form. Cluster cooling flows may also be influenced by TC, as shown by Fabian, Nulsen & Canizares (1991) and Pistinner & Shaviv (1996). TC is also important in regulating the heat transfer on the surfaces of Type I x-ray bursts, see Schatz et al (1999). Yokoyama & Shibata (1997) showed the importance of anisotropic TC in magnetic reconnection problems of relevance to solar physics. CM analyzed the dependence of classical and saturated TC on the direction of the magnetic field. Using results from Spitzer (1962), Balbus (1986) argued that heat conduction orthogonal to magnetic field lines is strongly suppressed, making TC anisotropic in such situations.

Since magnetic fields permeate all the astrophysical systems mentioned above, we wish to study the role of anisotropic TC in the presence of magnetic fields. One of the goals of this paper is to obtain an efficient methodology for the time-implicit treatment of TC with and without the presence of magnetic fields. While time-explicit methods have been used in the numerical treatment of TC (Marcolini et al 2005 and Piontek & Ostriker 2004), such methods require unusually large number of sub-cycled, time-explicit time steps making them unwieldy for problems involving large meshes or strong temperature gradients. The present paper rectifies this situation. The other goal of this paper is to use the present time-implicit methods for TC to study the multi-dimensional evolution of SNRs in unmagnetized and magnetized environments. A corresponding one-dimensional study had been presented in SC92 and SC93. SNRs are the dominant source of energy injection in the turbulent ISM and their evolution influences several observable features of our ISM. As a result, our decision to train our numerics on this important problem is well-justified. The temperatures and densities that are reached in the evolution of SNRs are also some of the most extreme in astrophysics, thus providing a stringent test of the numerics.

Woltjer (1972) introduced a simplified paradigm for the evolution of SNRs, suggesting that they pass through four distinct phases. Over time, we have come to know those phases as “free expansion”, “Sedov, adiabatic blast wave”, “radiative snowplow” and “dispersal”. Because the emissivity of SNRs fades as they evolve, observers usually focus on the first ~100,000 years in the life of a remnant. It is, therefore, very interesting to study the evolution of SNRs during that epoch in their evolution. TC could affect the early time (Sedov phase) evolution of remnants, changing the propagation of the inward-facing shock. Since the inward-facing shock plays a very important role in particle acceleration, see Jun & Jones (1999), a correct treatment of TC could change the amount of particle-acceleration that takes place in SNRs. TC can also influence the later evolution (snowplow phase) of SNRs by transporting energy within the remnant, see SC92 and Cox et al (1999). In doing so, it can change the x-ray emission characteristics of the hot gas bubbles within SNRs. Even past the first ~100,000 years during which SNRs are observable, they continue to influence the dynamics of the ISM and its evolution. The present paragraph has shown the importance of TC in determining the evolution of individual SNRs. In subsequent paragraphs we discuss the physics of SNRs, highlighting the role of TC in those systems.

An examination of SNR morphologies shows that they come in three important morphological types, see Jones et al (1998) for a review. Classical shell-type SNRs have x-ray bright shells which are also bright in the radio. Typical shell-type remnants include Tycho’s SNR, SN 1006 and the Cygnus Loop. In such remnants, the interaction of the SNR’s shock wave with the ISM is responsible for the non-thermal synchrotron processes which give rise to the radio emission. The x-ray brightness arises due to the thermal bremmstrahlung emission from the shell. Plerionic remnants form a second class of remnants. They include Crab-like sources where the central pulsar is thought to produce a wind resulting in centrally concentrated radio and x-ray luminosity, see Hester et al (2002). Mixed morphology (MM henceforth) remnants have been identified by Rho & Petre (1998, RP henceforth) as being a third, intermediate class of remnant. RP find that ~ 8% of all Galactic SNRs fall in this class and prominent examples include W44, W28, Kes 27 and 3C 391. These SNRs have a limb-brightened radio shell and a centrally-concentrated x-ray flux. It is also worth noting that RP exclude remnants that have a compact source in the x-ray or radio from membership in this class, thus ensuring that the centrally-concentrated x-ray flux is a result of thermal emission from hot gas. While several models have been presented for the formation of MM SNRs, the most prominent, competing, explanations have been provided by Cox et al (1999) and WL. On observational grounds, RP98 noted that MM SNRs tend to occur in denser parts of the ISM while shell-type SNRs occur in less dense parts of the ISM. The denser ISM may be distributed as small cloudlets that get engulfed by the remnants' shock resulting in the model of WL or it could consist of a larger scale density gradient resulting in the model of SC92 and Cox et al (1999).

Cox et al (1999) realized that a SNR evolving with TC in an over-dense ISM with a density gradient could explain W44, a MM SNR. Velazquez et al (2004) included the anisotropy in the thermal conduction introduced by the magnetic field, but they did not include the Lorentz force of the magnetic field on the gas. In this limit, stirring of the magnetic field by turbulent motions that develop behind the supernova shock greatly reduce the effect of thermal conduction. Without the feedback of the magnetic field on the gas, however, this can only be regarded as a lower limit on the degree of thermal conduction. Fully magnetohydrodynamic (MHD) models for the multi-dimensional evolution of SNRs that include TC have been recently presented in Tilley, Balsara & Howk (2006, TBH henceforth). They showed that SNRs expanding into a dense medium become center-bright during the course of their observable evolution. Such was not the case for SNRs expanding into a tenuous ISM. TC cools the hot gas bubbles more efficiently, bringing the temperature down into the regimes where radiative cooling becomes more efficient. TBH found that this effect is enhanced for SNRs in denser ISMs.

Cox et al (1999) and Shelton et al (1999) had to draw on a density gradient in the ISM to produce the elongated morphology of W44. Tilley & Balsara (2006, TB henceforth) showed that an elongated morphology could also be reproduced by the presence of a net magnetic field, though this does not explain the one-sided enhancement in W44's emissivity. It is possible to envision a range of input physics for such models, including non-equilibrium ionization, effect of the circumstellar medium, cosmic ray acceleration, turbulence in the ISM and so on. In this paper we take a first step in that direction by presenting simulations which include magnetic fields, TC and equilibrium cooling. It is realized that non-equilibrium ionization is an important ingredient in explaining the high-stage ions and will be included in a subsequent round of simulations for a future paper. The data that we generate for the high-stage ions in this paper will, therefore, serve as a foil for comparison with subsequent work that includes non-equilibrium ionization. RP find that shell-type SNRs tend to occur in tenuous environments above the Galactic mid-plane while MM SNRs tend to occur closer to the Galactic mid-plane where the density is higher. For this reason, we explore the evolution of SNRs in ISMs with a wide range of densities. Kawasaki et al (2005) have proposed on the basis of measured over-ionizations in the majority of MM SNRs that in the presence of thermal conduction, all supernova remnants pass from a shell phase to a thermally composite phase at some point during their evolution. The present simulations will give us a chance to evaluate that hypothesis.

WL developed a similarity solution for the evolution of SNRs in a cloudy ISM. The shock-cloud crushing observed in Vela and IC333 might be taken as providing observational support for the model of WL. RP also found that MM SNRs tend to occur preferentially closer to molecular clouds, providing further support for WL's model. The clouds in WL's model were assumed to be small, dense and uniformly distributed in the ISM and they were assumed to evaporate once they were struck by the remnants' outer shock with a certain time-dependent scaling. Once the clumps get entrained in a SNR shock, they increase the mass-loading within the SNR's cavity. TC implicitly plays an important role in the model of WL because the blending of the crushed, cold clouds into the hot, post-shock environment occurs because of TC. Cloud evaporation, therefore, cools down the hot bubble in the remnant's interior while also increasing its density. WL studied the radial density and temperature profiles of SNRs as a function of the mass-loading of the ISM with clouds. It is possible to design numerical experiments to test the model of WL but we defer such an exercise to a subsequent paper.

Recently, Pedlar, Muxlow & Wills (2003) have also observed individual SNRs as well as superbubbles in the central star-bursting region of M82. Rieke et al (1980) and Blom, Paglione and Carramiñana (1999) find observational evidence that suggests that the ISM of M82 might be substantially denser than the Galactic ISM at the solar circle. The above papers also suggest that the magnetic fields in those environments might be similarly stronger. The role of massive stars in sculpting the ISMs of such starburst galaxies has also been reviewed in Veilleux, Cecil & Bland-Hawthorn (2005). It is also possible that if the ISM of M82 has fluctuations in density that exceed ten times the mean then low mass star formation through shell fragmentation could also take place in such a system, see Elmegreen (1994), Salvaterra, Ferrara & Schneider (2004) and references contained therein. For these reasons, an examination of SNR evolution in such extreme situations will also be undertaken here.

In Section II we describe the equations of MHD with anisotropic thermal conduction along field lines. In that section we also describe our solution methodology and tests. In Section III we describe the models that were run. In Section IV we describe the evolution and morphology of SNRs in the presence of thermal conduction. Section V presents conclusions.

## II) Governing Equations and Semi-Implicit Time Update for Thermal Conduction

In the ensuing four sub-sections we describe four things. In Sub-section II.1 we describe the MHD equations and their extension for thermal conduction. In Sub-section II.2 we describe the treatment of the anisotropic energy fluxes that describe TC in the classical and saturated limits, showing how the equations change character in either limit. In Sub-section II.3 we present the second order accurate semi-implicit time-stepping strategy.

## II.1) MHD Equations

We solve the equations of MHD in cylindrical geometry. Radiative cooling from Figure 1 of MacDonald & Bailey (1981) is included. This curve follows the equilibrium ionization cooling curve of Raymond, Cox & Smith (1976) for $10^6$ K < T < $10^8$ K; the nonequilibrium curve for isochoric cooling derived by Shapiro & Moore (1976) from $10^4$ K < T < $10^6$ K; and then is smoothly extrapolated to zero at $10^2$ K. We also include a diffuse heating term to represent processes such as heating by cosmic rays and photoelectric heating by starlight. We set the heating per unit mass to be constant in space and time. Anisotropic thermal conduction that transitions smoothly from classical TC to saturated TC has also been included using descriptions provided in Spitzer (1962), CM and Balbus (1986). The continuity equation is given by:

$$\frac{\partial \rho}{\partial t} + \frac{1}{r} \frac{\partial \left( r \rho v_r \right)}{\partial r} + \frac{1}{r} \frac{\partial \left( \rho v_\theta \right)}{\partial \theta} + \frac{\partial \left( \rho v_z \right)}{\partial z} = 0 \tag{2.1}$$

The r-component of the momentum equation becomes:

$$\frac{\partial (\rho v_r)}{\partial t} + \frac{1}{r} \frac{\partial}{\partial r}\left[ r \left( \rho v_r^2 - \frac{B_r^2}{4\pi} \right) \right] + \frac{1}{r} \frac{\partial}{\partial \theta}\left[ \rho v_r v_\theta - \frac{B_r B_\theta}{4\pi} \right] + \frac{\partial}{\partial z}\left[ \rho v_r v_z - \frac{B_r B_z}{4\pi} \right] + \frac{\partial}{\partial r}\left[ P + \frac{\mathbf{B}^2}{8\pi} \right] = \frac{\rho v_\theta^2}{r} - \frac{B_\theta^2}{4\pi r} \tag{2.2}$$

The θ-component of the momentum equation, when written out in an angular momentum conserving form, becomes:

$$\frac{\partial(\mathrm{r}\rho \mathrm{v}_\theta)}{\partial \mathrm{t}} + \frac{1}{\mathrm{r}}\frac{\partial}{\partial \mathrm{r}}\left[\mathrm{r}^2\left(\rho \mathrm{v}_\mathrm{r}\mathrm{v}_\theta - \frac{\mathrm{B}_\mathrm{r}\,\mathrm{B}_\theta}{4\,\pi}\right)\right] + \frac{1}{\mathrm{r}}\frac{\partial}{\partial\,\theta}\left[\mathrm{r}\left(\rho \mathrm{v}_\theta^2 - \frac{\mathrm{B}_\theta^2}{4\,\pi} + \mathrm{P} + \frac{\mathbf{B}^2}{8\,\pi}\right)\right]$$
$$+\frac{\partial}{\partial \mathrm{z}}\left[\mathrm{r}\left(\rho \mathrm{v}_\theta \mathrm{v}_\mathrm{z} - \frac{\mathrm{B}_\theta\,\mathrm{B}_\mathrm{z}}{4\,\pi}\right)\right] = 0 \qquad (2.3)$$

The z-component of the momentum equation becomes:

$$\frac{\partial(\rho \mathrm{v}_\mathrm{z})}{\partial \mathrm{t}} + \frac{1}{\mathrm{r}}\frac{\partial}{\partial \mathrm{r}}\left[\mathrm{r}\left(\rho \mathrm{v}_\mathrm{r}\,\mathrm{v}_\mathrm{z} - \frac{\mathrm{B}_\mathrm{r}\,\mathrm{B}_\mathrm{z}}{4\,\pi}\right)\right] + \frac{1}{\mathrm{r}}\frac{\partial}{\partial\,\theta}\left[\rho \mathrm{v}_\theta\,\mathrm{v}_\mathrm{z} - \frac{\mathrm{B}_\theta\,\mathrm{B}_\mathrm{z}}{4\,\pi}\right]$$
$$+\frac{\partial}{\partial \mathrm{z}}\left[\rho \mathrm{v}_\mathrm{z}^2 - \frac{\mathrm{B}_\mathrm{z}^2}{4\,\pi} + \mathrm{P} + \frac{\mathbf{B}^2}{8\,\pi}\right] = 0 \qquad (2.4)$$

The energy equation is given by:

$$\frac{\partial\,\varepsilon}{\partial \mathrm{t}} + \frac{1}{\mathrm{r}}\frac{\partial}{\partial \mathrm{r}}\left[\mathrm{r}\left(\left(\varepsilon + \mathrm{P} + \mathbf{B}^2/8\pi\right)\mathrm{v}_\mathrm{r} - \mathrm{B}_\mathrm{r}\left(\mathbf{v}\cdot\mathbf{B}\right)/4\pi\right)\right]$$
$$+\frac{1}{\mathrm{r}}\frac{\partial}{\partial\,\theta}\left[\left(\varepsilon + \mathrm{P} + \mathbf{B}^2/8\pi\right)\mathrm{v}_\theta - \mathrm{B}_\theta\left(\mathbf{v}\cdot\mathbf{B}\right)/4\pi\right]$$
$$+\frac{\partial}{\partial \mathrm{z}}\left[\left(\varepsilon + \mathrm{P} + \mathbf{B}^2/8\pi\right)\mathrm{v}_\mathrm{z} - \mathrm{B}_\mathrm{z}\left(\mathbf{v}\cdot\mathbf{B}\right)/4\pi\right] = \mathrm{n}_\mathrm{H}\,\Gamma - \mathrm{n}_\mathrm{e}\,\mathrm{n}_\mathrm{H}\,\Lambda(\mathrm{T}) - \nabla\bullet\mathbf{F}_\mathrm{net}$$
$$(2.5)$$

And the induction equation is:

$$\frac{\partial\mathbf{B}}{\partial \mathrm{t}} - \nabla\times\left(\mathbf{v}\times\mathbf{B}\right) = 0 \qquad (2.6)$$

Here ρ is the fluid’s density, ( $v_r$ , $v_\theta$ , $v_z$ ) represent the components of the velocities along the r, θ and z-axes, ( $B_r$ , $B_\theta$ , $B_z$ ) represent the components of the magnetic field along the axes and $\varepsilon$ is the total energy density given by $\varepsilon = \rho\,\mathbf{v}^2/2 + \mathrm{P}/(\gamma - 1) + \mathbf{B}^2/8\pi$ where γ is the ratio of specific heats and P is the pressure. For later use, the temperature is denoted by T, the gas constant by R, the sound speed by $c_s$ and the reduced mass of the interstellar gas is taken to be $\mu$ . $\Gamma$ and $\Lambda(T)$ are heating and cooling terms. $n_e$ and $n_H$ are the number densities of electrons and hydrogen atoms in the gas and can be related to the density $\rho$ . The flux of thermal conduction, $\mathbf{F}_{net}$ , in eqn. (2.5) will be the object of our study in the next sub-section.

The MHD equations are solved explicitly using the RIEMANN code which uses a dimensionally unsplit scheme. The scheme uses spatially second order accurate total variation diminishing (TVD) interpolation and temporally second order accurate Runge-Kutta time stepping along with a divergence-free reconstruction of the magnetic field as shown in Balsara (2001a). The MHD algorithms have been described in Roe & Balsara (1996), Balsara (1998a,b), Balsara & Spicer (1999a,b) and Balsara (2004). Various stringent MHD tests have been described in Balsara (1998b), Balsara & Spicer (1999b), Balsara (2004) and Balsara & Kim (2004). The code has been applied to SNR-related problems in Balsara, Benjamin & Cox (2001), Balsara et al (2004), Mac Low et al (2005) and Balsara & Kim (2005). It has also been applied to numerous other astrophysical problems.

For this work, the MHD equations have to be solved in conjunction with the equations of thermal conduction. The next sub-section will show that thermal conduction can behave like a parabolic operator in the classical limit and like a hyperbolic operator in the saturated limit. If we consider just the $10^8$ K gas in a SNR and evaluate the explicit time step that is required with a typical mesh size of 0.25 pc, we find that the explicit time step needs to be less than 100 yr. Thus an explicit strategy for the temporal update of thermal conduction is out of question, especially if we intend to evolve SNRs over millions of years. For that reason, we carry out a detailed analysis of the flux of thermal conduction in the next sub-section. This analysis will then lead us to an implicit solution strategy for thermal conduction along with a few beneficial schemes for interleaving such an implicit solution strategy with an explicit strategy for the rest of the MHD equations. This will be described, along with numerical tests, in the next section.

### II.2) Understanding the Physical Behavior of the Flux of Thermal Conduction

The treatment of the flux of thermal conduction in eqn. (2.5) requires sensitive treatment since it changes character in different limits. An analytic discussion of these limits is contained in CM and Balbus (1986). In this section we present a complementary treatment which is suitable for numerical implementation. In the presence of magnetic fields, the thermal conduction front is restricted to propagate in the direction of the magnetic field. As a result, it is very helpful to define the unit vector along the direction of the magnetic field by $\mathbf{b} = \mathbf{B}/|\mathbf{B}|$ . Following Spitzer (1962) the classical flux for TC can be written in a vectorial notation as:

$$\mathbf{F}_{\mathrm{class}} = -\ a\, T^{5/2}\ \mathbf{b}\ \left(\mathbf{b} \bullet \nabla\ T\right) \qquad (2.7)$$

where, for the temperature regime being simulated, CM and SC92 show that it is appropriate to set a = 6X10$^{-7}$ cgs units. A vectorially consistent extension of the equations in Balbus (1986) shows that the saturated flux for TC can be written as:

$$\mathbf{F}_{\mathrm{sat}} = -\ 5\ \phi\ \rho\ c_s^3\ \mathrm{sgn}\left(\mathbf{b} \bullet \nabla\ T\right)\ \mathbf{b} \qquad (2.8)$$

Balbus & McKee (1982) suggest using $\phi = 0.3$ . CM show that the choice of $\mathbf{F}_{\mathrm{class}}$ or $\mathbf{F}_{\mathrm{sat}}$ depends strongly on the ratio $F_{\mathrm{class}}/F_{\mathrm{sat}}$ . When the ratio is less than unity, one should choose $\mathbf{F}_{\mathrm{class}}$ from eqn. (2.7) ; but when the ratio exceeds unity, one should choose $\mathbf{F}_{\mathrm{sat}}$ from eqn. (2.8). To prevent an abrupt transition from one limit to the other, CM also present a flux limiter. The form of the flux limiter is not very important, since SC92 have shown that alternative flux limiters seem to work about as well. The flux limiters presented in CM and SC92 use a single structure for the flux. However, since we know from Balbus (1986) that the choice of flux causes the equations to change character, it helps to use a form of limiter that enables us to disentangle $\mathbf{F}_{\mathrm{class}}$ and $\mathbf{F}_{\mathrm{sat}}$ . The advantage of such a formulation is that it permits us to use discretizations for $\mathbf{F}_{\mathrm{class}}$ and $\mathbf{F}_{\mathrm{sat}}$ that are truer to their parabolic or hyperbolic natures respectively. We thus write:

$$\mathbf{F}_{net} = G(F_{class}/F_{sat})\,\mathbf{F}_{class} + \left[1 - G(F_{class}/F_{sat})\right]\mathbf{F}_{sat} \qquad (2.9)$$

where G(x) can take on several functional forms, a couple of which are given below:

$$G(x) = \max\left\{0, \min\left[1, 1 - 0.5\,x\right]\right\} \qquad (2.10)$$

or

$$G(x) = \max\left\{0, \min\left[1, 1.5 - x\right]\right\} \qquad (2.11)$$

We now turn our attention to understanding the nature of $\mathbf{F}_{class}$ . To see that it is parabolic, take $\mathbf{b} = \mathbf{r}$ , where $\mathbf{r}$ is the unit vector along the r-axis. Ignoring the velocity and magnetic field terms in eqn. (2.5) then allows us to write it in the limit of classical TC as:

$$\frac{R\,\rho}{\mu(\gamma-1)}\,\frac{\partial T}{\partial t} = \frac{1}{r}\,\frac{\partial}{\partial r}\left(r\,a\,T^{5/2}\,\frac{\partial T}{\partial r}\right) \qquad (2.12)$$

which clearly illustrates that the equation is parabolic. Eqn. (2.7) also shows that the heat flux is restricted to diffuse in the direction of the magnetic field. The diffusion coefficient transverse to the direction of the magnetic field is zero. The flux of classical TC can be written in cylindrical geometry as:

$$\begin{aligned}\mathbf{F}_{class} = -\,a\,T^{5/2}\,\Big\{\,&\mathbf{r}\left(b_r^2\frac{\partial T}{\partial r} + \frac{b_r\,b_\theta}{r}\,\frac{\partial T}{\partial \theta} + b_r\,b_z\,\frac{\partial T}{\partial z}\right)\\ &+ \boldsymbol{\theta}\left(b_r\,b_\theta\frac{\partial T}{\partial r} + \frac{b_\theta^2}{r}\,\frac{\partial T}{\partial \theta} + b_\theta\,b_z\,\frac{\partial T}{\partial z}\right)\\ &+ \mathbf{z}\left(b_r\,b_z\frac{\partial T}{\partial r} + \frac{b_\theta\,b_z}{r}\,\frac{\partial T}{\partial \theta} + b_z^2\,\frac{\partial T}{\partial z}\right)\Big\}\end{aligned} \qquad (2.13)$$

Similar expressions for Cartesian and Spherical geometries are given in Appendix A. Further insight into the nature of the flux can be gained by setting $b_\theta = 0$ and restricting the temperature gradient to be non-zero only along the r-axis in eqn. (2.13). We then get:

$$F_{\mathrm{class,\,r}} = -\, a\, T^{5/2}\, b_r^2 \frac{\partial T}{\partial r} \quad \text{and} \quad F_{\mathrm{class,\,z}} = -\, a\, T^{5/2}\, b_r\, b_z \frac{\partial T}{\partial r} \qquad (2.14)$$

Notice that the net flux of heat in the presence of a magnetic field is smaller than the non-magnetized case by a factor of $b_r$ in the limit of classical TC. Thus the presence of the magnetic field has indeed diminished the classical heat flux, as expected. The structure of $F_{\mathrm{class,\,r}}$ in eqn. (2.14) can be intuitively understood because it contributes a heat flux in the direction of the temperature gradient. The structure of $F_{\mathrm{class,\,z}}$ in eqn. (2.14) is harder to understand until one realizes that the field channels the heat flux in its own direction, thus producing a heat flux in the z-direction. The r-component of the classical heat flux in eqn. (2.14) is reduced by $b_r^2$ relative to the case without magnetic field. This reduction is consistent with the trend noted in CM.

We now turn our attention to understanding the nature of $\mathbf{F}_{\mathrm{sat}}$ . To see that it is hyperbolic, we write eqn. (2.8) as:

$$\mathbf{F}_{\mathrm{sat}} = -\, 5\, \phi\, \rho \left( \frac{\gamma R}{\mu} \right)^{3/2} T^{3/2}\, \mathrm{sgn}\left( \mathbf{b} \bullet \nabla T \right)\, \mathbf{b} \qquad (2.15)$$

Ignoring the velocity and magnetic field terms in eqn. (2.5) then allows us to write it in the limit of saturated TC as:

$$
\frac{R\,\rho}{\mu(\gamma-1)}\;\frac{\partial T}{\partial t}+\frac{1}{r}\;\frac{\partial}{\partial r}\left(-\,r\,b_r\;5\;\phi\;\rho\left(\frac{\gamma\,R}{\mu}\right)^{3/2}\;\mathrm{sgn}\left(\mathbf{b}\bullet\nabla\,T\right)\;T^{3/2}\right)
$$

$$
+\;\frac{1}{r}\;\frac{\partial}{\partial\,\theta}\left(-\,b_\theta\;5\;\phi\;\rho\left(\frac{\gamma\,R}{\mu}\right)^{3/2}\;\mathrm{sgn}\left(\mathbf{b}\bullet\nabla\,T\right)\;T^{3/2}\right) \tag{2.16}
$$

$$
+\;\frac{\partial}{\partial\,z}\left(-\,b_z\;5\;\phi\;\rho\left(\frac{\gamma\,R}{\mu}\right)^{3/2}\;\mathrm{sgn}\left(\mathbf{b}\bullet\nabla\,T\right)\;T^{3/2}\right)=0
$$

To clarify the structure of eqn. (2.16) we define the following auxiliary variables:

$$
d_{r,\theta,z}\;=\;-\;b_{r,\theta,z}\;\gamma\;(\;\gamma\;-\;1)\;5\;\phi\left(\frac{\gamma\,R}{\mu}\right)^{1/2}\;\mathrm{sgn}\left(\mathbf{b}\bullet\nabla\,T\right) \tag{2.17}
$$

In the limit where the density is constant in time (a choice that can easily be effected by operator-splitting the MHD update from the update of the thermal conduction), eqn. (2.16) can now be rewritten as:

$$
\frac{\partial T}{\partial t}\;+\;\left(\frac{2}{3}\;d_r\;T^{1/2}\right)\frac{\partial T}{\partial r}\;+\;\left(\frac{2}{3}\;d_\theta\;T^{1/2}\right)\frac{1}{r}\;\frac{\partial T}{\partial\,\theta}\;+\;\left(\frac{2}{3}\;d_z\;T^{1/2}\right)\frac{\partial T}{\partial z}\;+\;\frac{d_r\;T^{3/2}}{r}\;=0 \tag{2.18}
$$

Eqn. (2.18) shows the temperature is transported advectively, illustrating the hyperbolic nature of saturated heat conduction. We realize that the signal speeds in the three directions in eqn. (2.18) are given by $\frac{3}{2}\;T^{1/2}\;\left(d_r\;,\;d_\theta\;,\;d_z\right)$. We see, therefore, that the propagation speed for the conduction front in the limit of saturated TC is comparable to the sound speed. If the problem is dominated by saturated TC alone, it can even be solved entirely by time-explicit upwind techniques. In practice, classical TC will also be important so that the problem has to be solved using iterative, implicit techniques that retain saliency in the parabolic and hyperbolic limits. Because the fractional importance of $\mathbf{F}_{class}$ and $\mathbf{F}_{sat}$ in eqn. (2.9) can change with each successive iteration of an iterative method, it is not acceptable to freeze the saturated TC and treat it time-explicitly in the general case. Newton-Krylov Multigrid methods have been known to work well on such problems and so they will form an important part of the solution strategy that we will describe later. If the temperature gradient is restricted to the r-direction in eqn. (2.16) we see that the saturated flux is proportional to $b_r$ , in keeping with the observation in CM.

The discussion in the previous two paragraphs has shown us that we should choose numerical discretizations for $\mathbf{F}_{class}$ and $\mathbf{F}_{sat}$ in eqn. (2.9) that are tailored to the parabolic and hyperbolic natures of the respective fluxes. Since parabolic equations respond best to zone-centered discretizations, we will choose such discretizations for $\mathbf{F}_{class}$ . Note though, that $\mathbf{F}_{class}$ has a structure similar to a Laplacian operator with variable coefficients when the magnetic field is absent. That structure is destroyed when anisotropies related to the magnetic field are introduced in eqn. (2.13). These anisotropies also make the equations diagonally sub-dominant, making BiCGStab-based solution techniques ineffective. However, GMRES-based solution techniques still remain salient in this situation and we will resort to such techniques for our implicit, iterative solvers. Saad (1996) has provided a nice introduction to BiCGStab and GMRES-based solution techniques. GMRES solvers are also dramatically accelerated when coupled to Multigrid methods and so we will use GMRES as a smoother within a Multigrid scheme, see Balsara (2001b). We next describe favorable methods for discretizing $\mathbf{F}_{sat}$ in eqn. (2.9).

A solution technique for saturated TC emerges when one studies the Burgers-like equation:

$$\frac{\partial T}{\partial t} + \frac{\partial}{\partial z}\left(d_z\ T^{3/2}\right) = 0 \qquad (2.19)$$

We see that the solution should always respect the restriction $T > 0$ , which is achieved by using TVD techniques to evaluate the variable T at zone boundaries. We also see that while the signal propagation in eqn. (2.19) depends on the sign of $d_z$ at the zone boundary, the two waves emanating from a zone boundary always travel in the same direction because T > 0. To evaluate the flux in eqn. (2.19) we, therefore, only need to use the monotonic temperature from the upwind direction. The above discussion makes it easy to see how one should evaluate the fluxes in eqn. (2.16). Thus by labeling zone centers by ( i, j, k) and the upper r-boundary in that zone by ( i+1/2, j, k) we get:

$$F_{sat,\, i+1/2,\, j,\, k} = -\, b_{r,\, i+1/2,\, j,\, k}\; 5\, \phi\, \rho_{i+1/2,\, j,\, k} \left( \frac{\gamma\, R}{\mu} \right)^{3/2} \operatorname{sgn}\left( \mathbf{b} \bullet \nabla\, T \right)_{i+1/2,\, j,\, k} \left( T_{i+1/2,\, j,\, k} \right)^{3/2} \qquad (2.20)$$

where

$$T_{i+1/2,\, j,\, k} = \begin{cases} T_{i,\, j,\, k} + 0.5\, \text{MinMod}\left( T_{i+1,\, j,\, k} - T_{i,\, j,\, k} \,,\, T_{i,\, j,\, k} - T_{i-1,\, j,\, k} \right) & \text{when } d_{x,\, i+1/2,\, j,\, k} \geq 0 \\ T_{i+1,\, j,\, k} - 0.5\, \text{MinMod}\left( T_{i+1,\, j,\, k} - T_{i,\, j,\, k} \,,\, T_{i+2,\, j,\, k} - T_{i+1,\, j,\, k} \right) & \text{otherwise} \end{cases} \qquad (2.21)$$

The MinMod function is a slope limiter that is traditionally used in TVD schemes. Other slope limiters may also be used. Notice from eqns. (2.16) and (2.20) that the discretization is conservative. The thermal conduction is treated in a fully implicit fashion while the MHD part is treated explicitly. In Sub-section II.3 we will present implicit-explicit (IMEX, henceforth) time-stepping strategies that allow one to retain second order accuracy while making such a split. The implicit thermal conduction operator is solved without any operator splitting between the classical and saturated fluxes.

## II.3) Time-Stepping Strategy

From eqns. (2.1) to (2.6) we see that the MHD parts are best solved with a time-explicit, second order accurate Runge-Kutta scheme. We want to retain that simplicity in our solution strategy even when the thermal conduction terms in eqn. (2.5) are included. However, the thermal conduction needs to be treated with a time-implicit or semi-implicit scheme to overcome time-step restrictions that might arise from a time-explicit treatment of parabolic terms. It is also advantageous to have an update strategy for the thermal conduction terms that is temporally second order accurate. For that reason, we draw on IMEX schemes that were introduced by Rosenbrock (1963) and further catalogued in the text of Dekker & Verwer (1984). The papers by Zhong (1996) and Verwer et al (1999) have also proved useful. Within the context of such schemes one formally writes eqns. (2.1) to (2.6) as:

$$\frac{\partial u}{\partial t} = f(u) + g(u) \tag{2.22}$$

Here "u" is an eight component vector that consists of the mass density, the three momentum densities, the total energy density and the three components of the magnetic field. The gradients of the MHD fluxes and the MHD source terms are written as f(u). The eight component vector g(u) has zeros in all its components but for the fifth, which consists of the divergence of the flux of thermal component. Only g(u) is treated implicitly while f(u) is treated explicitly. To build an IMEX scheme, we want to treat f(u) explicitly while treating g(u) implicitly. In Sub-section II.3.a we present an IMEX extension of the well-known Heun scheme. In Sub-section II.3.b we present a lesser-known but more powerful scheme known as the additive semi-implicit Runge-Kutta (ASIRK) scheme. In Sub-section II.3.c we present results of a von Neumann stability analysis of the competing schemes, which then permits us to choose one of them as being superior.

**II.3.a) IMEX Extension of the Heun Scheme**

A very simple, fully-implicit extension of the Heun scheme might be made as follows:

$$u^{**} = u^n + \frac{1}{2}\Delta t\Big( f(u^n) + g(u^{**})\Big) \tag{2.23}$$

$$u^{n+1} = u^n + \Delta t \Big(f(u^{**}) + g(u^{n+1}) \Big) \tag{2.24}$$

In the above equations, $u^n$ is the eight component solution vector at the beginning of the timestep, $u^{n+1}$ is the same at the end of the timestep and u** is the solution at an intermediate time level. Such a scheme, while fully implicit in the thermal conduction terms, is also only first order accurate in the time update from the g(u) term. Its temporal accuracy is, therefore, restricted to first order.

It is easy to make a second order semi-implicit extension of the previous scheme. It consists of the following scheme:

$$u^{**} = u^n + \frac{1}{2}\Delta t\left( f(u^n) + g(u^{**})\right) \tag{2.25}$$

$$u^{n+1} = u^n + \Delta t \left( f(u^{**}) + \frac{1}{2}g(u^{n+1}) + \frac{1}{2}g(u^n) \right) \tag{2.26}$$

The above semi-implicit scheme is temporally second order accurate. Von Neumann stability analysis also shows it to be unconditionally stable, as we will show in Sub-section II.3.c. It is helpful to make explicit the solution strategy for eqns. (2.25) and (2.26). Realize that for all but the fifth component of eqn. (2.25) it is possible to carry out a time-explicit update to get the density ρ** the velocity vector **v**** and the magnetic field vector **B**** . The fifth component of eqn. (2.25) can then be simplified to get an equation for the evolution of the temperature T** which can be written explicitly as:

$$\left[\frac{R\,\rho^{**}}{\mu(\gamma-1)}\right] T^{**} - \frac{\Delta t}{2}\,\nabla\bullet\left(\mathbf{F}^{**}_{net}\right) = u^n_5 - \rho^{**}\left(\mathbf{v}^{**}\right)^2/2 - \left(\mathbf{B}^{**}\right)^2/(8\pi) + \frac{\Delta t}{2}\, f_5(u^n) \tag{2.27}$$

Note that T** implicitly appears in $\mathbf{F}^{**}_{net}$ . Once T** has been evaluated using the Krylov Multigrid method, $u^{**}_5$ can be evaluated, completing the update of the fifth component of eqn. (2.25). This completes the first stage in the two-stage IMEX Runge-Kutta update.

Using all but the fifth component of eqn. (2.26) it is now possible to evaluate f(u**). We use f(u**) to carry out a time-explicit update and, therefore, get the density $\rho^{n+1}$ the velocity vector $\mathbf{v}^{n+1}$ and the magnetic field vector $\mathbf{B}^{n+1}$ . The fifth component of eqn. (2.26) can now be written as an implicit equation for $T^{n+1}$ to get:

$$\left[\frac{R\,\rho^{n+1}}{\mu(\gamma-1)}\right] T^{n+1} - \frac{\Delta t}{2}\,\nabla\bullet\left(\mathbf{F}_{net}^{n+1}\right) = u_5^n - \rho^{n+1}\left(\mathbf{v}^{n+1}\right)^2/2 - \left(\mathbf{B}^{n+1}\right)^2/(8\pi) + \Delta t\, f_5(u^{**}) + \frac{\Delta t}{2}\, g(u^n) \tag{2.28}$$

Once $T^{n+1}$ has been evaluated using the Krylov Multigrid method, $u_5^{n+1}$ can be evaluated, giving us the vector of flow variables at the next time level. This completes our description of the second order accurate, IMEX time-update strategy that is closest to the well-known Heun scheme.

### II.3.b) Additive Semi-implicit Runge-Kutta (ASIRK) Scheme

This scheme, as described in eqn. (28) of Zhong (1996) goes as follows:

$$\begin{aligned} k_1 &= \Delta t\left\{f\left(u^n\right) + g\left(u^n + a_1\, k_1\right)\right\} \\ k_2 &= \Delta t\left\{f\left(u^n + b_{21}\, k_1\right) + g\left(u^n + c_{21}\, k_1 + a_2\, k_2\right)\right\} \\ u^{n+1} &= u^n + \omega_1\, k_1 + \omega_2\, k_2 \end{aligned} \tag{2.29}$$

The coefficients are given by:

$$\omega_1 = \frac{1}{2} \;;\; \omega_2 = \frac{1}{2} \;;\; b_{21} = 1 \;;\; a_1 = a_2 = 1 - \frac{1}{\sqrt{2}} \;;\; c_{21} = \sqrt{2} - 1 \tag{2.30}$$

Eqn. (2.29) is difficult to interpret for a real astrophysical application, especially when the operator is not a simple diffusion operator and strong non-linearities are involved, so we do that for the reader. We form the intermediate variables u* and $u^\dagger$ as follows:

$$u^* = u^n + k_1 = u^n + \Delta t \left\{ f(u^n) + g(u^\dagger) \right\} \tag{2.31}$$

$$\begin{aligned} u^\dagger &= u^n + a_1\, k_1 \\ &= a_1\, u^* + (1 - a_1)\, u^n \\ &= u^n + a_1\, \Delta t \left\{ f(u^n) + g(u^\dagger) \right\} \end{aligned} \tag{2.32}$$

We see from the above two equations that as soon as $f(u^n)$ is evaluated, all but the fifth component of u* and $u^\dagger$ are known. Thus with the density $\rho^\dagger$, velocity vector $\mathbf{v}^\dagger$ and magnetic field vector $\mathbf{B}^\dagger$ known, our task is to evaluate the temperature $T^\dagger$ using the Krylov Multigrid method for the following equation:

$$\left[ \frac{R\, \rho^\dagger}{\mu(\gamma - 1)} \right] T^\dagger - a_1\, \Delta t\, \nabla \bullet \left( \mathbf{F}^\dagger_{net} \right) = u_5^n - \rho^\dagger \left( \mathbf{v}^\dagger \right)^2 / 2 - \left( \mathbf{B}^\dagger \right)^2 / (8\pi) + a_1\, \Delta t\, f_5(u^n) \tag{2.33}$$

Once $T^\dagger$ is known, all the components of the intermediate state u* can be evaluated using eqn. (2.31). This completes the first stage in the two-stage IMEX Runge-Kutta update.

Using u* we can now evaluate f(u*). As before, we define the auxiliary variable $u^/$ as:

$$\begin{aligned} u^/ &= u^n + c_{21}\, k_1 + a_1\, k_2 \\ &= u^n + c_{21} \left( u^* - u^n \right) + 2\, a_1 \left[ u^{n+1} - \frac{1}{2}\left( u^n + u^* \right) \right] \\ &= 2\, a_1\, u^{n+1} + \left( c_{21} - a_1 \right) u^* + \left( 1 - c_{21} - a_1 \right) u^n \\ &= u^n + c_{21} \left( u^* - u^n \right) + a_1\, \Delta t \left\{ f(u^*) + g(u^/) \right\} \end{aligned} \tag{2.34}$$

The last of the equations in eqn. (2.34) shows us that once f(u*) is evaluated, all components of $u^{/}$ except for the fifth component are known. Thus the density $\rho^{/}$ , velocity vector $\mathbf{v}^{/}$ and magnetic field $\mathbf{B}^{/}$ are known. The temperature $T^{/}$ can be found by using the Krylov Multigrid method for the following equation:

$$\left[\frac{R\,\rho^{/}}{\mu(\gamma-1)}\right] T^{/} - a_1\,\Delta t\,\nabla\bullet\left(\mathbf{F}^{/}_{net}\right) =$$
$$u_5^n + c_{21}\,(u_5^* - u_5^n) - \rho^{/}\left(\mathbf{v}^{/}\right)^2/2 - \left(\mathbf{B}^{/}\right)^2/(8\pi) + a_1\,\Delta t\,f_5(u^*) \qquad (2.35)$$

Using $T^{/}$ from the above equation, we can find $u^{/}$ . Using $u_5^{/}$ we can now find $u_5^{n+1}$ by using the relation

$$u_5^{n+1} = \frac{1}{2\,a_1}\left[u_5^{/} - \left(c_{21} - a_1\right)u_5^* - \left(1 - c_{21} - a_1\right)u_5^n\right] \qquad (2.36)$$

This completes our description of the second order accurate, ASIRK time-update strategy.

**II.3.c) vonNeumann Stability Analysis**

vonNeumann showed that numerical schemes for solving partial differential equations can be analyzed for their stability as well as their ability to reproduce the results of the actual partial differential equation (PDE). It is not always possible to formulate such an analysis when the PDE does not have a beneficial structure or when the coefficients have strong non-linearities. Examining eqns. (2.7) and (2.8) we see that the coefficients in our problem do have strong non-linearities. Examining eqn. (2.13) shows that our PDE does not have a simple structure. It is, however, possible to make the simplifying assumption that in the non-magnetic limit the thermal conduction operator is isotropic and, therefore, simple. We also assume that instead of having coefficients in eqn. (2.7) that are strongly dependent on temperature, we can assume a simple, constant value K for isotropic thermal conduction. With this simplifying assumption we get $\mathbf{F}_{net} = \mathbf{F}_{class} = -K\,\nabla T$ . We also assume that the hyperbolic part f(u) in eqn. (2.22) can be set to zero. We further assume that only the temperature has one-dimensional fluctuations that are of the form $e^{i k x}$ . vonNeumann stability analysis then permits us to evaluate the amplification of these fluctuations when the actual PDE is used. It also enables us to evaluate the amplification of these fluctuations when the IMEX Heun scheme from Sub-section II.3.a is used and when the ASIRK scheme from Sub-section II.3.b is used. The factor by which the fluctuations are amplified is known as the amplification factor. The detailed formulae for the amplification factors are given in Appendix B.

To carry out a vonNeumann stability analysis, we assume a mesh with zone size $\Delta x$ and a time-step $\Delta t$. The dimensionless variable $\mu = K\,\Delta t / \Delta x^2$ parametrizes the size of the timestep $\Delta t$, with $\mu = 1$ being the limit for an explicit scheme. As good implicit scheme should remain stable for all values of $\mu$ . We wish to study the amplification factor as a function of ($k\,\Delta x$) and for $\mu$ = 0.5, 4.0, 10.0 and 50.0 . These values of $\mu$ give us insights that are applicable to the full range of $\mu$ , which can extend from zero to infinity. Figs. 1a, 1b, 1c and 1d show the amplification factor for $\mu$ = 0.5, 4.0, 10.0 and 50.0 respectively as a function of ($k\,\Delta x$). The solid line shows the exact result from the PDE, the dashed line shows the amplification factor for the ASIRK scheme from Sub-section II.3.b and the dotted line shows the amplification factor for the IMEX Heun scheme from Sub-section II.3.a. From Fig. 1a we see that for small values of $\mu$ the ASIRK and IMEX Heun schemes both track the exact amplification factor from the PDE quite well, indicating that close to the explicit limit, all implicit schemes also produce good results. From Fig. 1b we see that for $\mu = 4.0$, which is just a somewhat larger than the explicit limit for $\mu$, the IMEX Heun scheme shows a larger deviation from the exact amplification factor from the PDE, whereas the ASIRK scheme does quite well. From Figs. 1c and 1d we see that for even larger values of m the ASIRK scheme continues to track the exact amplification factor from the PDE, while this is not so for the IMEX Heun scheme. We see therefore, that while the IMEX Heun scheme is easier to code up, the ASIRK scheme from Sub-section II.3.b is the scheme of choice for the full range of values of $\mu$. We note though that all schemes presented in Sub-section II.3 are indeed unconditionally stable for all values of $\mu$.

## III) Code Tests

In this Section we describe three sets of code tests. The first code test pertains to the Field instability in the presence of TC (Field 1965). The second test evaluates the code's ability to anisotropically conduct thermal energy along field lines. The third test compares our results of a supernova exploding in a uniform medium to those of Cioffi, McKee & Bertschinger (1988, CMB henceforth).

### III.1) Field Instability

Field (1965) evaluated the growth rate for thermal instabilities in the presence of thermal conduction, atomic cooling and diffuse heating. This was done in the limit of a constant coefficient for thermal conduction, K. We use his results to test our code. The problem consists of an initially stationary patch of the interstellar medium with a mean density $\rho_0$ = 1 amu/cm and a mean temperature $T_0$ = 2531.65 K. The diffuse heating is made to balance radiative cooling in the unperturbed medium. The cooling function was taken from Wolfire et al (2003). The parameters chosen correspond to a medium that is susceptible to thermal instability with this particular choice of cooling function. We imposed sinusoidal fluctuations with wavenumber *k* and growth rate "n" in the pressure, density and velocity in one-dimensional slab geometry. The fluctuations were of the form:

$$\rho(x,t) = \rho_0 + \rho_1 \, e^{i k x + n t}$$
$$P(x,t) = P_0 + P_1 \, e^{i k x + n t}$$
$$v_x(x,t) = v_{x,1} \, e^{i k x + n t}$$

The above fluctuations can be substituted into the hydrodynamic equations and a dispersion relation obtained for them, as was done by Field (1965). The dispersion relation is given by:

$$n^3 + n^2(\gamma\text{-}1)\frac{T_0}{P_0}\left(\rho_0^2\frac{\partial\Lambda(T)}{\partial T} + K\,k^2\right) + n\frac{\gamma P_0 k^2}{\rho_0} + k^2(\gamma\text{-}1)\left(T_0\rho_0\frac{\partial\Lambda(T)}{\partial T} + \frac{K\,k^2 T_0}{\rho_0} - \rho_0\Lambda(T)\right) = 0$$

The above dispersion relation gives growing modes for certain combinations of density, temperature, wavenumber and conduction coefficient and we use those growing modes to test our numerics. The dispersion relation can be solved analytically and an eigenvector of fluctuations can be obtained for each combination of n and $k$ . The resulting evolution of eigenmodes in the linear regime is given by:

$$\rho(x,t) = \rho_0 + \rho_1\, e^{n\,t} \cos(k\,x)$$

$$P(x,t) = P_0 - \rho_1 \frac{n^2}{k^2} e^{n\,t} \cos(k\,x)$$

$$v_x(x,t) = -\rho_1 \frac{n}{\rho_0\,k} e^{n\,t} \sin(k\,x)$$

The above fluctuations were initialized with a 5% fluctuation in the density. For each choice of wavenumber we ensured that we resolved each wavelength with 100 zones. The system of equations was then evolved by our numerical code. The subsequent value for the r.m.s. fluctuation in the density as a function of time was used to obtain the numerically generated growth rates. Fig. 2 shows the numerically generated growth rates for various values of wavenumber $k$ and three values of the thermal conduction coefficient K. The growing modes from the dispersion relation are also shown for comparison. We observe that in each case the numerical code was able to match the analytic calculation of the growth rate to better than 7.4% in all cases.

**III.2) Anisotropic Thermal Conduction Along Magnetic Field Lines**

The conduction of heat along a closed loop of magnetic field presents another good test of anisotropic thermal conduction. Such a test has been examined in detail by Sharma & Hammet (2007) and is based on earlier work by Parrish & Stone (2005). The problem is discretized on a two-dimensional Cartesian mesh spanning [-1,1]X[-1,1] with uniform zones in the x and y directions. The initial magnetic field was specified by a magnetic vector potential $A_z$ (x,y) = $(x^2 + y^2)^{1/2}$ . The temperature follows the specification of Parrish & Stone (2005) and is given by:

$$T(r,\theta) = \begin{cases} T^* & \text{if } (0.5 \le r \le 0.7) \text{ and } \left( \frac{11\pi}{12} \le \theta \le \frac{13\pi}{12} \right) \\ T_0 & \text{otherwise} \end{cases}$$

With $T_0$ = 10 and $T^*$ = 12 . A constant, unsaturated thermal conduction was used for this problem and the equation that was evolved was given by:

$$\frac{\partial T}{\partial t} = - \nabla \bullet \mathbf{F}_{class} \quad \text{with} \quad \mathbf{F}_{class} = - \eta_0 \, \mathbf{b} \left( \mathbf{b} \bullet \nabla T \right)$$

and $\eta_0$ = 0.01. The dynamical evolution is suppressed so that we only solve for the anisotropic heat conduction equation. The initial pulse of heat flows along the magnetic field lines till the temperature reaches a steady state where it becomes isothermal along field lines. In practice, we find that a final time of 200 code units is adequate for achieving steady state and we use that as our final time in the simulation.

Table 1 shows the L1 and L2 norms for the temperature variable as well as $T_{max}$ and $T_{min}$, the maximum and minimum values of the temperature at the stopping time of 200 time units. The results are shown from simulations that were run on meshes with 32, 64, 128, 256 and 400 zones on a side. Similar results have been presented by Sharma & Hammet (2007) and we see that our results are competitive with their best-case results on comparable meshes. Fig. 3 shows the temperature at the final time on a 256X256 zone mesh at a time of 300 code units. We see that the temperature is isothermal along circular rings, as expected.

### III.3) Comparison of the Radius-Time Relationship for a SNR with the Semi-Analytic Formulation of CMB

CMB presented a semi-analytic formulation for the evolution of a SNR in a uniform, unmagnetized medium. The variation of the radius of the outer shock with time, known as the r-t relationship, was also catalogued in CMB. Matching this r-t relationship with our code is a strong test of our numerics. For this test, we used a 200X200 zone mesh in cylindrical geometry. The zones were taken to have the same length and the computational domain covered a 100X100 pc region along the r and z axes. The computational domain was initialized with a uniform density of 1 amu/cm$^3$ and a temperature of 10,000 K. The central 5 pc were imparted an energy of $10^{51}$ ergs. Two-thirds of that energy was in the form of kinetic energy and one-third of it was in the form of thermal energy. Heating and cooling as described in Sub-section II.1 were used. In keeping with CMB's formulation, thermal conduction was neglected for this test problem. Fig. 4 shows the r-t relationship obtained from our numerical code, shown as a solid line, along with the predicted values from eqns. (3.26) and (3.32a) of CMB, shown as a dashed line. Fig. 4 shows that the mean difference between the analytic curve from CMB and the numerical results is less than 2.5%.

## IV) Description of Models, Limitations and Resolution Study

Sub-section IV.1 describes the models in some detail, sub-section IV.2 describes the limitations of the present models and sub-section IV.3 presents a convergence study.

### IV.1) Description of Models

Since these are the first of the multi-dimensional simulations that include anisotropic thermal conduction, we decided that it is important to carry out a large parameter survey with a somewhat limited input physics set. Thus the simulations are all 2.5 dimensional, though the code is three-dimensional. In this section, we describe the simulation setup as well as the effects that are excluded in this work.

The simulations were carried out on a cylindrical mesh with the toroidal direction suppressed. Since the outer shock plays a role in confining the hot gas bubble, we chose a computational domain that captured the shock through the duration of the simulation. Since the outer shock propagates at different speeds in ISMs with different densities, pressures and magnetic fields, this required choosing a different value for $R_{outer}$ , the outer boundary of the computational domain in the r and z directions. The zones were uniformly spaced along the r and z axes. The reverse shock in these SNR simulations has to bounce at the origin. In most instances the outer shock's structure can show mesh imprinting if the reverse shock's bounce at the origin is captured improperly. For that reason, $N_{zones}$ , the number of zones in the simulations was always chosen to overcome this mesh imprinting. The boundary conditions at the axis and the equator were chosen to be reflective and the outer boundaries were chosen to be continuative. In each case, the simulations were run until the hot gas bubble collapsed or until a final time of 10Myr was reached. The final time $t_{final}$ , out to which the simulations were run is also catalogued.

The ambient ISM was always chosen to be quiescent with a constant density $\rho_{ism}$ , a constant temperature $T_{ism}$ and a constant magnetic field $B_{ism}$ . The SNR was initialized as a sphere with a 10 zone radius on the computing grid. For an SNR propagating into a standard ISM, this corresponds to an ~ 80 year old remnant. The density in the SNR was initially fixed at the interstellar density. In each case, $10^{51}$ ergs was injected in the zones that were demarcated as initially belonging to the SNR. 1/3 of that energy was injected as thermal energy with the remaining 2/3 being kinetic. The velocity profile in the SNR was initially set to be linear in the radial coordinate. Radiative cooling from MacDonald & Bailey (1981) was used in all the runs. Some of the runs were repeated without TC to enable us to assess the role that TC plays in the simulations. Table 2 shows the parameters used for the runs.

The nomenclature of the runs is as follows: VL0, VL2, VL6 refer to very low density runs with $\rho_{ism}$ = 0.2 amu/cm$^3$ , magnetic fields of 0, 2 and 6 μGauss respectively and $10^{51}$ ergs of energy in the initial SNR. L0, L2, L6 refer to low density runs with $\rho_{ism}$ = 0.7 amu/cm$^3$ , magnetic fields of 0, 2 and 6 μGauss respectively and $10^{51}$ ergs of energy in the initial SNR. We also refer to these as the fiducial runs because they are done with ISM parameters that are currently considered standard. I0e1, I2e1, I6e1 refer to intermediate density runs with $\rho_{ism}$ = 1.0 amu/cm$^3$ , magnetic fields of 0, 2 and 6 μGauss respectively and $10^{51}$ ergs of energy in the initial SNR. H0, H6, H20 refer to high density runs with $\rho_{ism}$ = 5.0 amu/cm$^3$ , magnetic fields of 0, 6 and 20 μGauss respectively and $10^{51}$ ergs of energy in the initial SNR. VH0, VH6, VH20 refer to very high density runs with $\rho_{ism}$ = 20.0 amu/cm$^3$ , magnetic fields of 0, 6 and 20 μGauss respectively and $10^{51}$ ergs of energy in the initial SNR.

**IV.2) Limitations of the Models**

The physical limitations in these simulations and their consequences are as follows:

1) A single fluid approximation is used. As a result, the electron and ion temperatures are not decoupled. As shown by Cui & Cox (1992), this is not a bad approximation for the evaluation of the outer shock and hot gas cavity, especially when TC is used. For our choice of parameters, the neutral component is not significant but could be so if lower temperatures are used.

2) Our MHD approximation is based on collisional shocks. The mean free path in the ISM is rather large, resulting in collisionless shocks. However, MHD is a very useful working approximation for this work. Plasma effects can introduce larger fluctuations in the post-shock magnetic field as shown by Bell & Lucek (2001) but we do not resolve such effects here.

3) We do not include cosmic ray acceleration. Since up to 20% of the explosion energy might go into the acceleration of cosmic rays, Jun & Jones (1999), this would diminish the SNR expansion by that factor. Cosmic ray acceleration can also change the compression ratio in the outer shock, making it prone to Rayleigh Taylor instability as shown by Blondin & Ellison (2001). However, the fraction of the explosion energy that is imparted to the cosmic rays depends strongly on the details of the ISM turbulence and the structure of the magnetic field within it and we do not include those details in this study.

4) The effect of interstellar turbulence has not been included. The turbulence could cause the ISM to be clumpy and we implicitly assume that if the ISM is clumpy, the clumps are on such small scales that they can be modeled with an ISM with a higher mean density and lower mean temperature. See Balsara, Benjamin & Cox (2001) for simulations that include ISM turbulence.

5) Our ejecta do not include metallicity effects from the exploding star. Kosenko (2006) shows that these do have an effect on the youngest SNRs. Once the Sedov phase has been reached, the ejecta have swept up an amount of interstellar matter that exceeds the mass in the ejecta themselves. As a result, we may be able to assume that the mean metallicity of the SNR is the metallicity of the ISM.

6) The pre-processing of the circumstellar medium, see Chevalier & Liang (1989), has been neglected. Dwarkadas (2005) has included such effects and they may be important during the early stages in the evolution of SNRs with core-collapse progenitors. This would be especially true for massive progenitors which blow off an extensive, dense circumstellar medium. It would not be important for SNRs with less massive core-collapse progenitors as well as for SNRs with white dwarf progenitors.

7) Dust and its effects have not been included. Dust could affect the cooling of the hot gas. Furthermore, dust can contain a significant fraction of interstellar silicon suppressing $Si^{+3}$ emission until the grains are destroyed.

8) We do not include non-equilibrium cooling. SC92, SC93 show the importance of non-equilibrium cooling in predicting the abundances of high-stage ions. This is a serious deficiency and will be rectified in a subsequent paper.

9) The ionizing radiation from a SNR shock could pre-ionize the interstellar medium that lies in front of the shock, see Chevalier & Dwarkadas (1995). Such an ionizing front of radiation could propagate a few pc into the ISM, making it important to include radiative effects.

### IV.3) Resolution Study

The simulations presented here are most interesting in determining the hot gas content of the ISM and also the x-ray emission characteristics of the SNR, see TB and TBH. TB showed that when predicting the fraction of gas that emits in high-stage ions, the volume of the gas in the SNR with temperatures in excess of $3X10^5$ K, $7.9X10^5$ K and $2.2X10^6$ K are most interesting. This is because these temperatures characterize various ionization stages of Oxygen that are observable by *FUSE* and *Chandra*. The numerical effects that could most likely influence the results are : a) the mesh size of the computation and b) the number of zones, in any one direction, over which the SNR is initialized. For that reason, we show in this sub-section that the evolution of the hot gas as a function of time is independent of both the above-mentioned effects.

We set up multiple copies of the run L2 with $R_{outer}$ = 200 pc on meshes having $N_{zones}$ = 192, 256 and 384. For the run with $N_{zones}$ = 192 zones we initialized the same supernova energy over 6 and 12 zones. Similarly, for the run with $N_{zones}$ = 256 zones we initialized the same supernova energy over 8 and 12 zones. For the run with $N_{zones}$ = 384 zones we initialized the same supernova energy over 12 zones. The runs with $N_{zones}$ = 192 and 256 with 12 initial zones in the SNR correspond to SNRs that are initially a little larger than the other runs and so we make a small allowance for them being marginally more evolved. As a point of reference, we also show the same run without TC on a mesh with $N_{zones}$ = 384 zones where we initialized the same supernova energy over 12 zones. Figs. 5a, 5b and 5c show a resolution study for the evolution of the volume of gas with T > $3X10^5$ K, T > $7.9X10^5$ K and T > $2.2X10^6$ K respectively as a function of time. We see that past the initial timestep, the volume of hot gas for all the runs that do include TC is convergent at all times. The only small differences we observe arise in Fig. 5b corresponding to T > $7.9X10^5$ K. The bounces that are observed for times > 500 kyrs correspond to the collapse of the hot gas bubble, which gets periodically shocked as the interstellar pressure squeezes the hot gas bubble. The temperature of $7.9X10^5$ K is particularly sensitive because this is also the mean temperature of the hot gas bubble at late epochs.

We also set up multiple copies of run H6 with $R_{outer}$ = 200 pc on meshes having $N_{zones}$ = 192, 256 and 384. For the run with $N_{zones}$ = 192 zones we initialized the same supernova energy over 6 and 12 zones. Similarly, for the run with $N_{zones}$ = 256 zones we initialized the same supernova energy over 8 and 12 zones. For the run with $N_{zones}$ = 384 zones we initialized the same supernova energy over 12 zones. As a point of reference, we also show the same run without TC on a mesh with $N_{zones}$ = 384 zones where we initialized the same supernova energy over 12 zones. Figs. 6a, 6b and 6c show a resolution study for the evolution of the volume of gas with T > 3X$10^5$ K, T > 7.9X$10^5$ K and T > 2.2X$10^6$ K respectively as a function of time. Even for this case we confirm that past the initial timestep, the volume of hot gas for all the runs that do include TC is convergent at all times. As before, the temperature of 7.9X$10^5$ K is particularly sensitive because this is also the mean temperature of the hot gas bubble at late epochs.

Since Figs. 5 and 6 also include the cases without TC, we see that the exclusion of thermal conduction causes the largest departures in the evolution of the hot gas especially for T > 7.9X$10^5$ K and T > 2.2X$10^6$ K. Furthermore, this departure sets in rather early, i.e. within 0.15 Myrs and 0.1 Myrs for the hottest gas in L2 and H6 respectively. It has been argued, Avillez & Breitschwerdt (2005), that the turbulent diffusivity of the ISM would disperse the hot gas bubbles of SNRs in less time than thermal conduction can operate, making the inclusion of thermal conduction inessential. TB and TBH show otherwise using filling factors and emission maps of simulated SNRs. Here we present an even stronger demonstration. The maximum diameters of the hot gas bubbles in Figs. 5 and 6 are ~120 pc and ~60 pc respectively. Using measures for the turbulent diffusivity of the ISM, see Balsara & Kim (2005) or Avillez & Mac Low (2002), we estimate the turbulent diffusivity of the ISM as being a small multiple of $10^{26}$ $cm^2$ / sec and we use 2X$10^{26}$ $cm^2$ / sec here. As a result, the hot gas bubbles will undergo turbulent dispersal in a small multiple of the diffusion time. For Figs. 5 and 6 the turbulent diffusion time is given by 5.2 Myr and 1.36 Myr respectively which is comparable to the life times of the bubbles.

## V) Evolution and Morphology of SNRs in the Presence of Thermal Conduction

In this Section we describe the results of our SNR simulations. Sub-section V.1 explains the role of TC in the evolution of SNRs and describes their dynamical evolution without and in the presence of anisotropic thermal conduction. Sub-section V.2 describes the evolution of various flow variables as a function of time. Sub-section V.3 makes a case study of mixed-morphology remnants.

### V.1) Dynamical Evolution of SNRs with Anisotropic Thermal Conduction

In this Sub-section we study the temperature and density evolution for various models of interest, thereby gaining insight into the dynamics of SNRs with TC. Since TB and TBH concentrated on early evolution of the outer shocks and hot gas bubbles, we orient a good fraction of this sub-section towards studying the somewhat later time evolution of the SNRs' evolution.

Fig. 7 shows density in color with velocity vectors overlaid and temperature in color with magnetic field lines overlaid for run VL2. Figs. 7a and 7b correspond to the density and temperature respectively without TC at 100 kyr while Figs. 7c and 7d show the same with TC at the same epoch. Figs. 7e and 7f show the density and temperature respectively without TC at 7.5 Myr while Figs. 7g and 7h show the same with TC at the same epoch. The tick marks correspond to 50 pc intervals. Comparing Figs. 7a with 7c for the density we see that the location of the outer shock in the equatorial direction is the same and is independent of the inclusion of TC. Figs. 7e and 7g show the same trend at late epochs. This can be understood by realizing that TC plays no role in the energy transport perpendicular to the direction of the magnetic field. In the poloidal direction, however, we see that the outer shock in Fig. 7c has propagated a little further than the outer shock in Fig. 7a. This can be understood by the fact that TC does play a role in the propagation of thermal energy along the magnetic field. Since the original magnetic field was parallel to the z-direction, TC introduces a small excess in the flow of thermal energy in that direction, causing the outer shock in Fig. 7c to propagate a little further along the z-axis than the outer shock in Fig. 7a. An examination of Figs. 7e and 7g shows that by late times, the outer shocks have propagated almost the same distance in the z-direction. This is because the late-time propagation of the outer shock takes place when the shock is in the snowplow phase, by which time the temperature in the dense post-shock shell is rather low. The coefficient of thermal conduction, which varies as $T^{2.5}$, is therefore much reduced. Thus the late-time evolution of the outer shock in both the r and z-directions is independent of the role of thermal conduction. Turning attention to the temperatures in Figs. 7b and 7d, we see that the temperature of the hot gas bubble in Fig. 7d is lower by more than half an order of magnitude. The energy transport that takes place due to TC in the gas that lies within the outer shock plays an important role in reducing the temperature in the hot gas bubble. As we will see in Sub-section IV.2, the thermal + magnetic pressure in the hot gas bubble evolves isobarically in the post-shock region. This is especially true after the onset of the snowplow phase. Hence a reduced temperature corresponds to an increased density. Because the emissivity is proportional to the square of the density and also because plasmas become strongly radiative for temperatures below $10^7$ K, the simulation with TC can radiate more efficiently, thus enabling it to cool faster. Figs. 7e to 7h show the late time evolution of the hot gas bubble. We see that the magnetic field has caused the hot gas bubble to become very elongated at late times. Fig. 7h shows that at late times most of the hot gas bubble has become almost an order of magnitude cooler than the one in Fig. 7f. Figs. 7g and 7h also show a very interesting epoch in the collapse of the hot gas bubble. We see that the plasma that is causing the hot gas bubble to collapse has also shock heated a small part of the hot gas bubble close to the equator. While such episodes of reheating can also be seen at later epochs in Figs. 5 and 6, it is interesting to notice that the reheating is particularly vigorous when the ISM density is low, as it is for run VL2.

Fig. 8 shows density in color with velocity vectors overlaid and temperature in color with magnetic field lines overlaid for run L2. Figs. 8a and 8b correspond to the density and temperature respectively without TC at 400 kyr while Figs. 8c and 8d show the same with TC at the same epoch. Figs. 8e and 8f show the density and temperature respectively without TC at 1.5 Myr while Figs. 8g and 8h show the same with TC at the same epoch. Figs. 8i and 8j show the distribution of the ejecta at 400 kyr and 1.5 Myr in the version of the run L2 that included TC. The tick marks correspond to 50 pc intervals. TB showed a similar inter-comparison at 60 kyr for a run with almost the same parameters, the only difference being a small change in the magnetic field strength. Since the run L2 also pertains to a rather low density, corresponding to our present Galactic

ISM's fiducial parameters, we see that it shows many of the same trends as VL2. The higher mean density in run L2 slows down the expansion of the outer shock so that in 400 kyr the outer shock has expanded less than the outer shock in run VL2 has expanded in 100 kyr. By comparing Fig. 8a to Fig. 8c and likewise Fig. 8b to Fig. 8d we realize that the temperature is much lower in the hot gas bubble and the density much higher when TC is included. Furthermore, with increasing ISM density we see that TC causes a larger change in the temperature of the hot gas bubble for a longer duration in the SNR's evolution, a trend also observable in Fig. 2b of TB. This trend can be understood by realizing that a SNR expanding into a denser environment enters its snowplow phase earlier. The colder shell of post-shock gas, therefore, provides a large reservoir of dense, lower temperature gas. When the SNR initially enters the snowplow phase, the temperature in the post-shock shell is not so low as to totally impede thermal conduction. As a result, a significant fraction of the thermal energy in the hot gas bubble can be carried by TC to the dense shell, which radiates it away very efficiently. Figs. 8e through 8h also show the same trends at a much later time. Comparing Figs. 8f and 8h we see that the temperature at later epochs in the evolution of the hot gas bubble can be lower when TC is included by more than an order of magnitude.

Figs. 8i and 8j also show a very interesting structure. Since those figures trace the ejecta, we see very clear evidence for Rayleigh Taylor fingering in the ejecta. We see that the Rayleigh Taylor fingering in the ejecta is such that some of the ejecta may come very close to the outer shock. While this may not have much of an effect in ISMs with current metallicities, the metal-bearing ejecta can play a substantial role in proto-galactic environments. Salvaterra, Ferarra & Schneider (2004) have presented one of the leading models for forming extremely metal poor low mass stars in environments that are dominated by SNe from Population III stars. The model calls for gravo-thermal instabilities in the dense shells of primordial SNRs. For the postulated thermal instabilities to work in the dense shells, one requires a good fraction of the metal-rich ejecta to be mixed in with the gas in the shell. It is only via mixing that the post-shock gas can be made metal rich, a result not achieved in a perfectly spherical calculation. The Rayleigh Taylor fingering, which indeed reaches all the way to the outer boundary of the shocked gas in Fig. 8i, gives us one possible way of understanding how the requisite mixing of metal-rich ejecta into the post-shock primordial gas might be achieved in the model of Salvaterra, Ferarra & Schneider (2004). A similar process could also occur in the SNR shock-triggered model for low mass star formation presented by Elmegreen (1994).

It is also very interesting to compare Fig. 8i to Fig. 8d and similarly to compare Fig. 8j to Fig. 8h. We see that the lines of magnetic field indeed wrap around the Rayleigh Taylor fingers in the ejecta. Thus the ejecta sculpt the structure of the magnetic field lines in the hot gas bubbles. The magnetic fields, in turn, guide the flow of heat in the hot cavities. While the temperature in the hot gas bubbles is relatively uniform, we also observe small fluctuations in temperature that are bounded by the kinks in the magnetic fields. Consequently, we see that parcels of gas within the hot gas bubble that are connected by the same field line can rapidly exchange thermal energy and can, therefore, reach the same temperature.

Fig. 9 shows density in color with velocity vectors overlaid and temperature in color with magnetic field lines overlaid for run H6. Figs. 9a and 9b correspond to the density and temperature respectively without TC at 200 kyr while Figs. 9c and 9d show the same with TC at the same epoch. Figs. 9e and 9f show the density and temperature respectively without TC at 1.5 Myr while Figs. 9g and 9h show the same with TC at the same epoch. The tick marks correspond to 50 pc intervals. Run H6 enters its snowplow phase very rapidly, soon after the first 20 kyr. As a result, Fig. 9d shows us that the hot gas bubble has cooled down very dramatically compared to Fig. 9b where TC was excluded. Figs. 9c and 9d also show a faint, inward-going shock that is being driven into the hot gas bubble at later times. Such shocks can be seen in all the simulations and the time in Figs. 9c and 9d was specifically chosen to highlight one such episode. Comparing Figs. 9c and 9d to Figs. 9g and 9h shows that the hot gas bubble is already collapsing by 1.5 Myr in this case where the ISM is very dense. Figs. 9g and 9h also show evidence for the Rayleigh Taylor fingering that was so dramatically visible in Figs. 8g through 8j. The hot gas bubble in Fig. 9h is not as elongated in the direction of the magnetic field as the hot gas bubble in Fig. 7h because the magnetic pressure does not dominate the gas pressure in run H6, while the converse is true for run VL2.

In this paragraph we describe results from run VH6. This corresponds to a very high density and very high pressure ISM, probably more representative of the ISMs in starburst galaxies like M82, Pedlar, Muxlow & Wills (2003), Rieke et al (1980) and Blom, Paglione and Carramiñana (1999). Since this class of simulation has not been reported in any of the previous papers in this series, i.e. TB or TBH, we show the data from this simulation at 20 kyr and 60 kyr. We will show in Sub-section IV.3 that such earlier epochs might be interesting in understanding the formation of MM SNRs in extreme environments. Fig. 10 shows density in color with velocity vectors overlaid and temperature in color with magnetic field lines overlaid for run VH6. Figs. 10a and 10b correspond to the density and temperature respectively without TC at 20 kyr while Figs. 10c and 10d show the same with TC at the same epoch. Figs. 10e and 10f show the density and temperature respectively without TC at 60 kyr while Figs. 10g and 10h show the same with TC at the same epoch. The tick marks correspond to 10 pc intervals. Run VH6 enters its snowplow phase very rapidly, soon after the first ~10 kyr. As a result, Fig. 10d shows us that even at this extremely early epoch the hot gas bubble has cooled down quite substantially compared to Fig. 10b where TC was excluded. We would, therefore, expect that run VH6 should be center-bright in x-rays at a very early age, an expectation that will be borne out in Sub-section V.3. Figs. 10e to 10h show that at somewhat later stages in the evolution of SNRs in very dense environments, the inclusion of TC causes the temperature of the hot gas bubble to be lower and the density to be higher. This is in keeping with the trend reported in our description of Figs. 7 through 10.

### V.2) Evolution of Flow Variables with Time

Figs. 11a to 11f show the density, temperature, r-velocity, magnetic pressure, thermal pressure and total (thermal + magnetic) pressure respectively as a function of radial coordinate in the equatorial plane at various times for run L2. Fig. 11g shows the evolution of the thermal energy, kinetic energy and magnetic energy as a function of time for run L2. We get qualitative agreement with the models of SC92; however, we see a lot of additional structure not apparent in those results that are due to the anisotropy that develops in two dimensions. We see that the shock wave strengthens as the shell enters the radiative stage, accompanied by a decrease in the temperature of the gas passing through the shock. The interior of the remnant has a largely uniform density and temperature profile, even though the magnetic field geometry prevents TC from acting directly in this direction. As a consequence, the interior of the remnant evolves largely isobarically. The high temperatures in the interior of the remnant cause the thermal pressure to dominate the total pressure throughout the majority of the remnant, and as a consequence the total pressure traces the thermal pressure quite well.

The magnetic field inside the remnant is initially fairly uniform as the bulk of the field is swept up by the outer shock, and follows the density structure closely. At later times, a complicated magnetic structure develops as compression and rarefaction waves in the interior of the remnant reorder the field, as seen in Fig. 11d. We see further signatures of these waves in the radial velocity field, see Fig. 11c. A smoothly expanding remnant would have a radial velocity profile that increases with radius; however, we see at all times deviations from this profile due to waves bouncing through the interior of the remnant superimposed on this smooth flow.

As stated in Section III.1, the initial supernova had a total energy of $10^{51}$ ergs, distributed such that one-third was contained in thermal energy and two-thirds in kinetic energy. As demonstrated in Fig. 11e, the interior pressure remains largely uniform, or even increases at early times, as the remnant expands in its Sedov phase. The additional thermal energy required for this configuration comes at the expense of the initial kinetic energy, with the conversion arising due to compressive heating of the ISM gas as it passes through the shock. We can trace this evolution in Fig. 11g, which plots the various energy components of the hot gas as a function of time. We see that the initial kinetic energy has quickly been converted into thermal energy, such that the thermal energy contains nearly the entire bulk of the initial $10^{51}$ ergs. During the Sedov phase the magnetic field is swept up by the shock; since the magnetic energy varies as the square of the magnetic field, the effect of this concentrated magnetic field is to increase the magnetic energy content with time. Only after the onset of the radiative snowplow phase, at around 100 kyr, does the thermal energy of the hot gas bubble begin to decrease. After the outer shock becomes radiative, we refer back to Fig. 11d to show that the bulk of the magnetic field is found in the cooling shell that is formed between the hot gas bubble and the outer shock. As a consequence, the magnetic energy content of the hot gas bubble decreases more quickly than the thermal and kinetic components at the beginning of the snowplow phase. At late stages, however, the magnetic energy begins to increase again due to the contraction of the hot gas bubble, as seen by comparing the magnetic pressure at 3 Myr to the magnetic pressure at 1 Myr in Fig. 11d. Because the temperature and thermal pressure remain largely constant at these times, the decrease in the volume of the hot gas bubble leads to a continued decrease in the thermal energy as well. We see the signature of waves being excited in the interior of the hot gas bubble at late times in the kinetic energy as a series of small peaks in the kinetic energy of the bubble.

### V.3) A Case Study of Mixed Morphology Remnants

Figs. 12a and 12b show the simulated x-ray brightness in soft (300-800 eV) and hard (1-5 keV) x-rays for run VL2 at a time of 20 kyr while Figs. 12c and 12d show the same for a time of 60 kyr. Figs. 12e, 12f, 12g and 12h show analogous information for run VH6. Analogous information for runs L2 and H6 is shown in TBH. Those authors found that run L2 could become centre-bright in hard x-rays after 60 kyr, but remained shell-bright in soft x-rays at these times. They also found that run H6 would become centre-bright in hard x-rays after only 20 kyr, while it was still shell-bright in soft x-rays; however, after 60 kyr this remnant was centre-bright in both soft and hard x-rays. As a consequence, TC acting alone in a magnetized medium could reproduce the observed morphological structures seen in mixed-morphology remnants (e.g. Jones et al. 1998). We expand on those results here by considering the expansion of remnants into even more extreme environments.

For run VL2, we can see many of the same trends that TBH found for run L2. In this case the remnant remains shell-bright in both soft and hard x-rays 20 kyr after the supernova event, as illustrated in Figs. 12a and 12b. However, even after 60 kyr the remnant is still shell-bright in both of these x-ray energies, a clear departure from the results of TBH. This can be understood by realizing the evolution timescale is related to onset of the snowplow phase. As the density is significantly lower in run VL2, the timescale for its evolution is much larger. Consequently, the remnant is kinematically younger at 60 kyr than a supernova remnant expanding into a denser ISM would be. The other extreme can be found in Figs. 12e-12h, with run VH6 expanding into a very dense ISM. While this remnant has not quite become centre-bright in soft x-rays after 20 kyr, we can see signs of the transition from shell-bright to centre-bright due to the relatively low contrast between the brightest parts of the shell and the centre. In hard x-rays this remnant is clearly centre-bright, as we would expect extrapolating from TBH. Similarly, Fig. 12g and Fig. 12h show that this remnant is centre-bright in both x-ray bands at an age of 60 kyr.

## VI) Conclusions

We have presented a numerical formulation of anisotropic thermal conduction that correctly accounts for the classical and saturated character of the TC operator. Furthermore, we have introduced a temporally second-order accurate implicit-explicit scheme for the time update of the TC terms in the MHD equations. We have verified our numerical implementation of anisotropic TC with the predicted behavior for three important test problems. The tests include a study of the growth rates for the Field (1965) thermal instability, anisotropic conduction of a temperature pulse along loops of magnetic field and the CMB expansion of a supernova remnant into a uniform medium.

A range of parameters has been explored with variations in ISM density, temperature and magnetic field strength being the principle parameters being varied. Convergence testing has also been presented to ensure that we are accurately tracking the proper physics with adequate resolution. This convergence study might also be helpful in global simulations of the ISM that include TC. We have demonstrated that anisotropic TC introduces a number of variations into the gross morphology of the remnant that have direct observational consequences. In particular, the heat transport along the field lines leads to significantly larger elongations parallel to the mean magnetic field. Furthermore, the temperature near the poles is warmer than the temperature at a similar distance along the equatorial plane, leading to x-ray morphologies that are brightest near the poles that would otherwise form a bright shell in the absence of TC. The isobaric evolution of the remnant also leads to increased x-ray brightness due to higher densities in the centers of the remnants expanding into denser media at ages of several tens of thousands of years, giving a possible explanation for the existence of mixed-morphology remnants.

We see evidence that the supernova ejecta develop Rayleigh-Taylor fingering at late stages in the evolution. This provides a mechanism for producing metal-rich gas at large distances from the initial explosion. This metal enrichment is an essential part of the process of forming extremely metal-poor stars. The ejecta also sculpt the magnetic field in the hot gas bubble, thereby regulating the flow of thermal energy in it.

**Acknowledgements**

We acknowledge interesting conversations with R.A. Benjamin and N. Lehner. DSB acknowledges support via NSF grants R36643-7390002, AST-005569-001, AST-0607731 and NSF-PFC grant PHY02-16783. DSB and JCH also acknowledge NASA grant HST-AR-10934.01-A. The majority of simulations were performed on PC clusters at UND but a few initial simulations were also performed at NCSA.

## References

Avillez, M.A., Mac Low M.M. 2002, ApJ, 581, 1047

Avillez, M.A. & Breitschwerdt, D., 2005, ApJ, 634, L65

Balbus, S.A. 1986, ApJ, 304, 787

Balbus, S.A. & McKee, C.F. 1982, ApJ, 252, 529

Balsara, D.S., 1998a, Ap.J. Supp., 116, 119

Balsara, D.S., 1998b, Ap.J. Supp., 116, 133

Balsara, D.S. & Spicer, D.S., 1999a, J. Comput. Phys., 148, 133

Balsara, D.S. & Spicer, D.S., 1999b, J. Comput. Phys., 149, 270

Balsara, D.S., 2001a, J. Comput. Phys., 174(2), 614

Balsara, D.S., 2001b, J. Quant. Spec. Rad. Transf., 69(6), 671

Balsara, D.S., Benjamin, R.A. & Cox, D.Q., 2001, ApJ, 563, 800

Balsara, D.S., 2004, Ap.J. Supp., 151(1), 149

Balsara, D.S., Kim, J.S., Mac Low, M.M. & Mathews, G.J., 2004, ApJ, 617, 339

Balsara, D.S. & Kim, J.S., 2004, ApJ, 602, 1079

Balsara, D.S. & Kim, J.S., 2005, ApJ, 634, 390

Begelman, M.C. & McKee, C.F., 1990, ApJ, 358, 375

Bell, A.R. & Lucek, S.G. 2001, MNRAS, 321, 433

Blom, J.J., Paglione, T.A.D. and Carramiñana, A. 1999, 516, 744

Blondin, J.M. & Ellison, D.C. 2001, ApJ, 560, 244

Borkowski, K. J., Balbus, S. A., & Fristrom, C. C. 1990, ApJ, 355, 501

Bowen. D.V. et al 2005, in Astrophysics in the Far Ultraviolet: Five years of Discovery with FUSE, ed G.Sonneborn, W.Moos & B.-G. Andersson (San Franscisco: ASP), in press (astro-ph/0410008)

Chevalier, R.A. 1975, ApJ, 200, 698

Chevalier, R.A. & Dwarkadas, V.V., 1995, ApJ, 452, L45

Chevalier, R.A. & Liang, E. 1989, ApJ 344, 332

Cioffi, D.F. McKee, C.F. & Bertschinger, E. 1988, ApJ, 334, 252 (CMB)

Cowie, L.L. & McKee, C.F. 1977, ApJ, 211, 135 (CM)

Cox, D.P. et al 1999, ApJ, 524, 179

Cui, W. & Cox, D.P. 1992, ApJ, 401, 206

Dekker, K. and Verwer, J.G., "Stability of Runge-Kutta Methods for Stiff Nonlinear Differential Equations", Elsevier-North Holland, Amsterdam (1984)

Dixon, W.V, Sankrit, R., Otte, B. 2006, ApJ, 647, to appear (astro-ph/0604408)

Dwarkadas, V.V. 2005, ApJ, 630, 892

Elmegreen, B.G. 1994, ApJ, 427, 384

Fabian, A.C. Nulsen, P.E. & Canizares, C.R. 1991, A&A Rev., 2, 191

Ferriere, K.,1998, ApJ, 503, 700

Field, G.B. 1965, ApJ, 142, 531

Forslund, D.W. 1970, J. Geophys. Res., 75, 17

Heitsch, F. et al 2005, ApJ, 633, L113

Hester, J.J. et al 2002, ApJ, 577, L49

Jones, T.W. et al 1998, PASP, 110, 125

Jun, B.I. & Jones, T.W. 1999, ApJ, 511, 774

Kawasaki, M., Ozaki, M., Nagase, F., Inoue, H. & Petre, R. 2005, ApJ, 631, 935

Klein, R.I., McKee, C.F. & Colella, P. 1994, ApJ, 420, 213

Kosenko, D.I. 2006, MNRAS, submitted, (astro-ph/0605349)

Maller, A.H. & Bullock, J.S. 2004, MNRAS, 355, 694

MacDonald, J. & Bailey, M.E. 1981, MNRAS, 197, 995

Mac Low, M.M., Balsara, D.S., Avillez, M. & Kim, J.S., 2005, ApJ, 626, 864

Marcolini, A., Strickland, D.K., D'Ercole, A., Heckman, T.M. & Hoopes, C.G. 2005, MNRAS, 362, 626

Oegerle, W.R., Jenkins, E.B., Shelton, R.L., Bowen, D.V. & Chayer, P. 2005, ApJ, 622, 377

Parrish, I.J. & Stone, J.M. 2005, ApJ, 633, 334

Pedlar, A., Muxlow, T. & Wills, K, 2003, RevMexAA, 15, 303

Piontek, R. & Ostriker, E. 2004, ApJ, 601, 905

Pistinner, S. & Shaviv, G. 1996, ApJ, 459, 147

Raymond, J.C., Cox, D.P. & Smith, B.W. 1976, ApJ, 204, 290

Rho, J. & Petre, R., 1998, ApJ, 503, L167 (RP)

Rieke, G. H., Lebofsky, M. J., Thompson, R. I., Low, F. J. & Tokunaga, A. T., 1980, ApJ, 238, 24

Roe, P.L. & Balsara, D.S., 1996, SIAM Journal of Applied Mathematics, 56, 57

Rosenbrock, H.H. 1963, SIAM Journal of Applied Mathematics, 5, 329

Saad, Y., Iterative Methods for Sparse Linear Systems, International Thompson Publishing Inc.

Sharma P. & Hammet, G.W. , 2007, J. Comput. Phys., 227, 123

Salvaterra, R. Ferrara, A. & Schneider, R. 2004, New Astronomy, 10, 113

Savage, B.D. & Lehner, N. 2005, ApJS, in press (astro-ph/0509458)

Schatz, H., Bildsten, L., Cumming, A. & Wiescher, M. 1999, ApJ, 524, 1014

Shapiro, P.R. & Moore, R.T. 1976, ApJ, ApJ, 207, 460

Shelton, R.L. et al 1999, ApJ, 524, 192

Slavin, J.D. & Cox, D.P. 1992, ApJ, 417, 187 (SC92)

Slavin, J.D. & Cox, D.P. 1993, ApJ, 392, 131 (SC93)

Spitzer, L. 1962, Physics of Fully Ionized Gases (2$^{nd}$ edition, New York Interscience)

Tilley, D.A. & Balsara, D.S. 2006, ApJLett., to appear (TB) (astro-ph/0604117)

Tilley, D.A., Balsara, D.S. & Howk, J.C. 2006, MNRAS, submitted (TBH) (astro-ph/0604474)

Veilleux, S., Cecil, G. & Bland-Hawthorn, J. 2005, Ann. Rev. Astron. Astroph., 43, 769

Velazquez, P.F., Martinell, J.J., Raga, A.C. & Giacani, E.B. 2004, ApJ, 601, 885

Verwer, J.G., Spee, E.J., Blom, J.G. and Hunsdorfer, W., 1999, SIAM J. Sci. Comput., 20(4), 1456

White, R.L. & Long, K.S. 1991, ApJ, 373, 567 (WL)

Wolfire, M.G., McKee, C.F., Hollenbach, D. & Tielens, A.G.G.M. 2003, ApJ, 587, 278

Woltjer, L. 1972, Ann. Rev. Astron. & Astrophys., 10, 129

Yokoyama, T. & Shibata, K. 1997, ApJ, 474, L61

Zhong, X. 1996, J. Comput. Phys., 128, 19

**Figure Captions**

Figs. 1a, 1b, 1c and 1d show the amplification factor for $\mu$ = 0.5, 4.0, 10.0 and 50.0 respectively as a function of ($k$ $\Delta$x). The solid line shows the exact result from the PDE, the dashed line shows the amplification factor for the ASIRK scheme from Sub-section II.3.b and the dotted line shows the amplification factor for the IMEX Heun scheme from Sub-section II.3.a.

Figure 2 shows the numerically generated growth rates for various values of wavenumber $k$ and three values of the thermal conduction coefficient K. The growing modes from the dispersion relation are also shown for comparison.

Fig. 3 shows the temperature at the final time for the anisotropic thermal conduction test problem on a 256X256 zone mesh at a time of 300 code units. There is a circular magnetic field resulting in the final temperature being isothermal along circles.

Figure 4 shows the r-t relationship obtained from our numerical code, shown as a solid line, along with the predicted values from eqns. (3.26) and (3.32a) of CMB, shown as a dashed line.

Figures 5a, 5b and 5c show a resolution study for the evolution of the volume of gas with $T > 3X10^5$ K, $T > 7.9X10^5$ K and $T > 2.2X10^6$ K respectively as a function of time for run L2.

Figures 6a, 6b and 6c show a resolution study for the evolution of the volume of gas with $T > 3X10^5$ K, $T > 7.9X10^5$ K and $T > 2.2X10^6$ K respectively as a function of time for run H6.

Fig. 7 shows density in color with velocity vectors overlaid and temperature in color with magnetic field lines overlaid for run VL2. Figs. 7a and 7b correspond to the density and temperature respectively without TC at 100 kyr while Figs. 7c and 7d show the same with TC at the same epoch. Figs. 7e and 7f show the density and temperature respectively without TC at 7.5 Myr while Figs. 7g and 7h show the same with TC at the same epoch. The tick marks correspond to 50 pc intervals.

Fig. 8 shows density in color with velocity vectors overlaid and temperature in color with magnetic field lines overlaid for run L2. Figs. 8a and 8b correspond to the density and temperature respectively without TC at 400 kyr while Figs. 8c and 8d show the same with TC at the same epoch. Figs. 8e and 8f show the density and temperature respectively without TC at 1.5 Myr while Figs. 8g and 8h show the same with TC at the same epoch. Figs. 8i and 8j show the distribution of the ejecta at 400 kyr and 1.5 Myr in the run that included TC. The tick marks correspond to 50 pc intervals.

Fig. 9 shows density in color with velocity vectors overlaid and temperature in color with magnetic field lines overlaid for run H6. Figs. 9a and 9b correspond to the density and temperature respectively without TC at 200 kyr while Figs. 9c and 9d show the same with TC at the same epoch. Figs. 9e and 9f show the density and temperature respectively without TC at 1.5 Myr while Figs. 9g and 9h show the same with TC at the same epoch. The tick marks correspond to 50 pc intervals.

Fig. 10 shows density in color with velocity vectors overlaid and temperature in color with magnetic field lines overlaid for run VH6. Figs. 10a and 10b correspond to the density and temperature respectively without TC at 20 kyr while Figs. 10c and 10d show the same with TC at the same epoch. Figs. 10e and 10f show the density and temperature respectively without TC at 60 kyr while Figs. 10g and 10h show the same with TC at the same epoch. The tick marks correspond to 10 pc intervals.

Figs. 11a to 11f show the density, temperature, r-velocity, magnetic pressure, thermal pressure and total (thermal + magnetic) pressure respectively as a function of radial coordinate in the equatorial plane at various times for run L2. Fig. 11g shows the evolution of the thermal energy, kinetic energy and magnetic energy as a function of time for run L2.

Figs. 12a and 12b show the simulated x-ray brightness in soft (300-800 eV) and hard (1-5 keV) x-rays for run VL2 at a time of 20 kyr while Figs. 12c and 12d show the same for a time of 60 kyr. Figs. 12e, 12f, 12g and 12h show analogous information for run VH6.

Table 1 shows the L1, L2 and  errors in the temperature for the anisotropic thermal conduction test problem.

| | L1 | L2 | $T_{min}$ | $T_{max}$ |
|---|---|---|---|---|
| 32 | 0.0357 | 0.0497 | 9.982 | 10.077 |
| 64 | 0.0303 | 0.0415 | 9.987 | 10.107 |
| 128 | 0.0201 | 0.0315 | 9.988 | 10.146 |
| 256 | 0.0133 | 0.0250 | 9.990 | 10.176 |
| 400 | 0.0114 | 0.0232 | 9.990 | 10.186 |

Table 2 shows the parameters for the simulations presented here.

| Run Name | $R_{outer}$ (pc) | $N_{zones}$ | $t_{final}$ (Myr) | $\rho_{ism}$ (amu/cm$^3$) | $T_{ism}$ (Kelvin) | $B_{ism}$ ($\mu$Gauss) |
|---|---|---|---|---|---|---|
| VL0 | 400 | 384 | 10 | 0.2 | 10,000 | 0 |
| VL2 | 400 | 384 | 10 | 0.2 | 10,000 | 2 |
| VL6 | 400 | 384 | 10 | 0.2 | 10,000 | 6 |
| L0 | 300 | 384 | 9 | 0.7 | 8,000 | 0 |
| L2 | 300 | 384 | 9 | 0.7 | 8,000 | 2 |
| L6 | 300 | 384 | 9 | 0.7 | 8,000 | 6 |
| H0 | 100 | 192 | 4 | 5.0 | 10,000 | 0 |
| H6 | 100 | 192 | 4 | 5.0 | 10,000 | 6 |
| H20 | 100 | 192 | 4 | 5.0 | 10,000 | 20 |
| VH0 | 100 | 192 | 2 | 20.0 | 10,000 | 0 |
| VH6 | 100 | 192 | 2 | 20.0 | 10,000 | 6 |
| VH20 | 100 | 192 | 2 | 20.0 | 10,000 | 20 |

**Appendix A**

In this appendix we provide the formulae for thermal conduction in Cartesian and Spherical geometries. For Cartesian geometries we have

$$\mathbf{F}_{class} = -\,a\,T^{5/2}\,\Big\{\,\mathbf{i}\left(b_x^2\frac{\partial T}{\partial x} + b_x\,b_y\,\frac{\partial T}{\partial y} + b_x\,b_z\,\frac{\partial T}{\partial z}\right) + \mathbf{j}\left(b_x\,b_y\frac{\partial T}{\partial x} + b_y^2\,\frac{\partial T}{\partial y} + b_y\,b_z\,\frac{\partial T}{\partial z}\right) + \mathbf{k}\left(b_x\,b_z\frac{\partial T}{\partial x} + b_y\,b_z\,\frac{\partial T}{\partial y} + b_z^2\,\frac{\partial T}{\partial z}\right)\Big\}$$

For Spherical coordinates we have

$$\mathbf{F}_{class} = -\,a\,T^{5/2}\,\Big\{\,\mathbf{r}\left(b_r^2\frac{\partial T}{\partial r} + \frac{b_r\,b_\theta}{r}\,\frac{\partial T}{\partial \theta} + \frac{b_r\,b_\phi}{r\sin\theta}\,\frac{\partial T}{\partial \phi}\right) + \boldsymbol{\theta}\left(b_r\,b_\theta\frac{\partial T}{\partial r} + \frac{b_\theta^2}{r}\,\frac{\partial T}{\partial \theta} + \frac{b_\theta\,b_\phi}{r\sin\theta}\,\frac{\partial T}{\partial \phi}\right) + \boldsymbol{\varphi}\left(b_r\,b_\phi\frac{\partial T}{\partial r} + \frac{b_\theta\,b_\phi}{r}\,\frac{\partial T}{\partial \theta} + \frac{b_\phi^2}{r\sin\theta}\,\frac{\partial T}{\partial \phi}\right)\Big\}$$

**Appendix B**

We provide explicit formulae for the amplification factors that are shown in Fig. 1. The amplification factor for the exact PDE, without the effect of discretization, is given by:

$$\lambda_{PDE}(k) = e^{-\mu\,(k\,\Delta x)^2}$$

The amplification factor for the IMEX Heun scheme from Sub-section II.3.a is given by:

$$\lambda_{IMEX\ Heun}(k) = \frac{[1-2\,\mu\,\psi]}{[1+2\,\mu\,\psi]}$$

The amplification factor for the ASIRK scheme from Sub-section II.3.b is given by:

$$\lambda_{\mathrm{ASIRK}}(k) = \frac{1}{2\,\mathrm{a}_1}\left\{\frac{\left(1 - 4\,(\mathrm{c}_{21}-\mathrm{a}_1)\,\mu\,\psi\right)}{\left(1 + 4\,\mathrm{a}_1\,\mu\,\psi\right)^2} - \frac{(\mathrm{c}_{21}-\mathrm{a}_1)}{\mathrm{a}_1}\,\frac{1}{\left(1 + 4\,\mathrm{a}_1\,\mu\,\psi\right)} - 2 + 2\,\mathrm{a}_1 + \frac{\mathrm{c}_{21}}{\mathrm{a}_1}\right\}$$

In the previous two equations we have:

$$\mu = \frac{\mathrm{K}\,\Delta\mathrm{t}}{\Delta\mathrm{x}^2} \quad ; \psi = \sin^2\left(k\,\Delta\mathrm{x}/2\right)$$

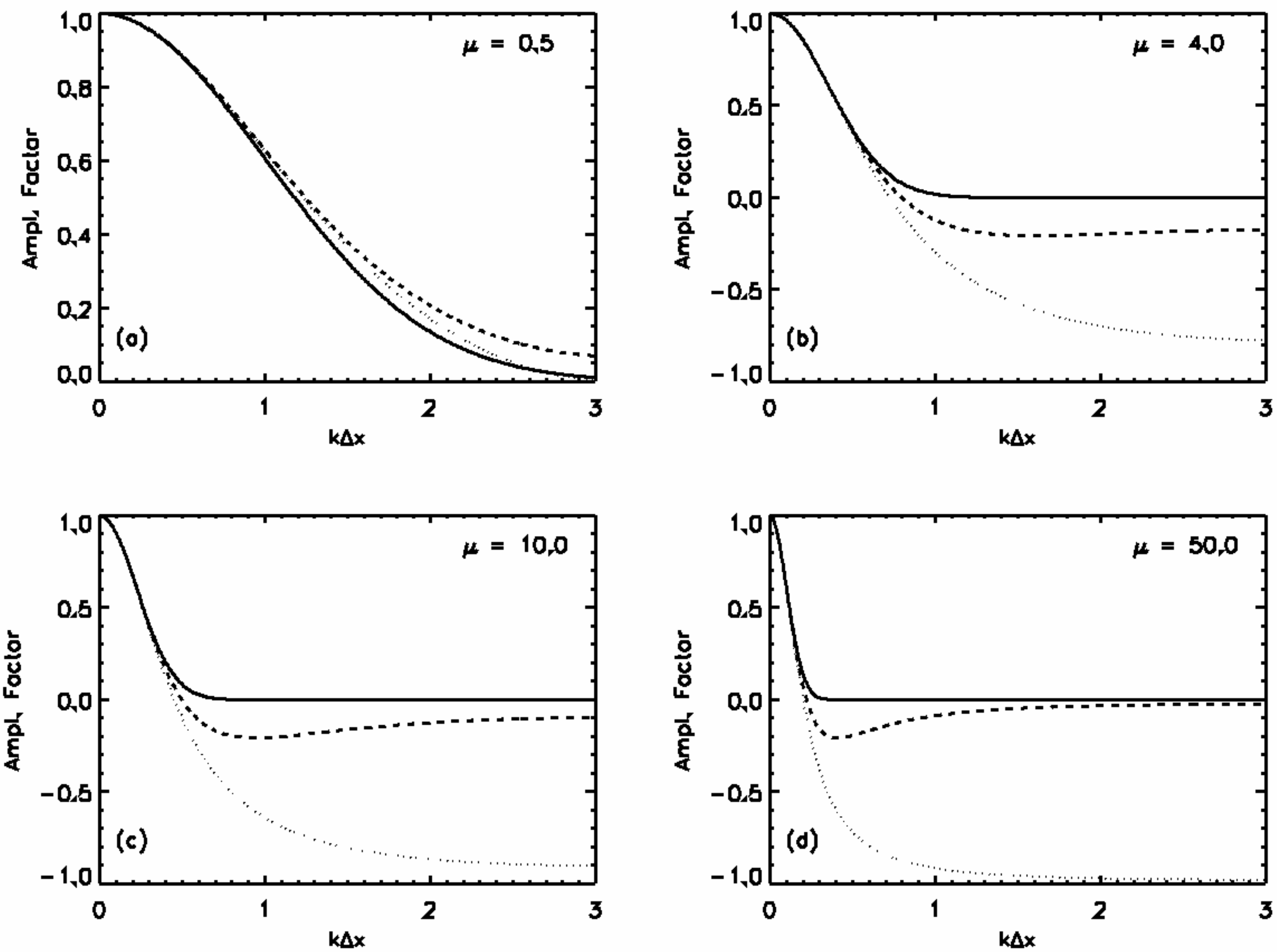

Fig. 1

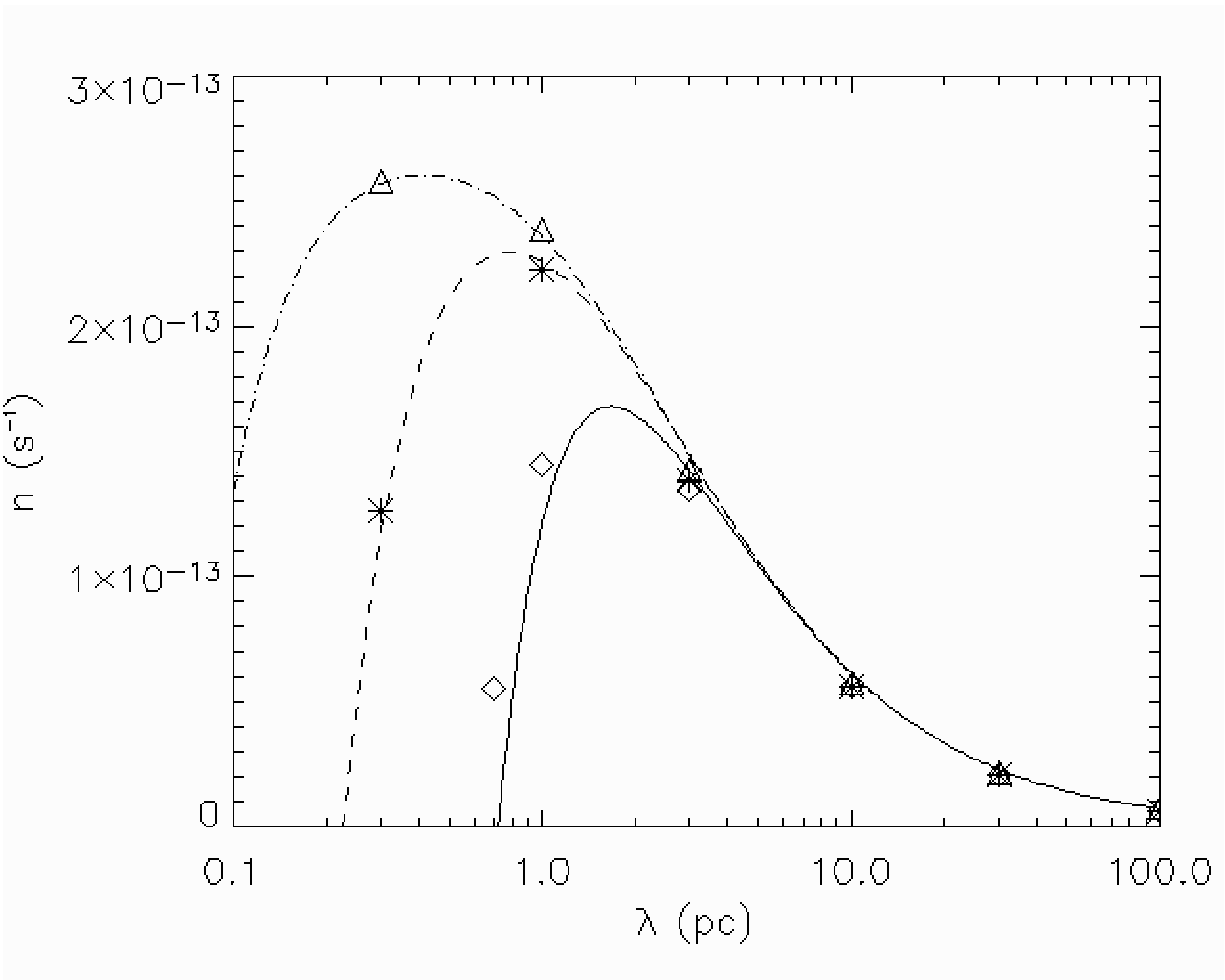

Fig. 2

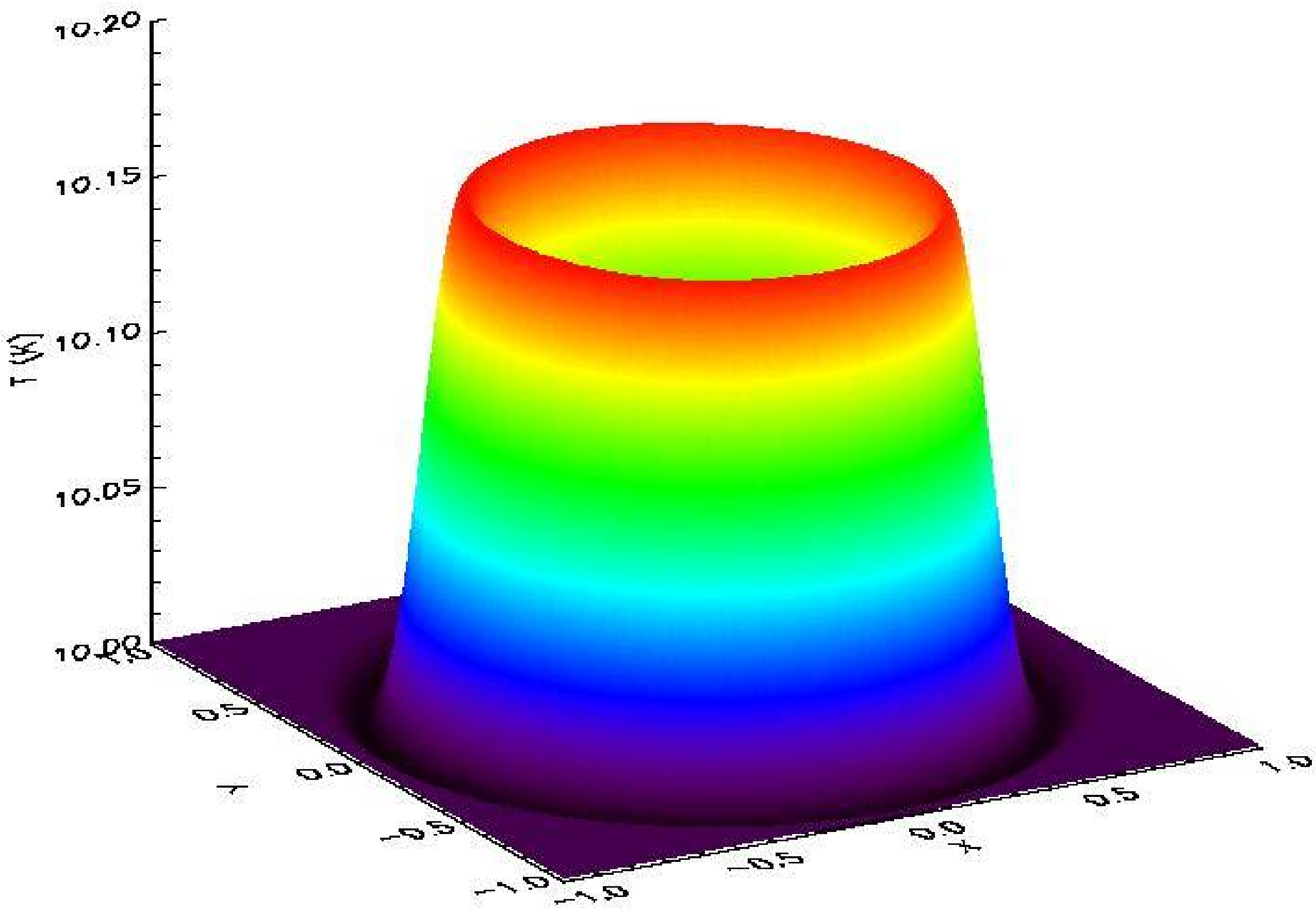

Fig. 3

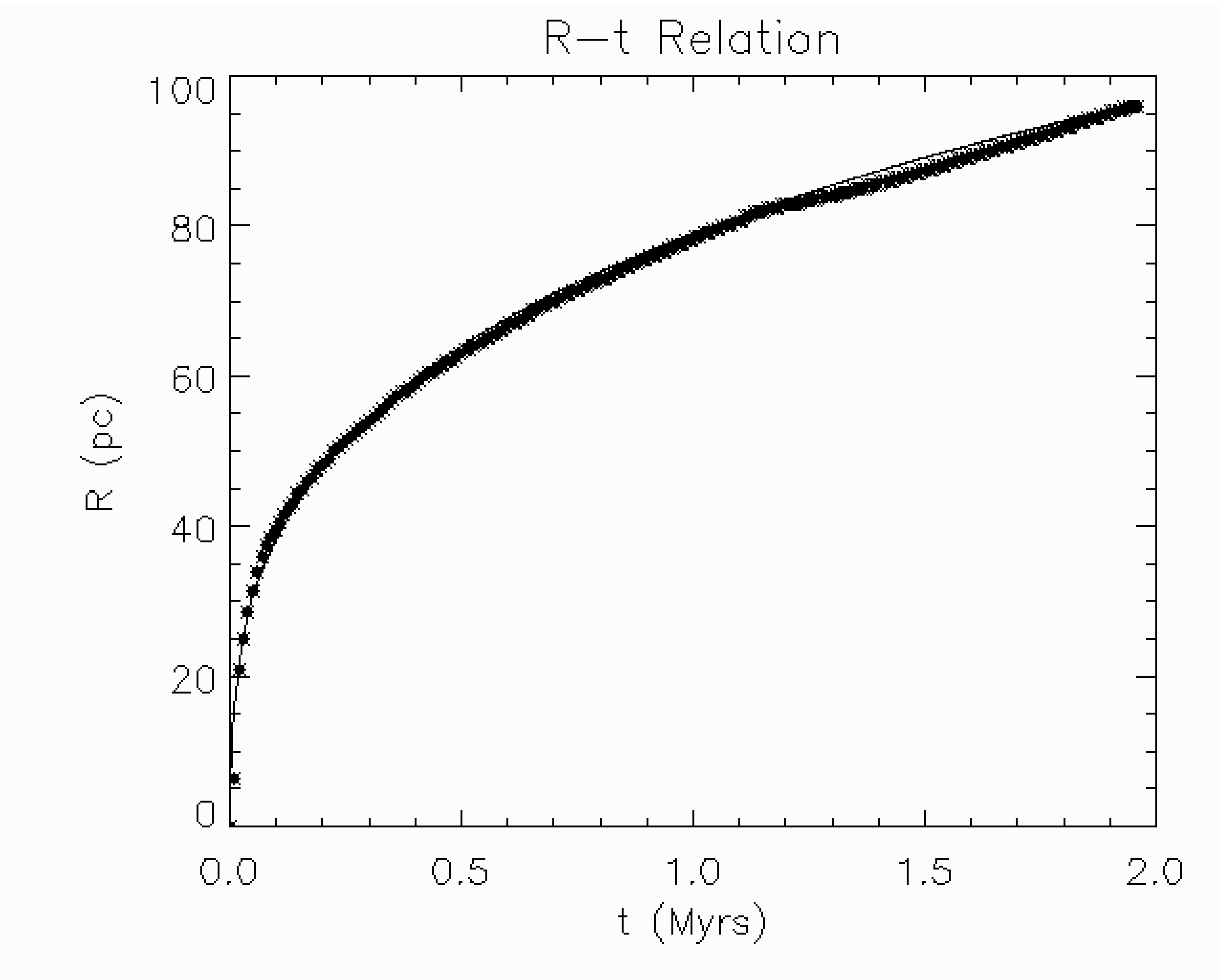

Fig. 4

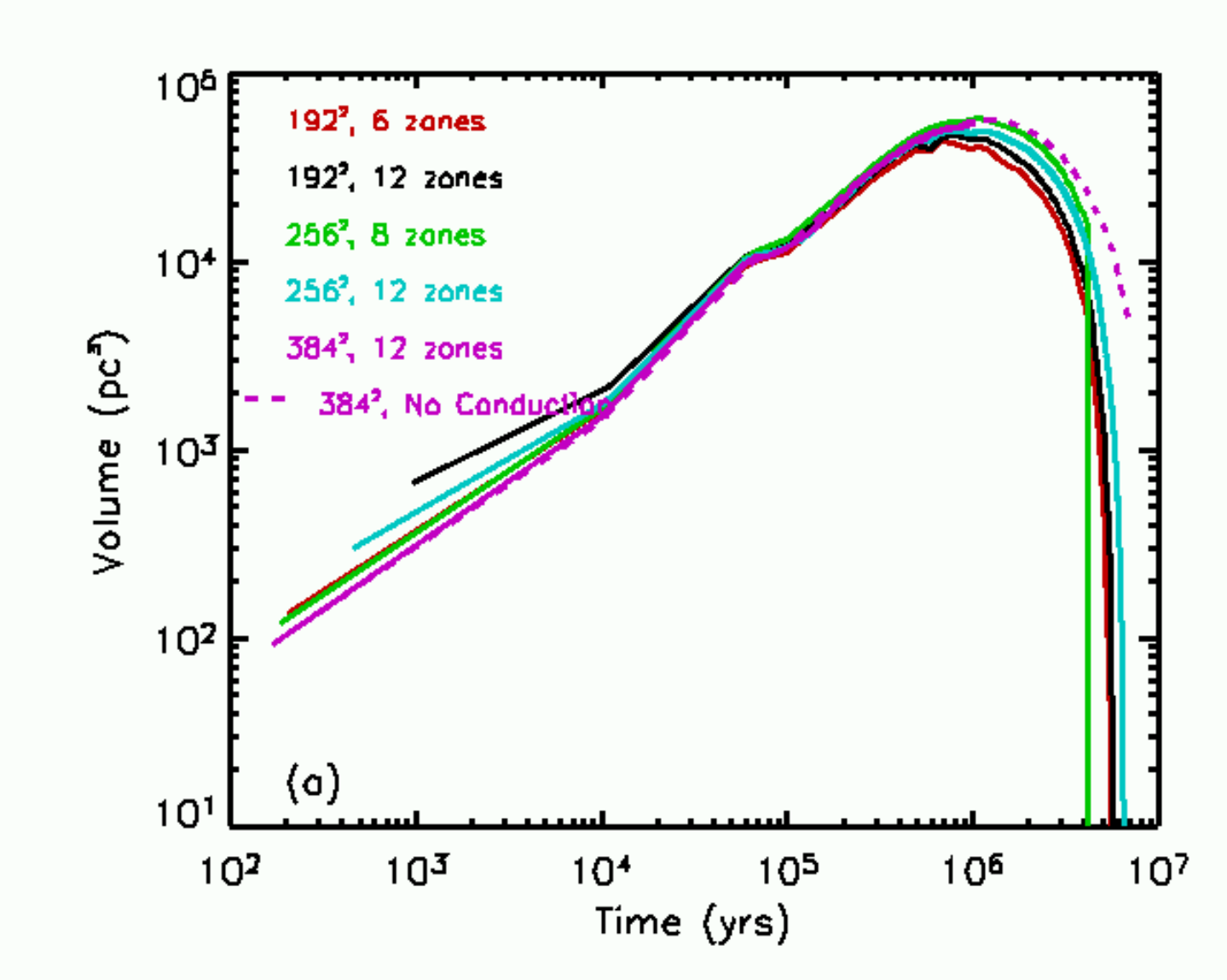

Fig. 5a

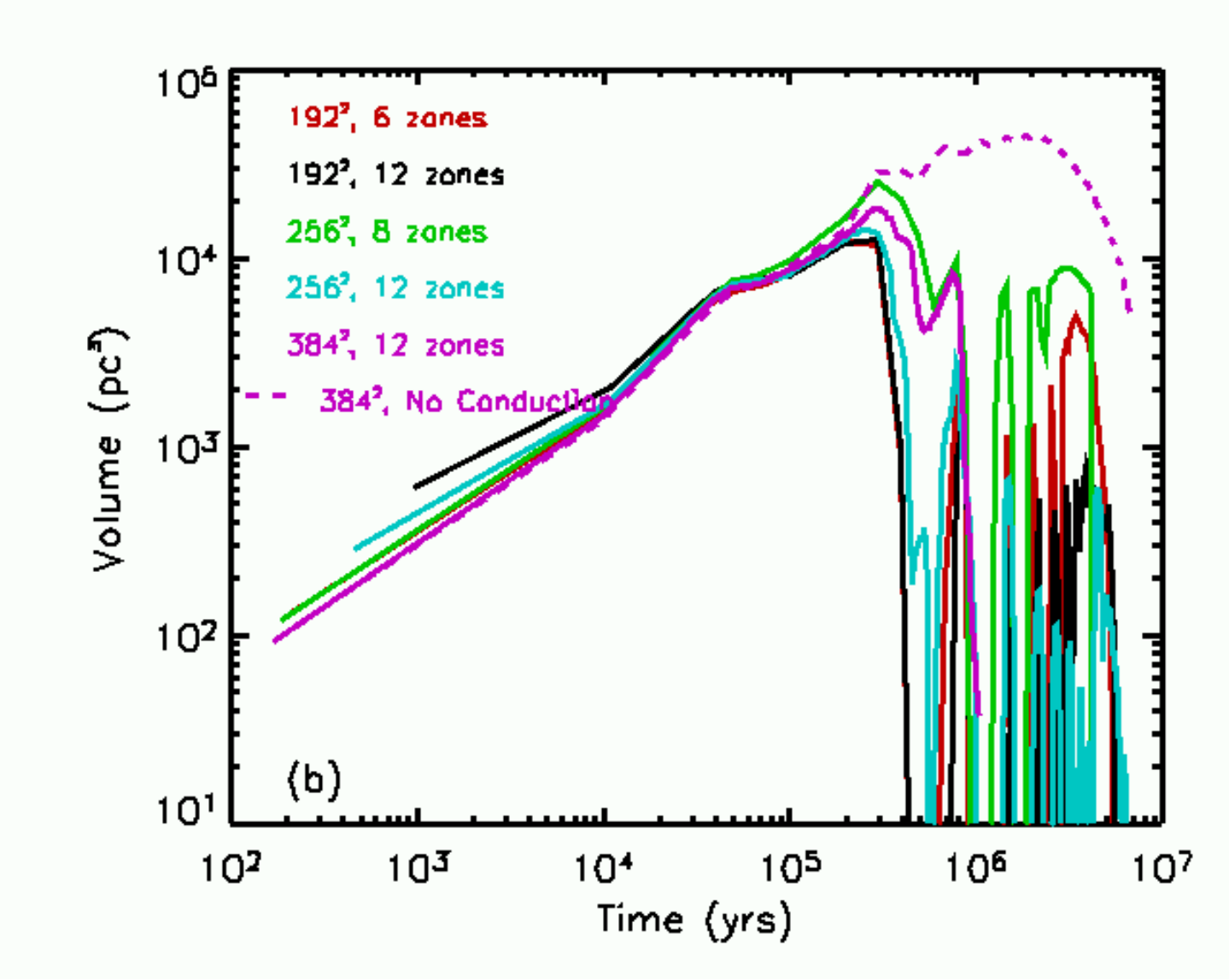

Fig. 5b

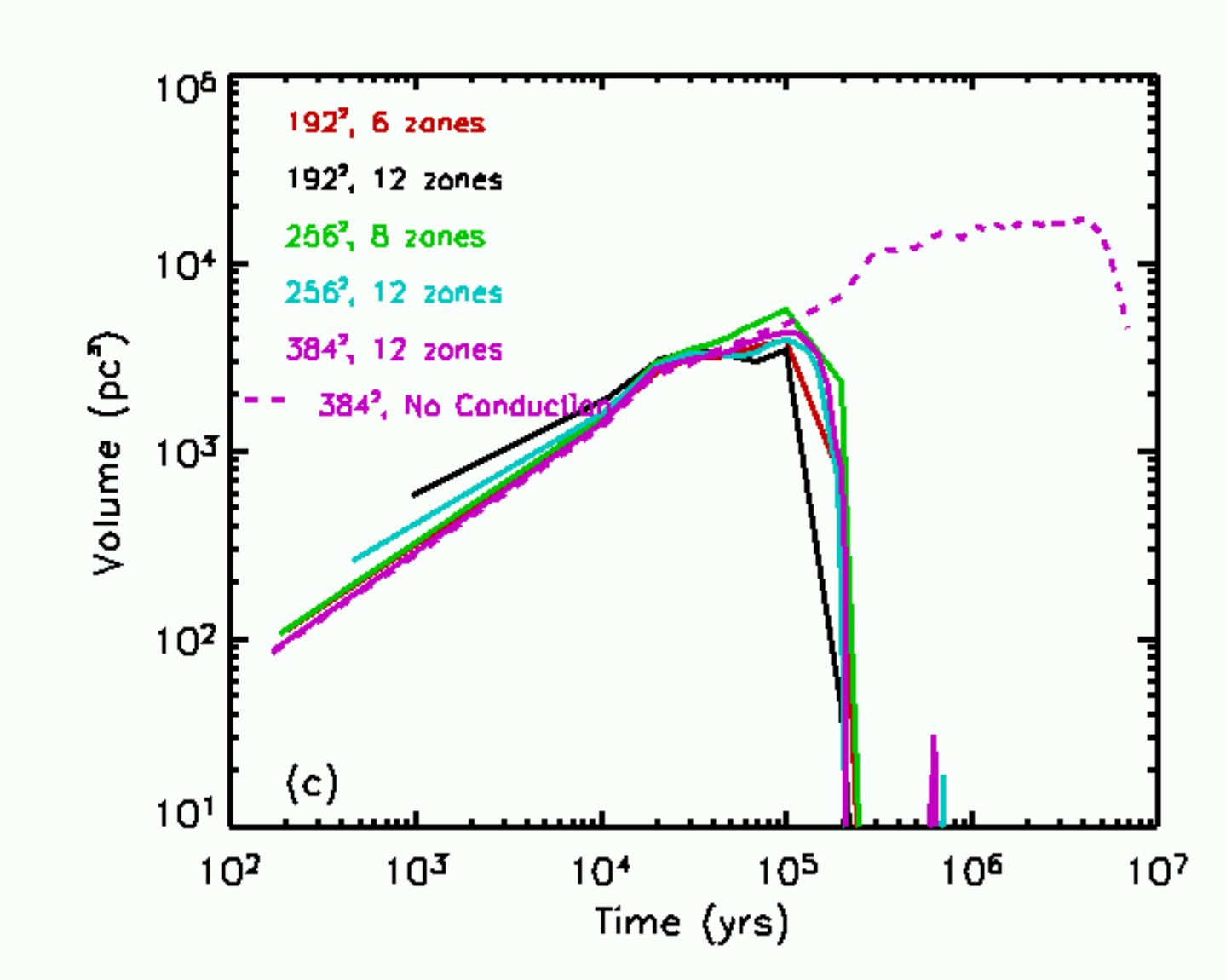

Fig. 5c

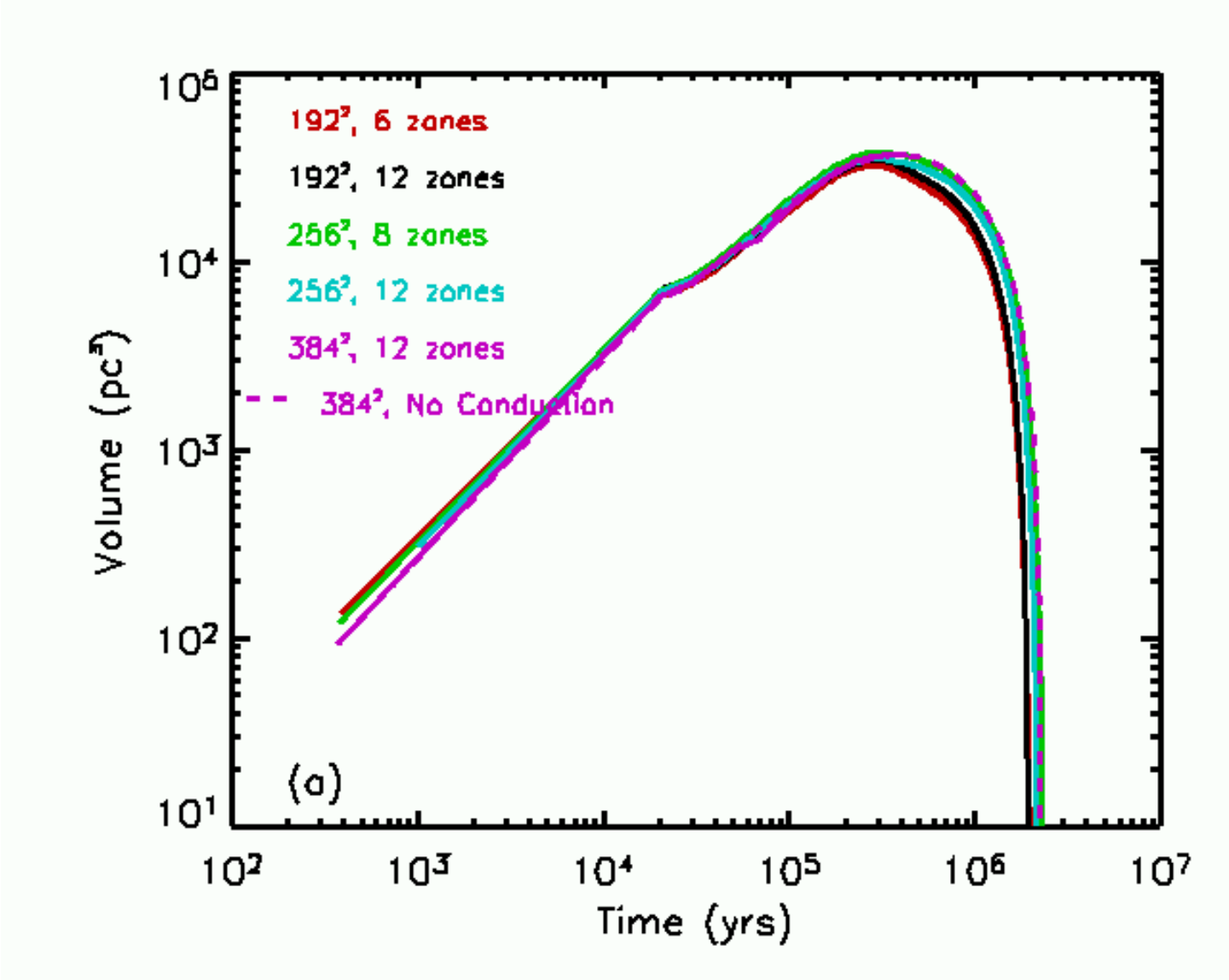

Fig. 6a

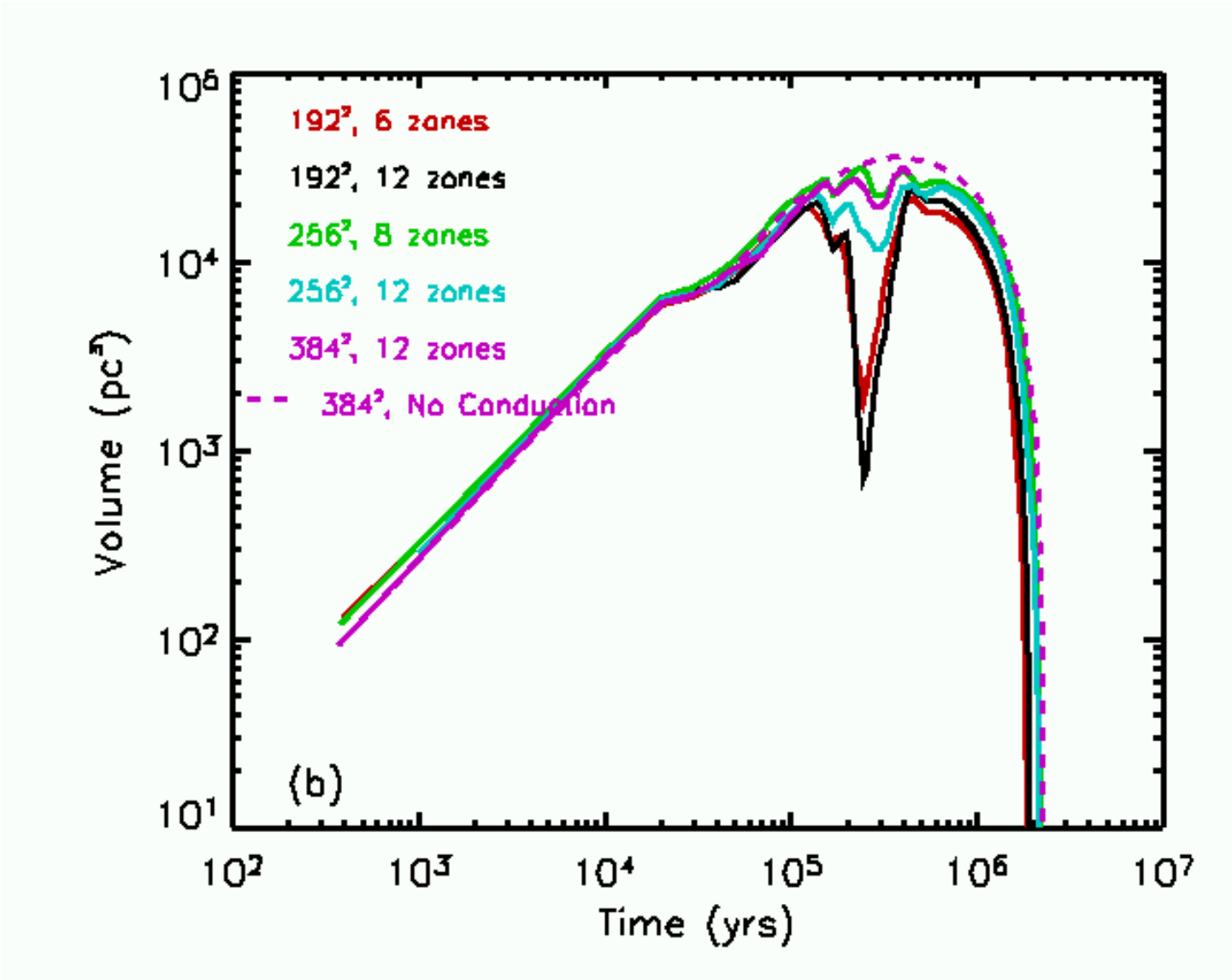

Fig. 6b

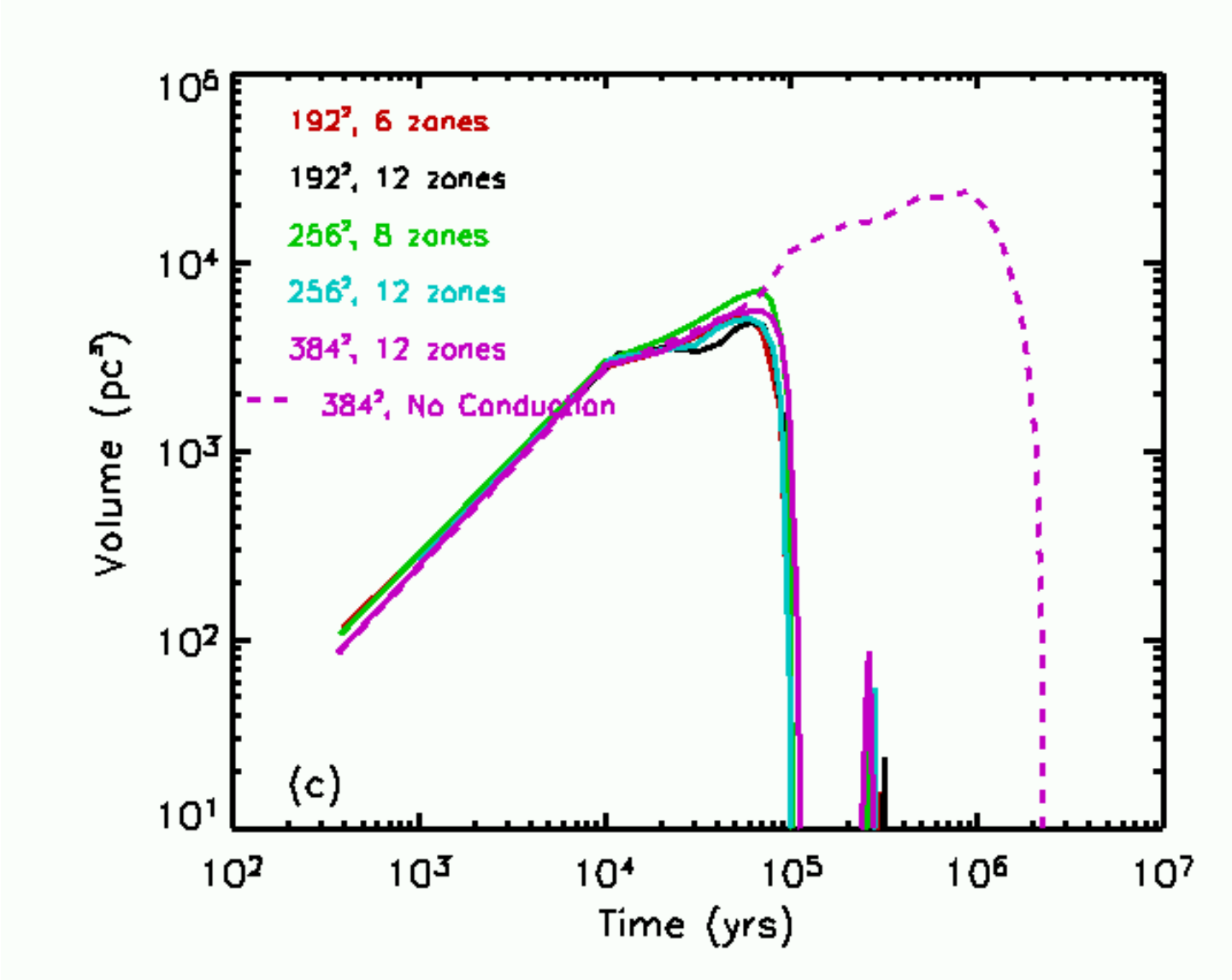

Fig. 6c

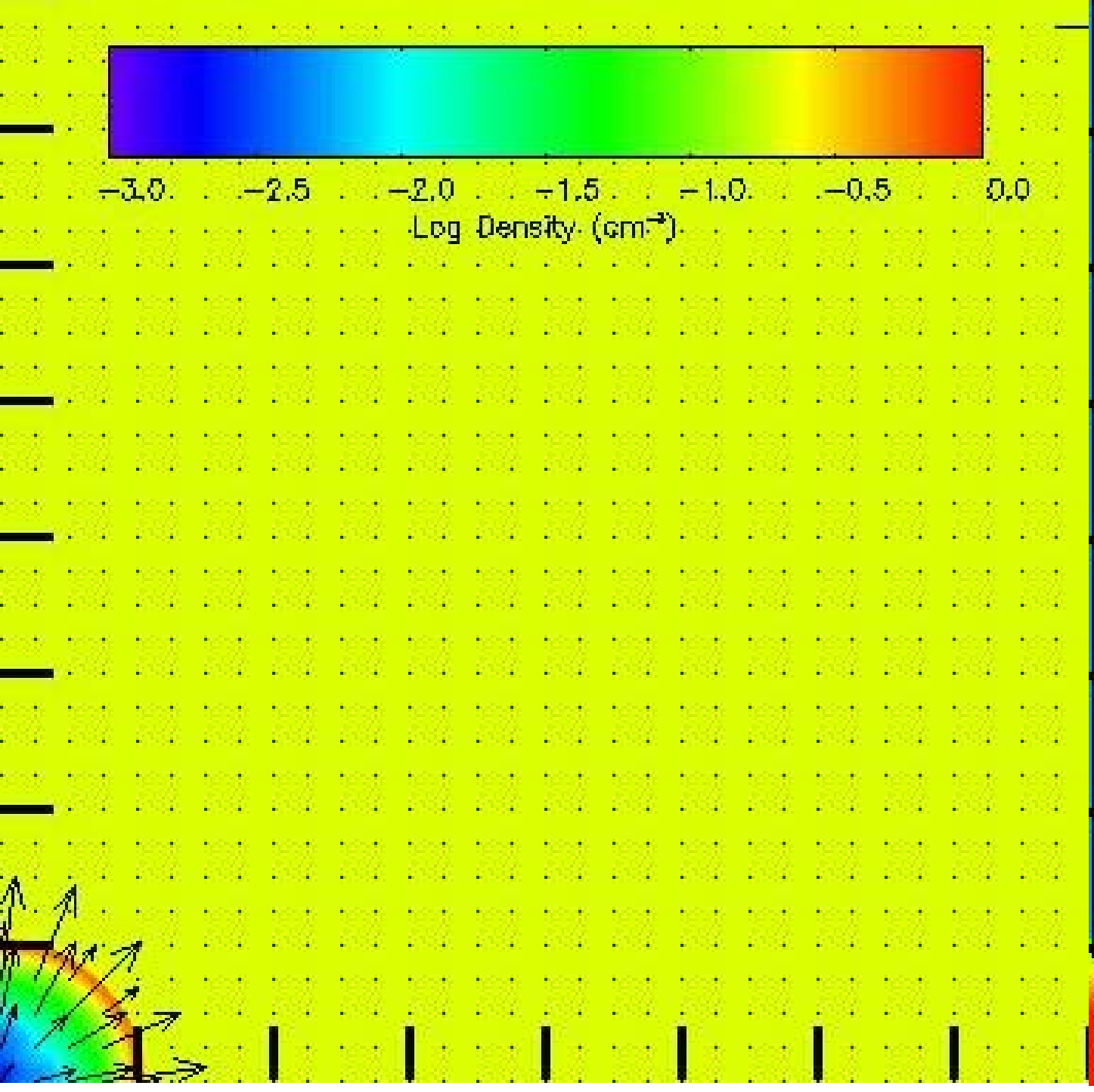

Fig. 7a

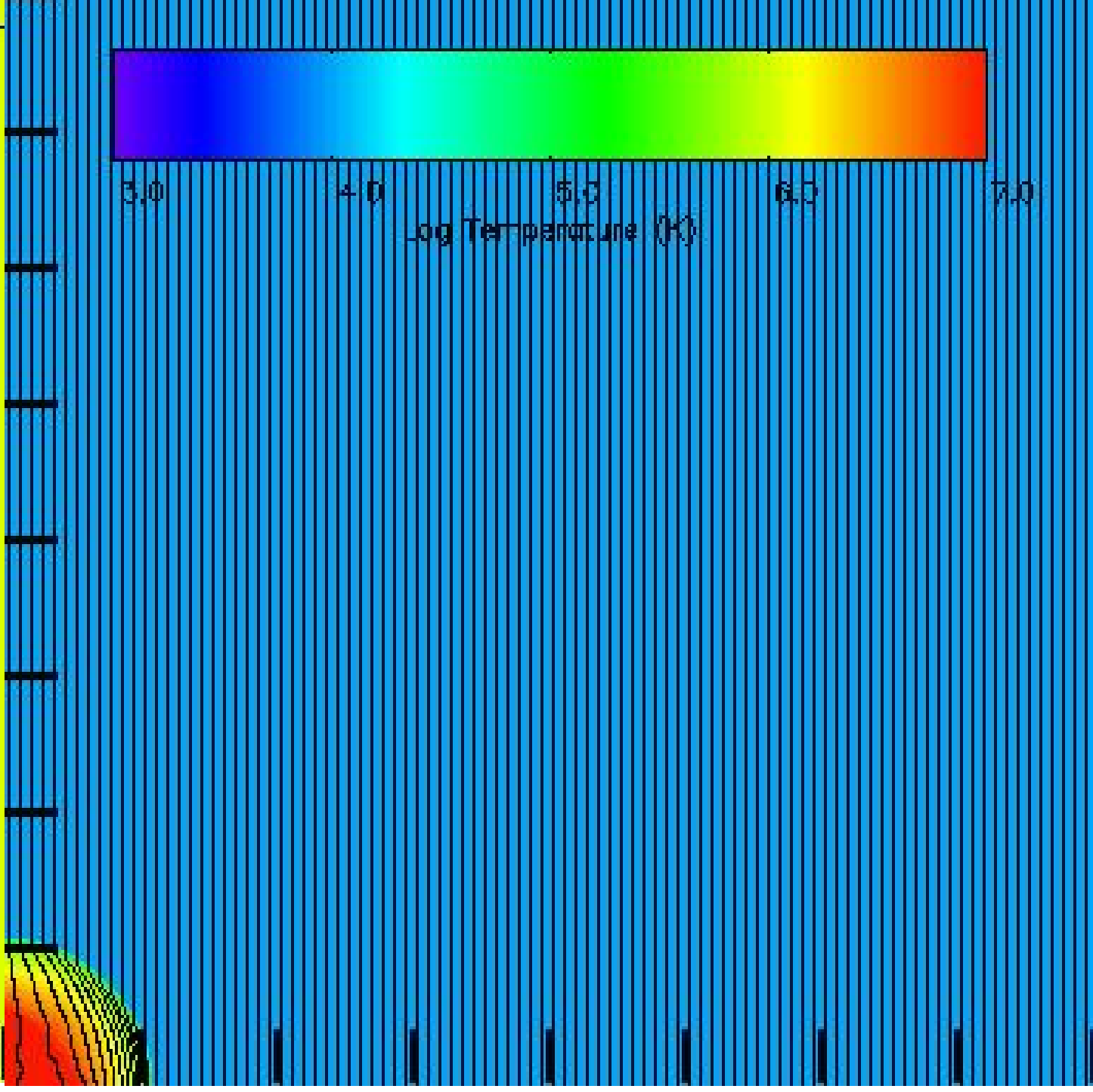

Fig. 7b

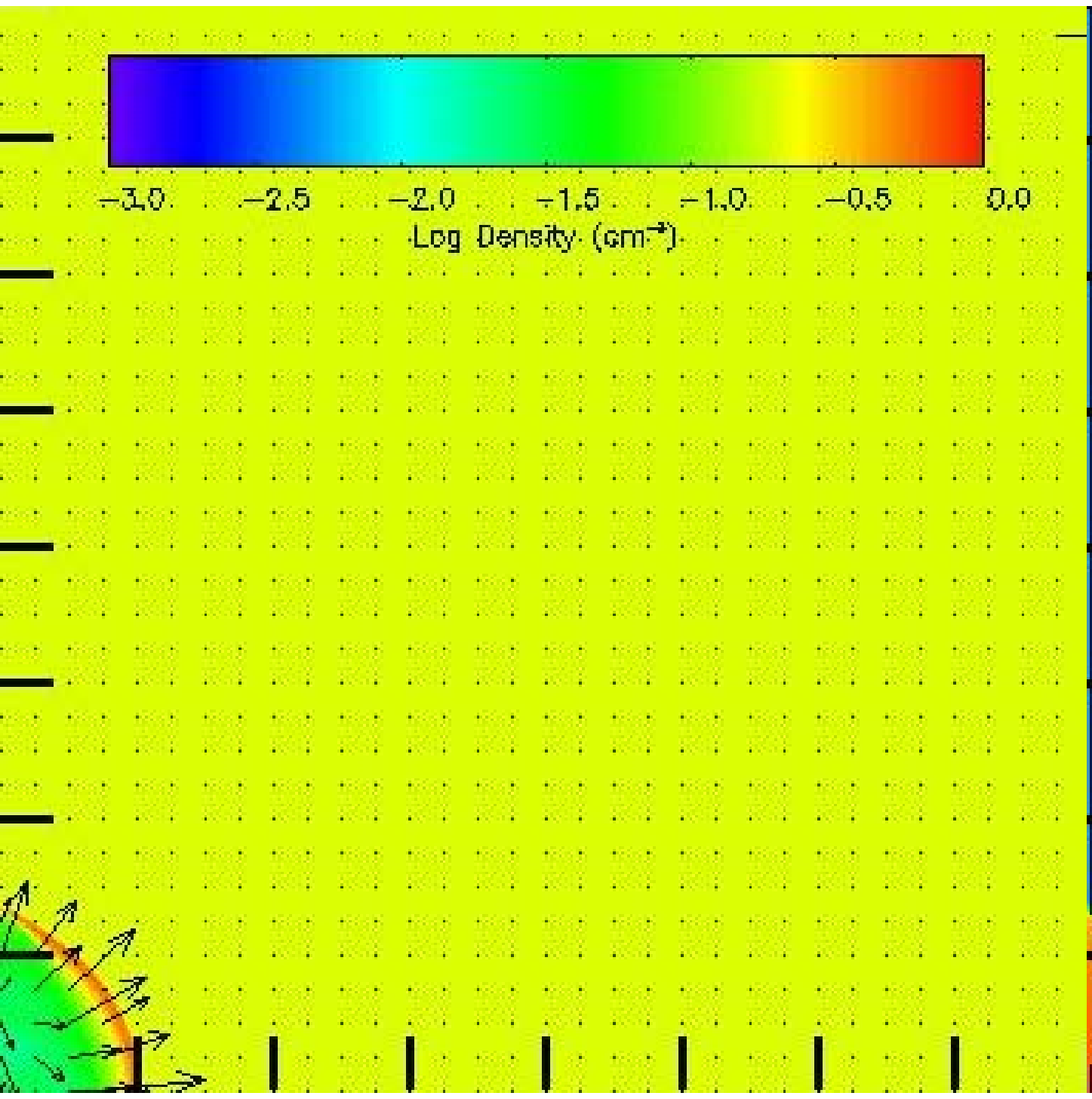

Fig. 7c

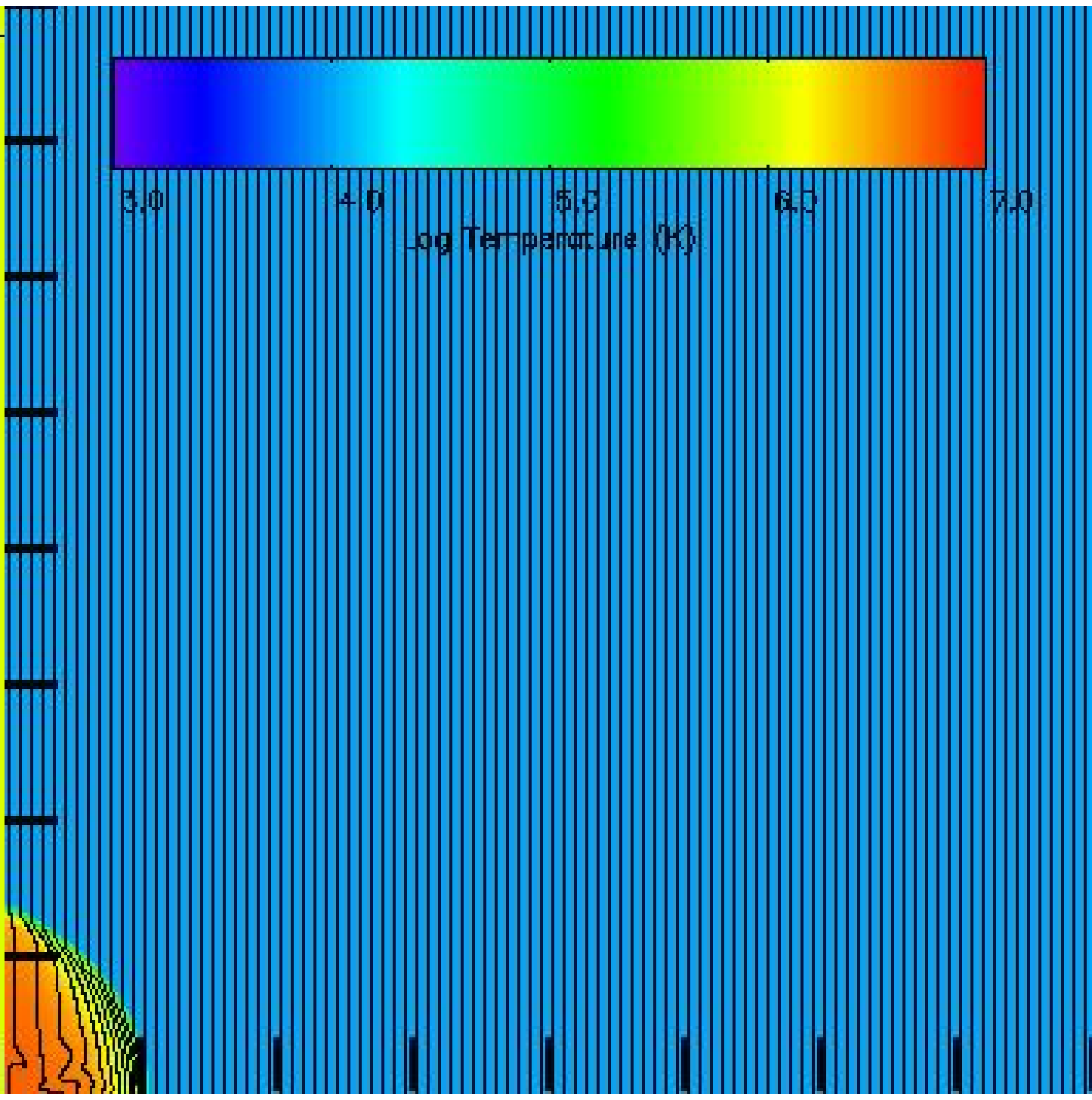

Fig. 7d

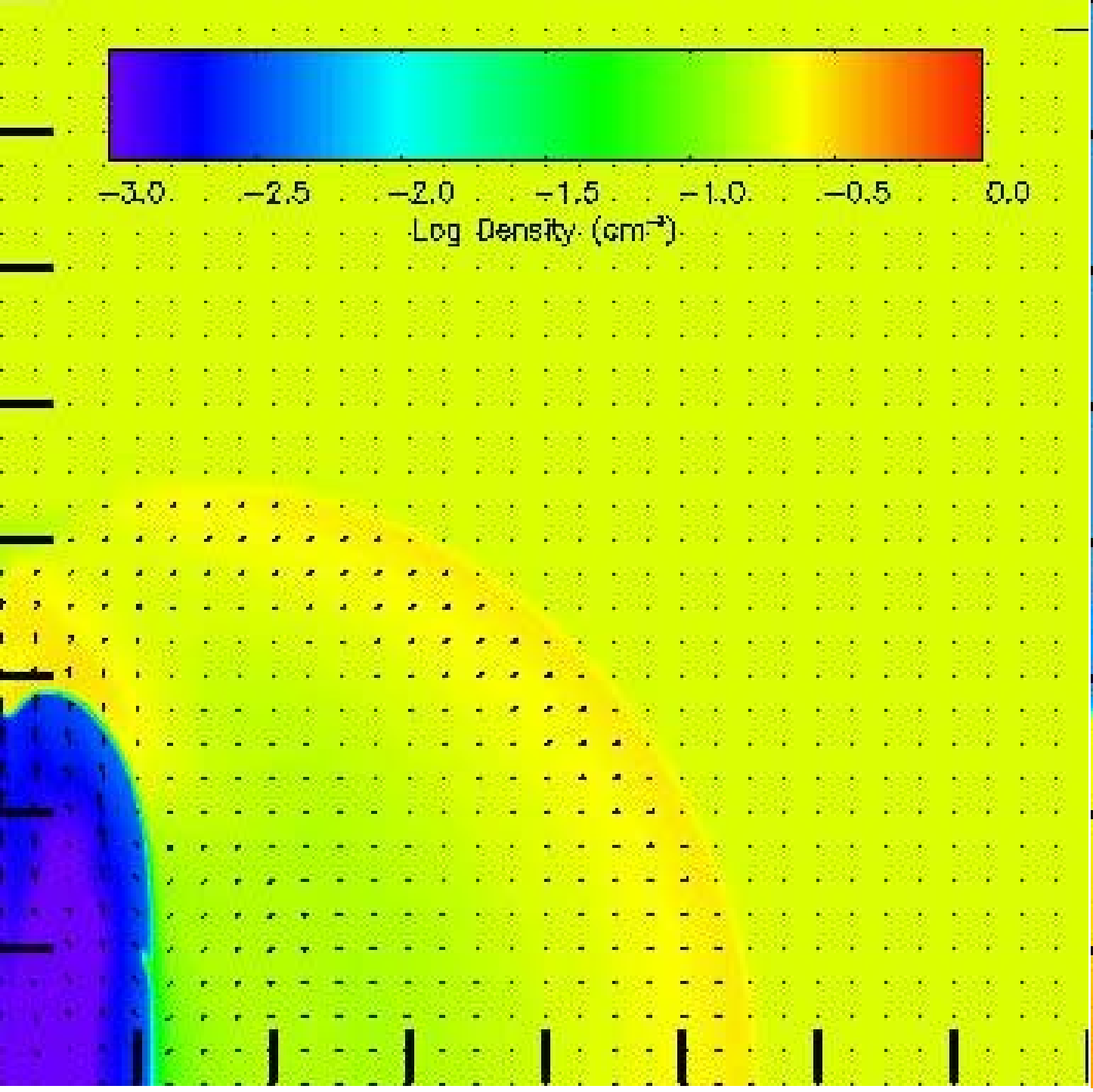

Fig. 7e

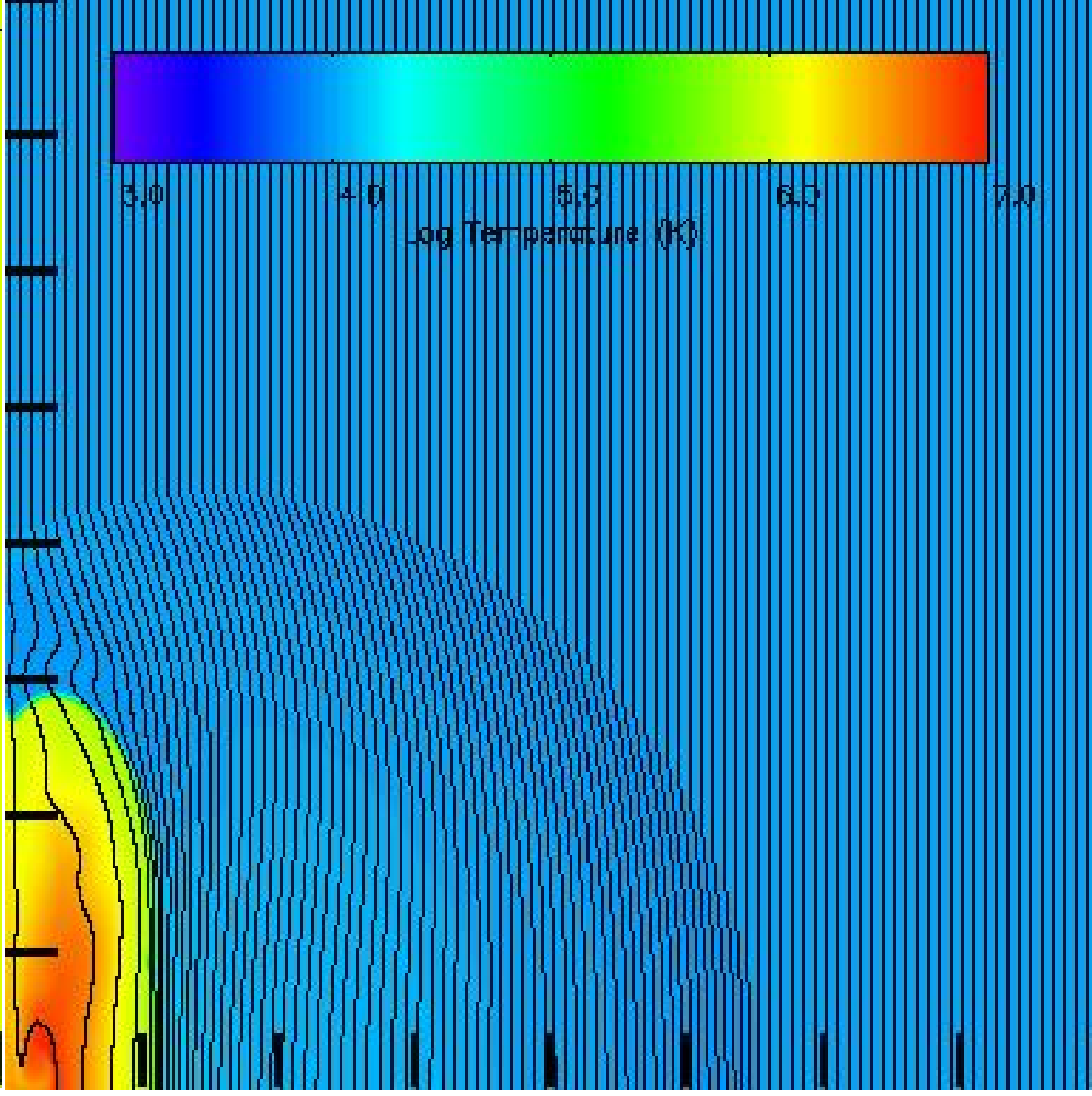

Fig. 7f

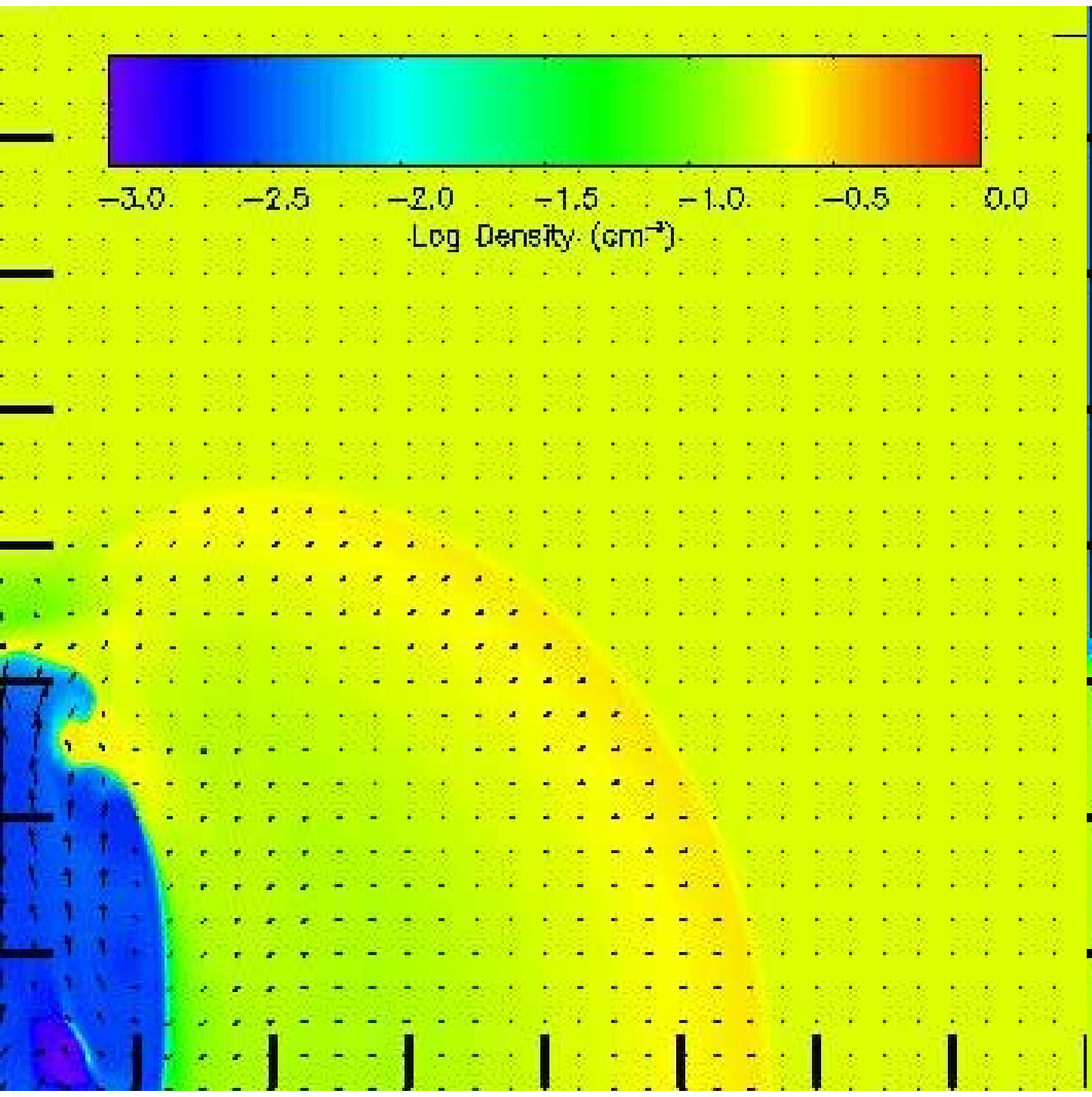

Fig. 7g

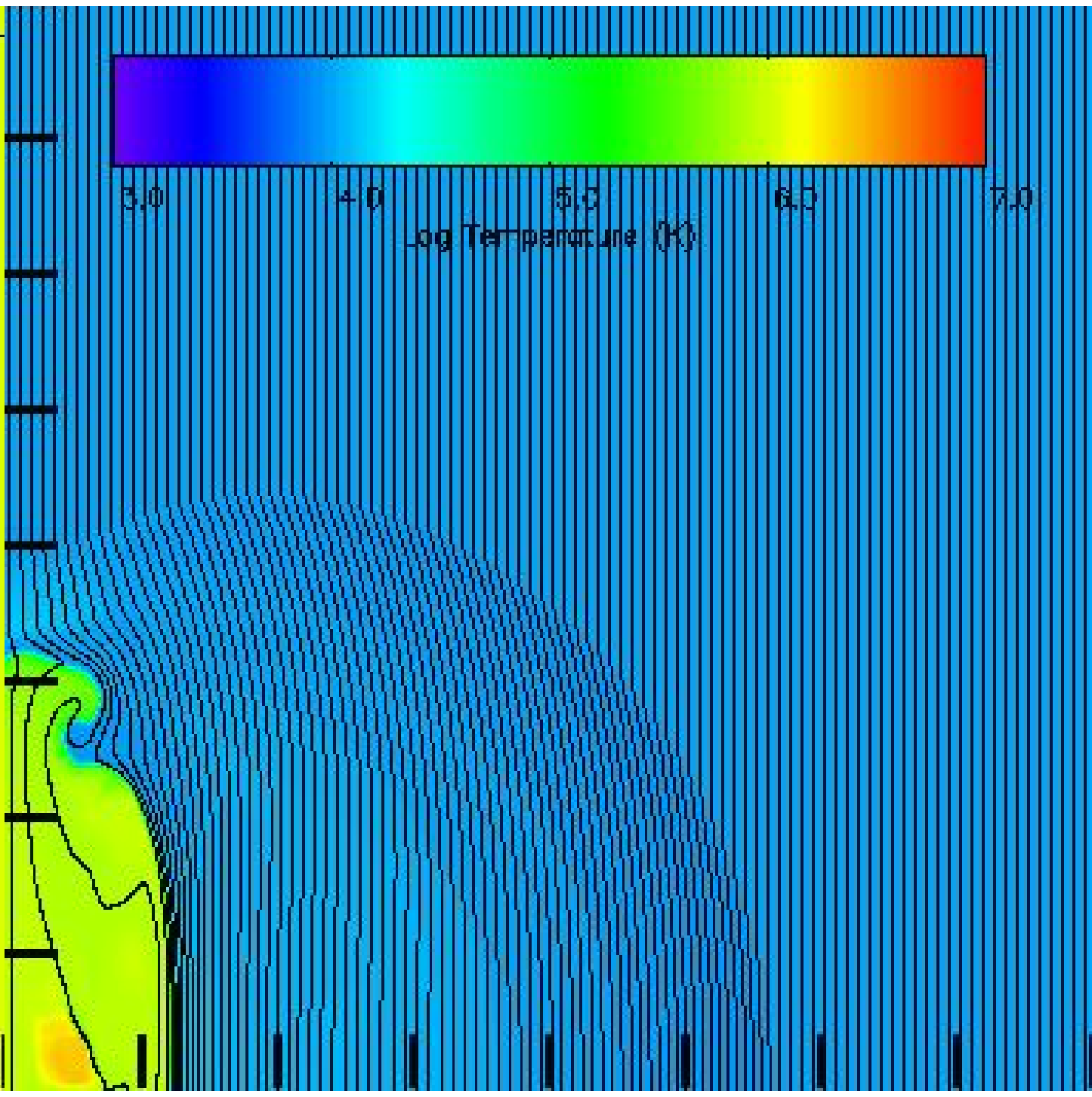

Fig. 7h

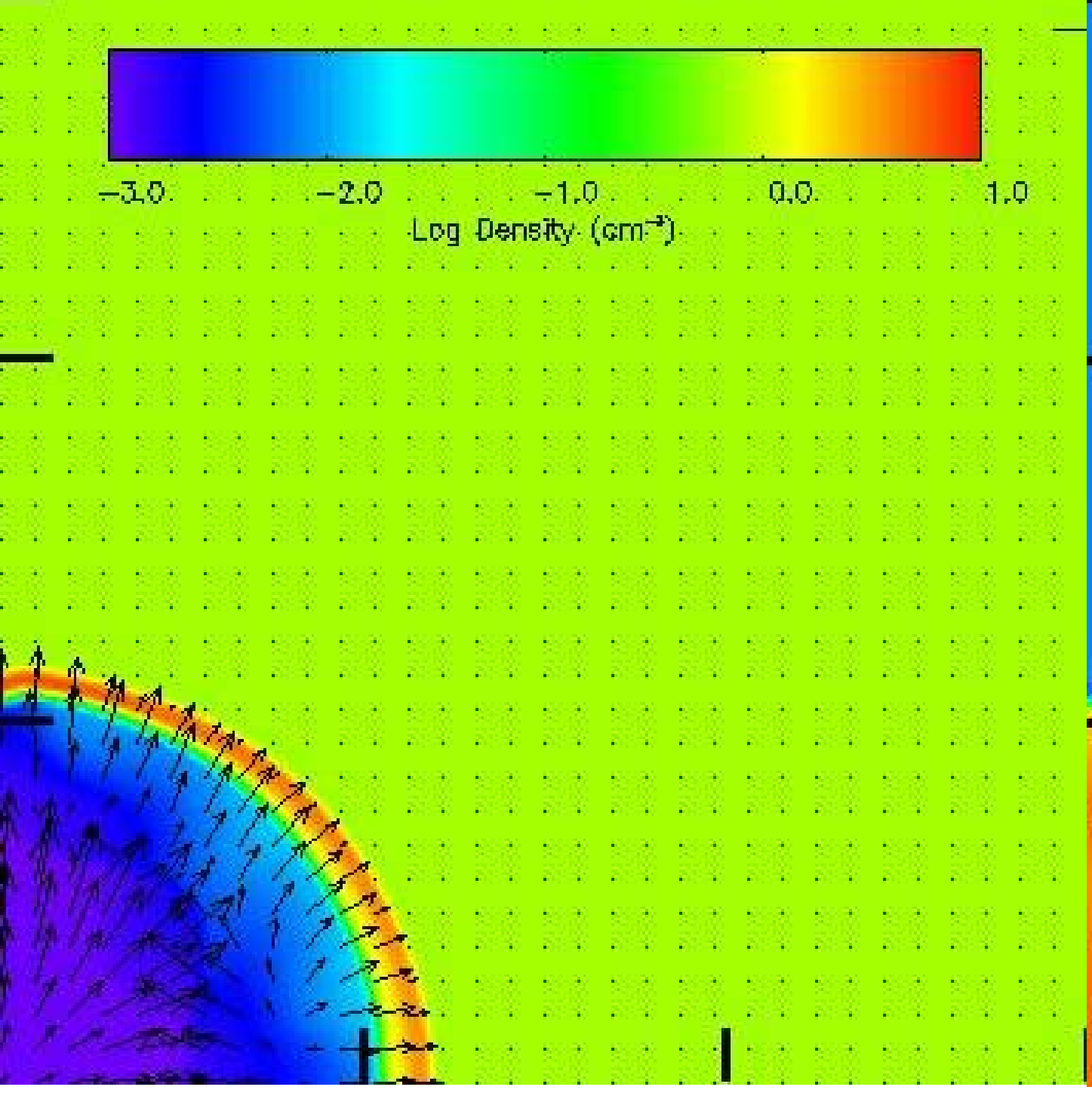

Fig. 8a

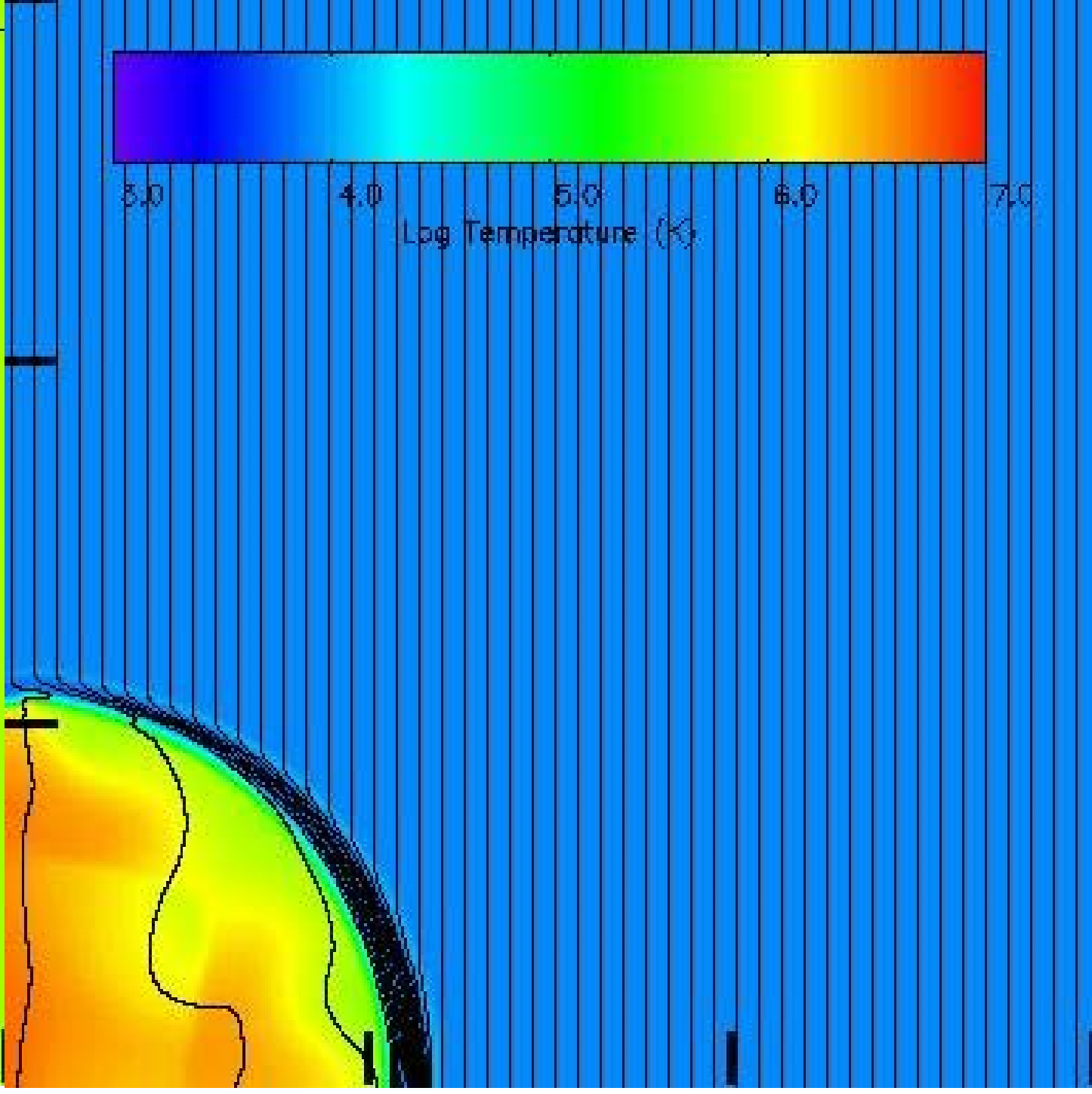

Fig. 8b

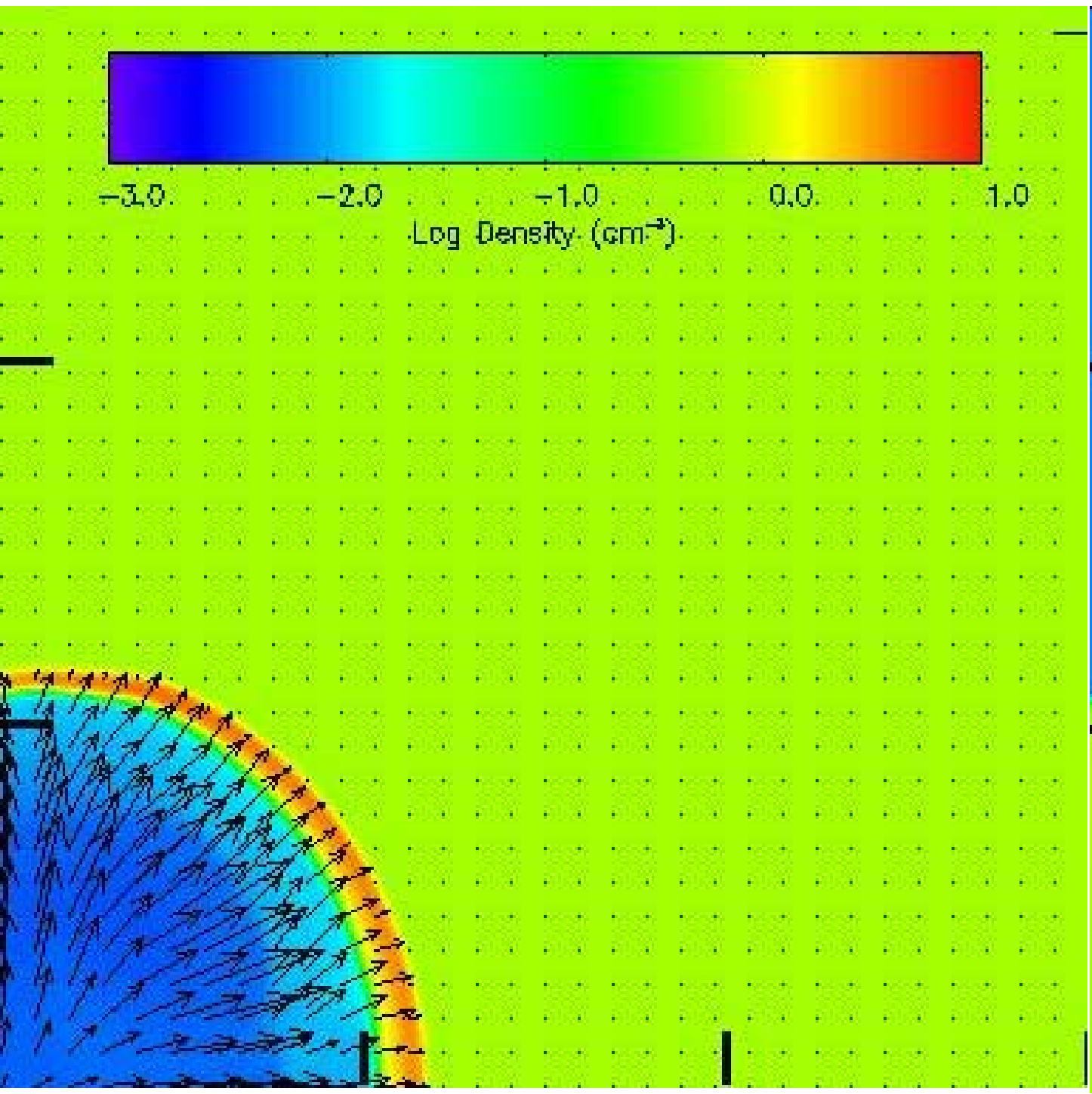

Fig. 8c

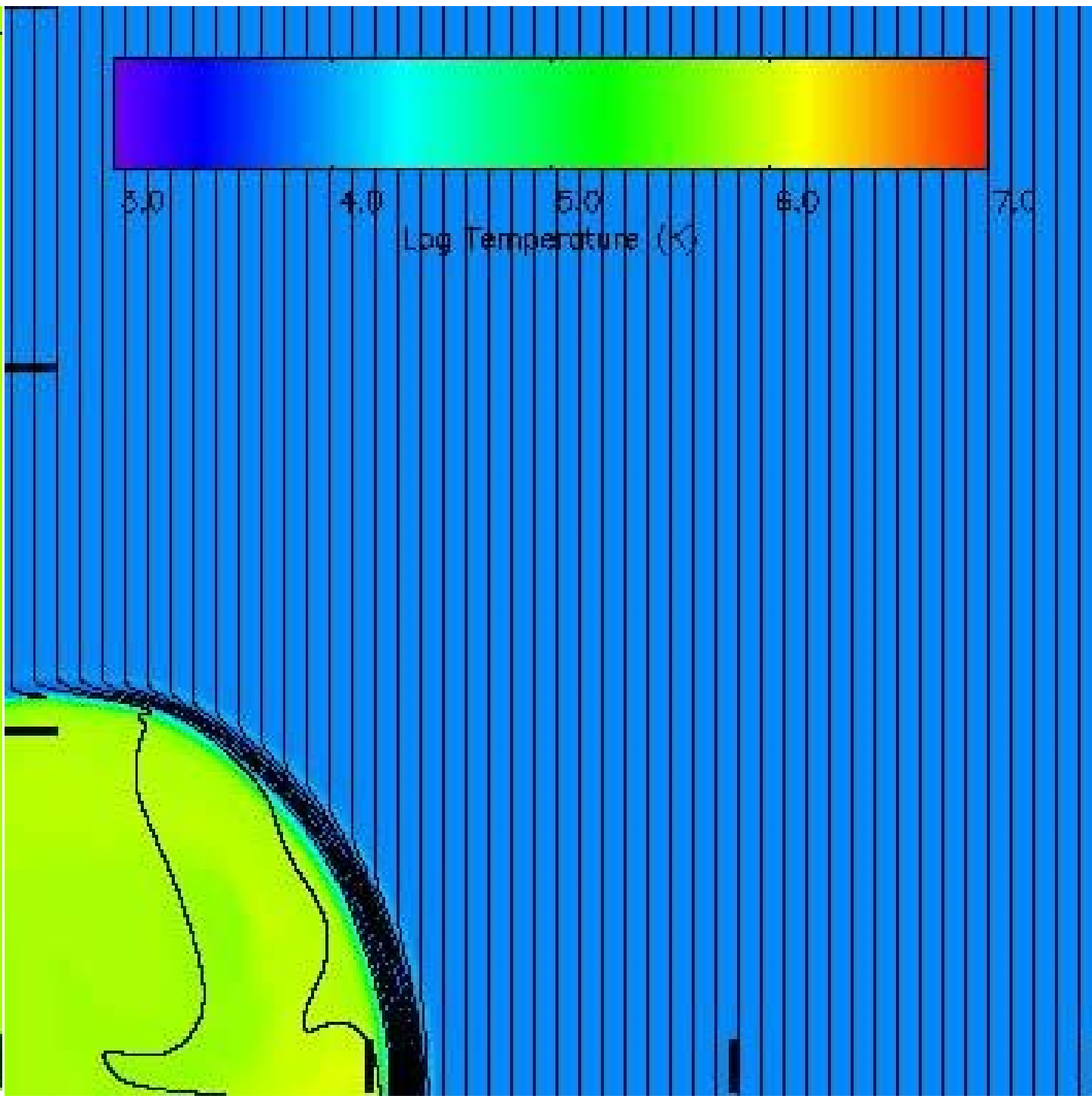

Fig. 8d

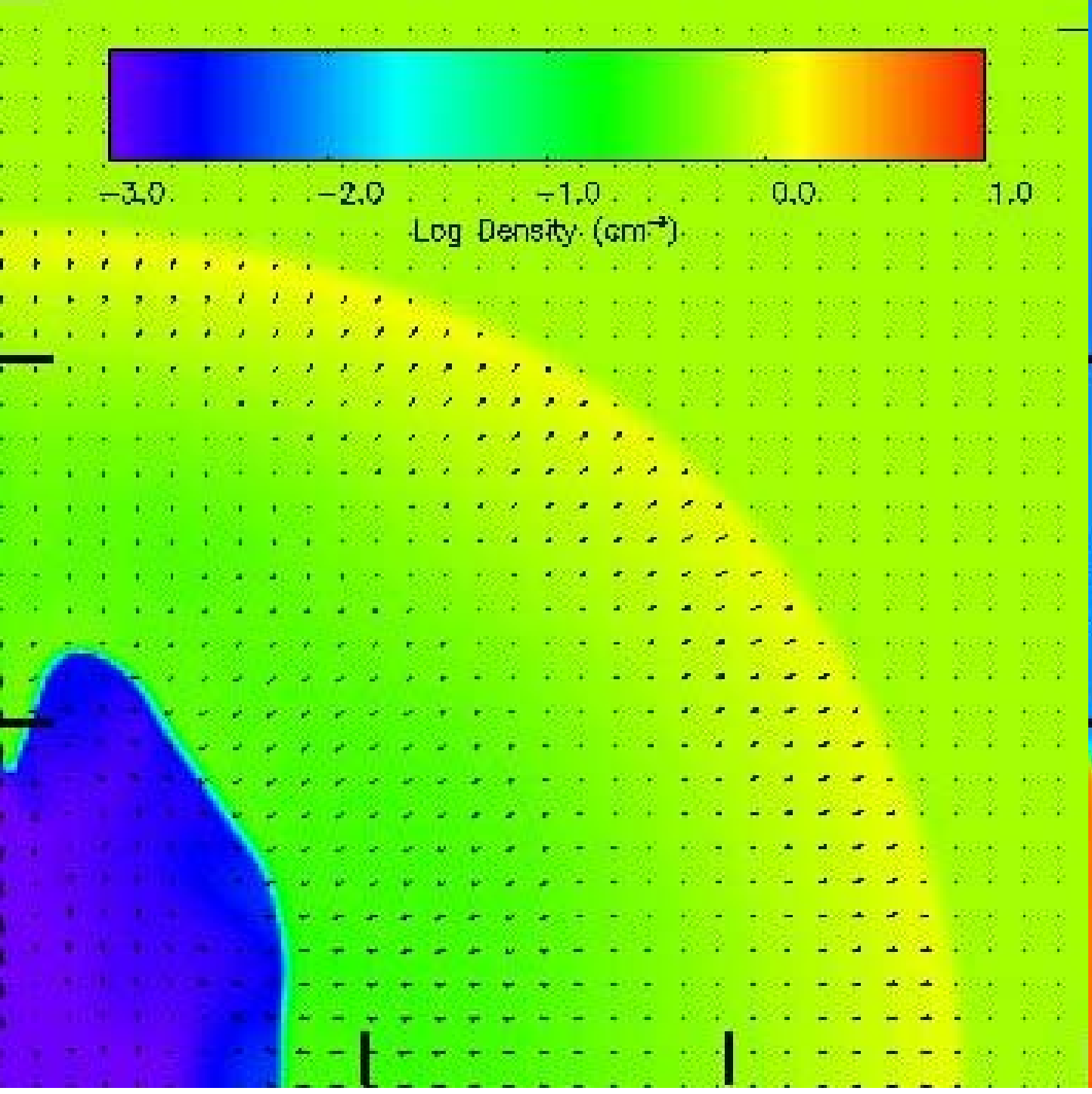

Fig. 8e

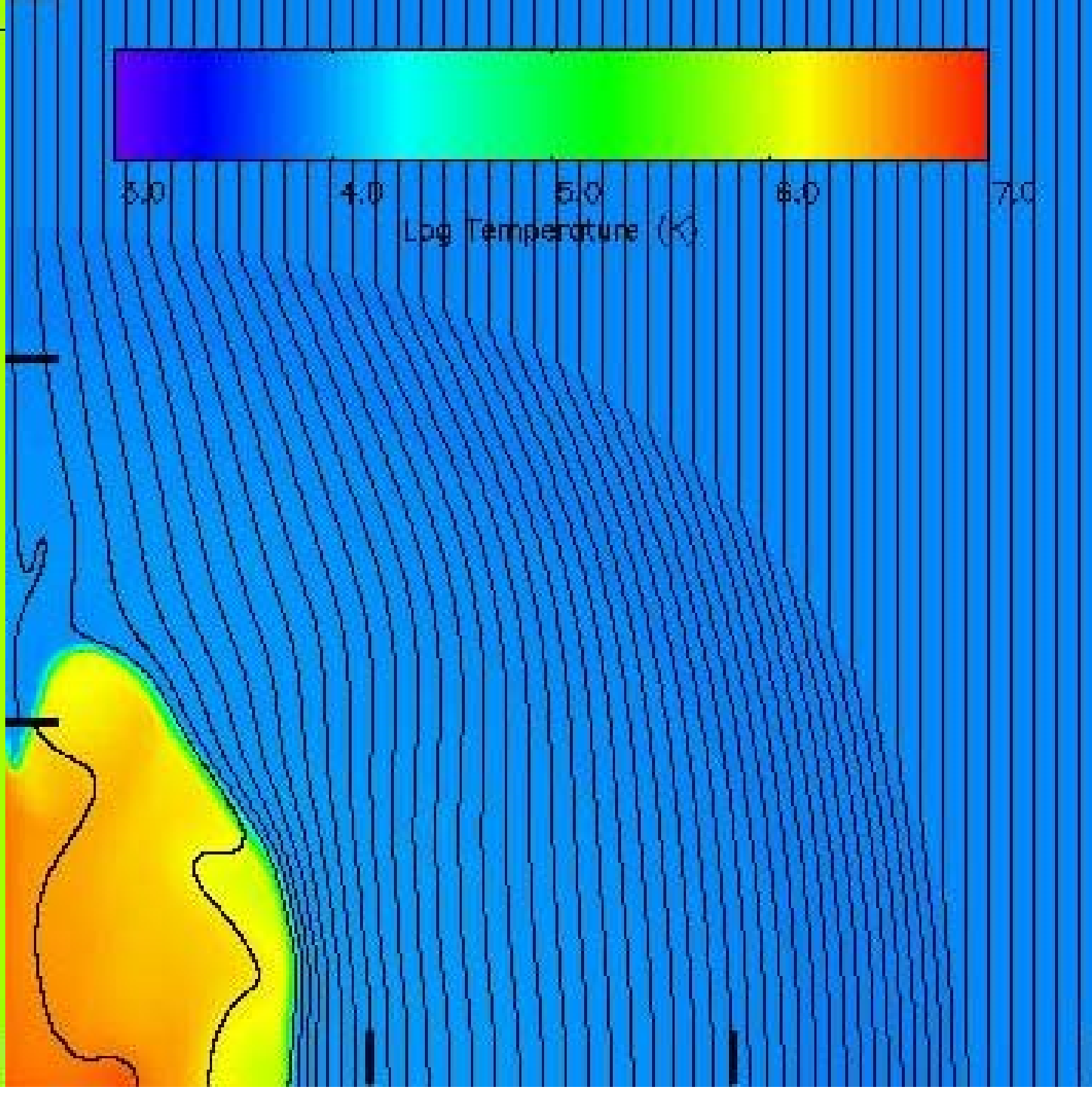

Fig. 8f

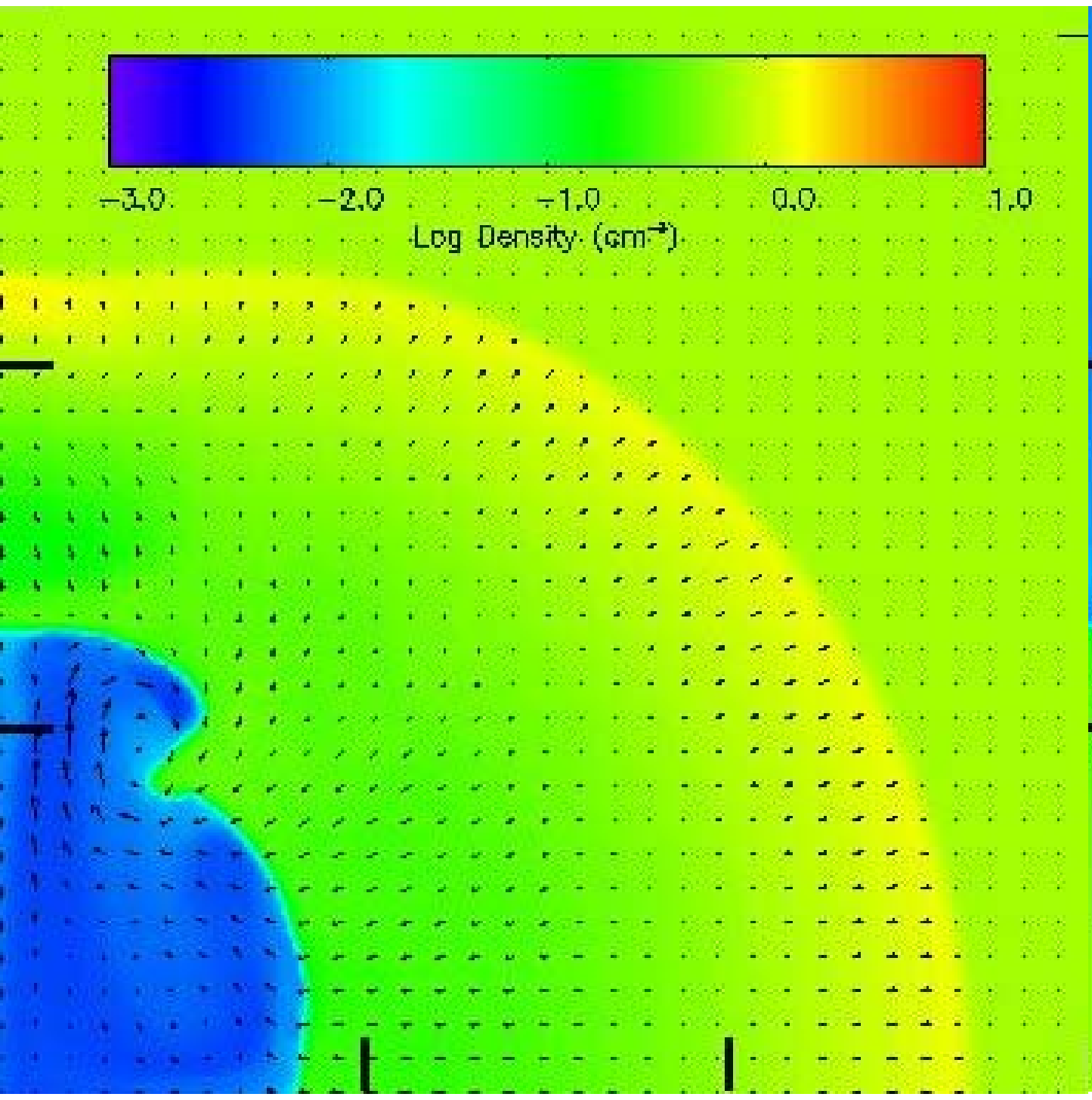

Fig. 8g

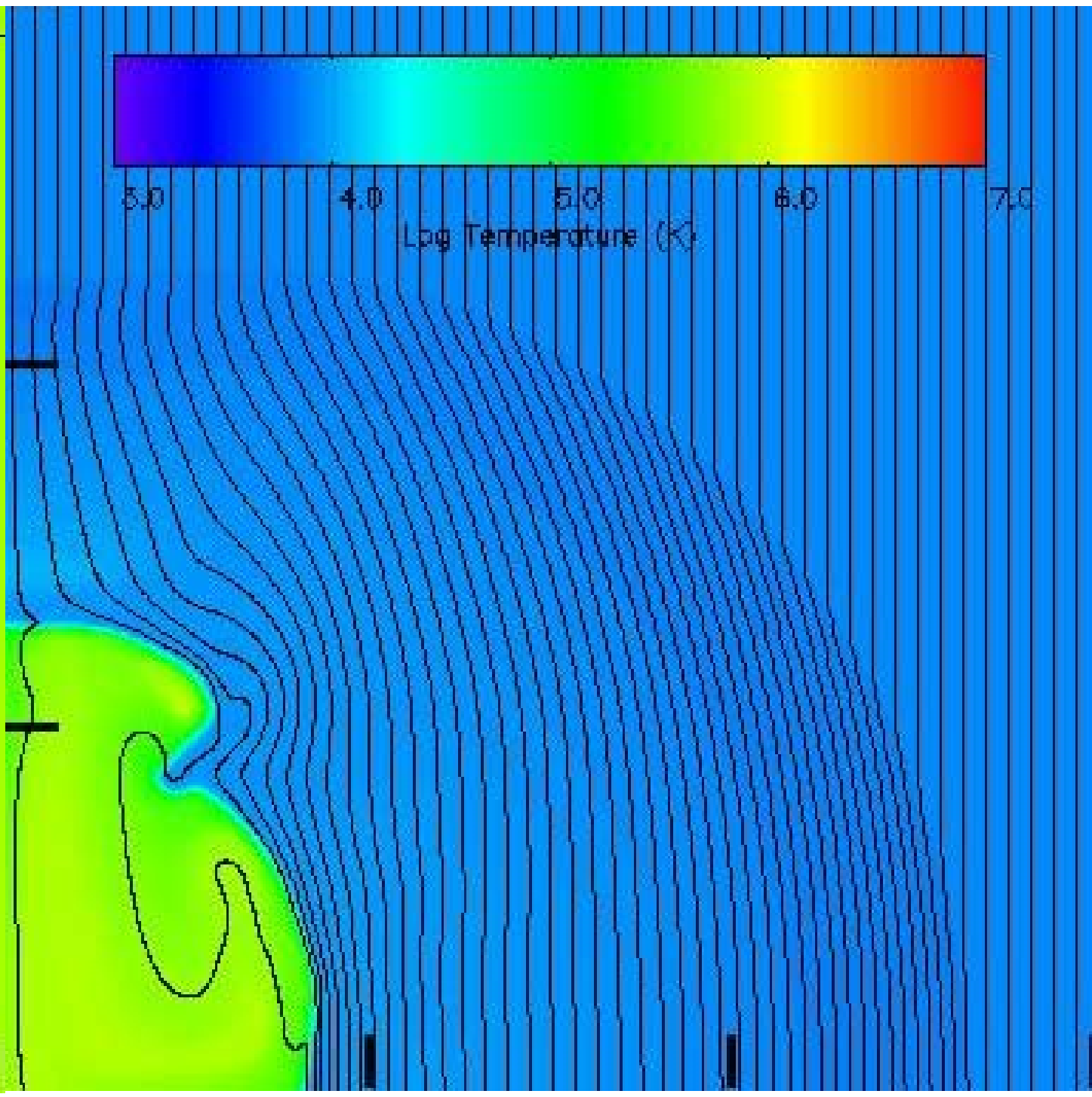

Fig. 8h

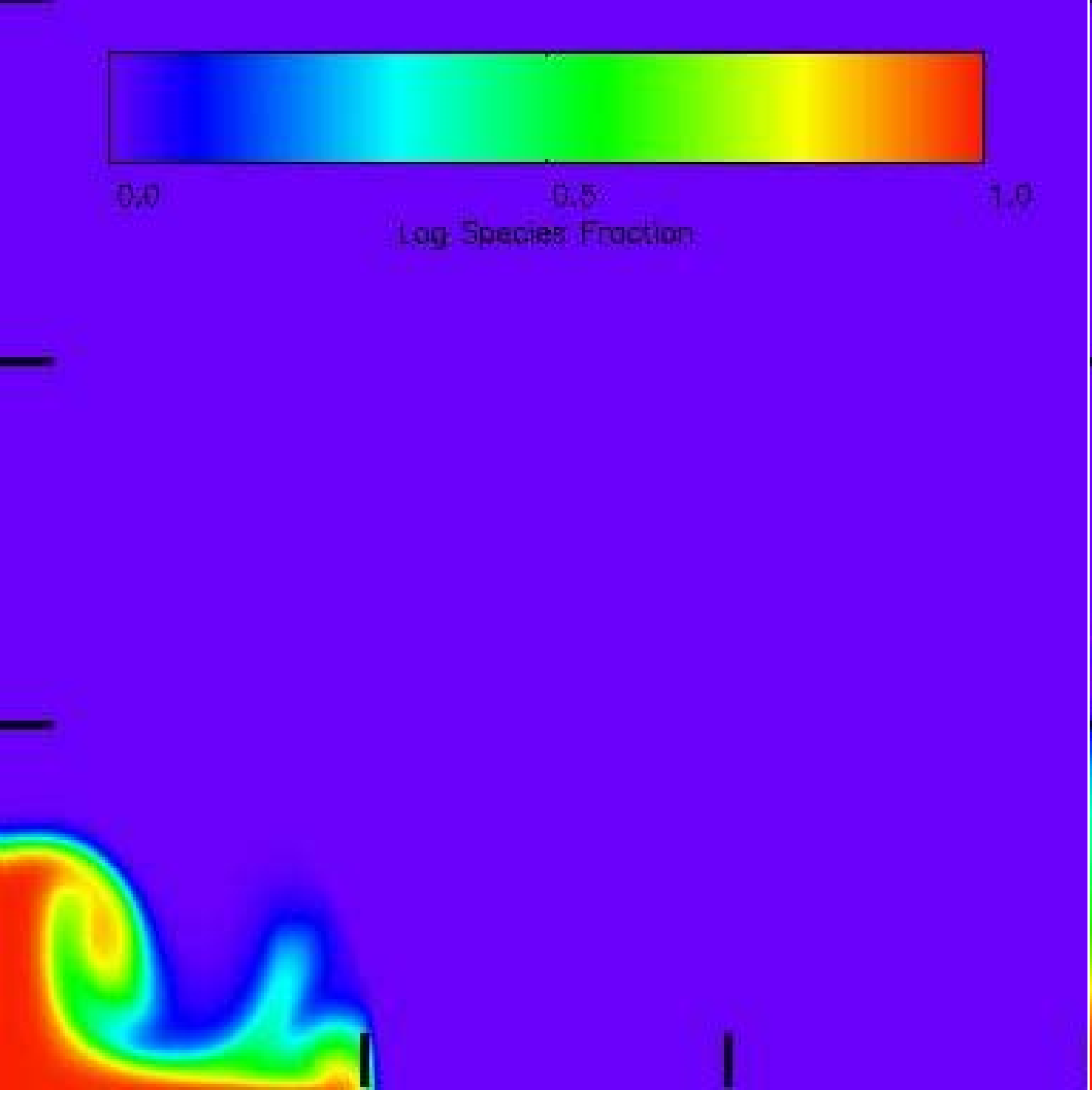

Fig. 8i

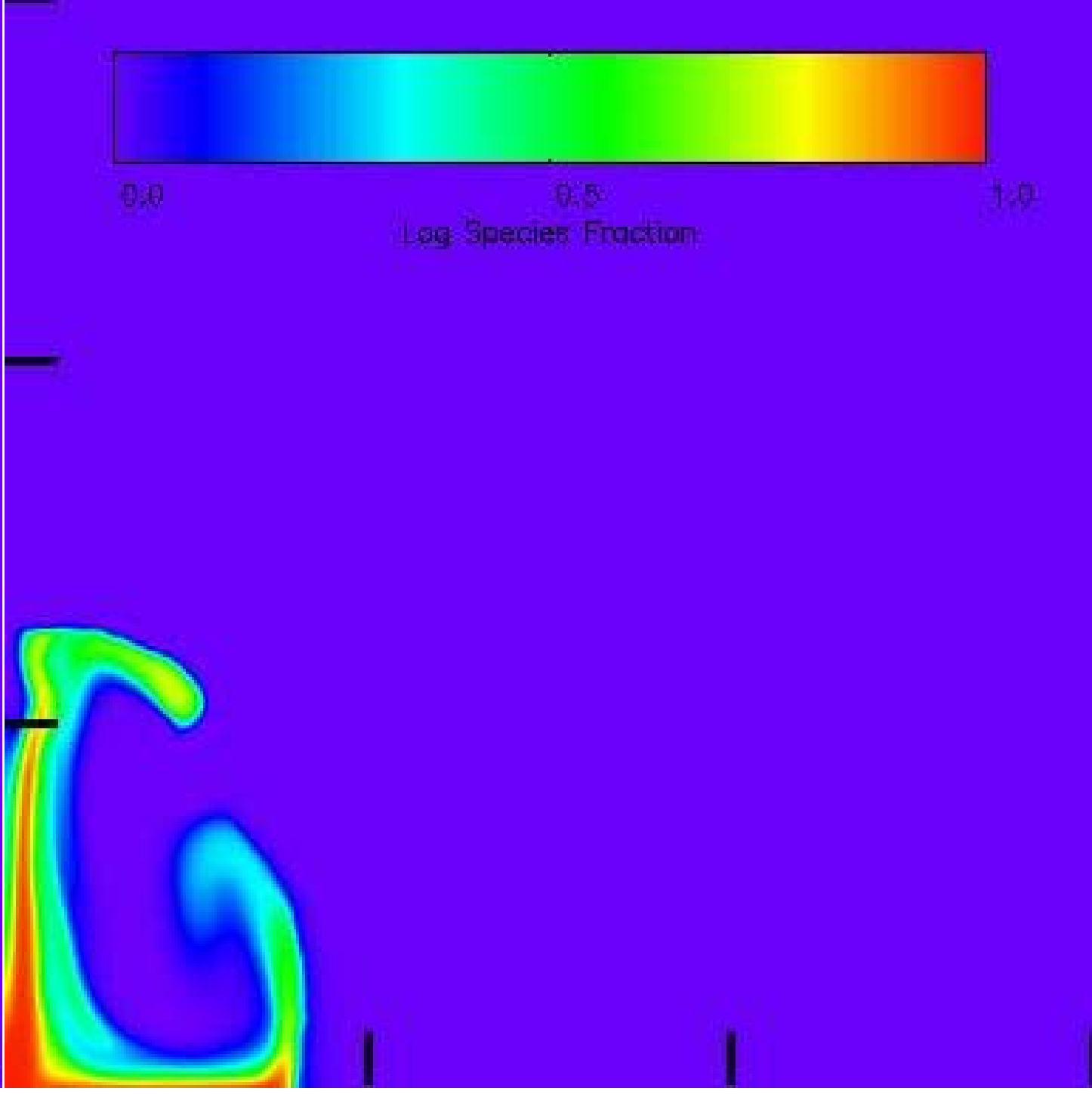

Fig. 8j

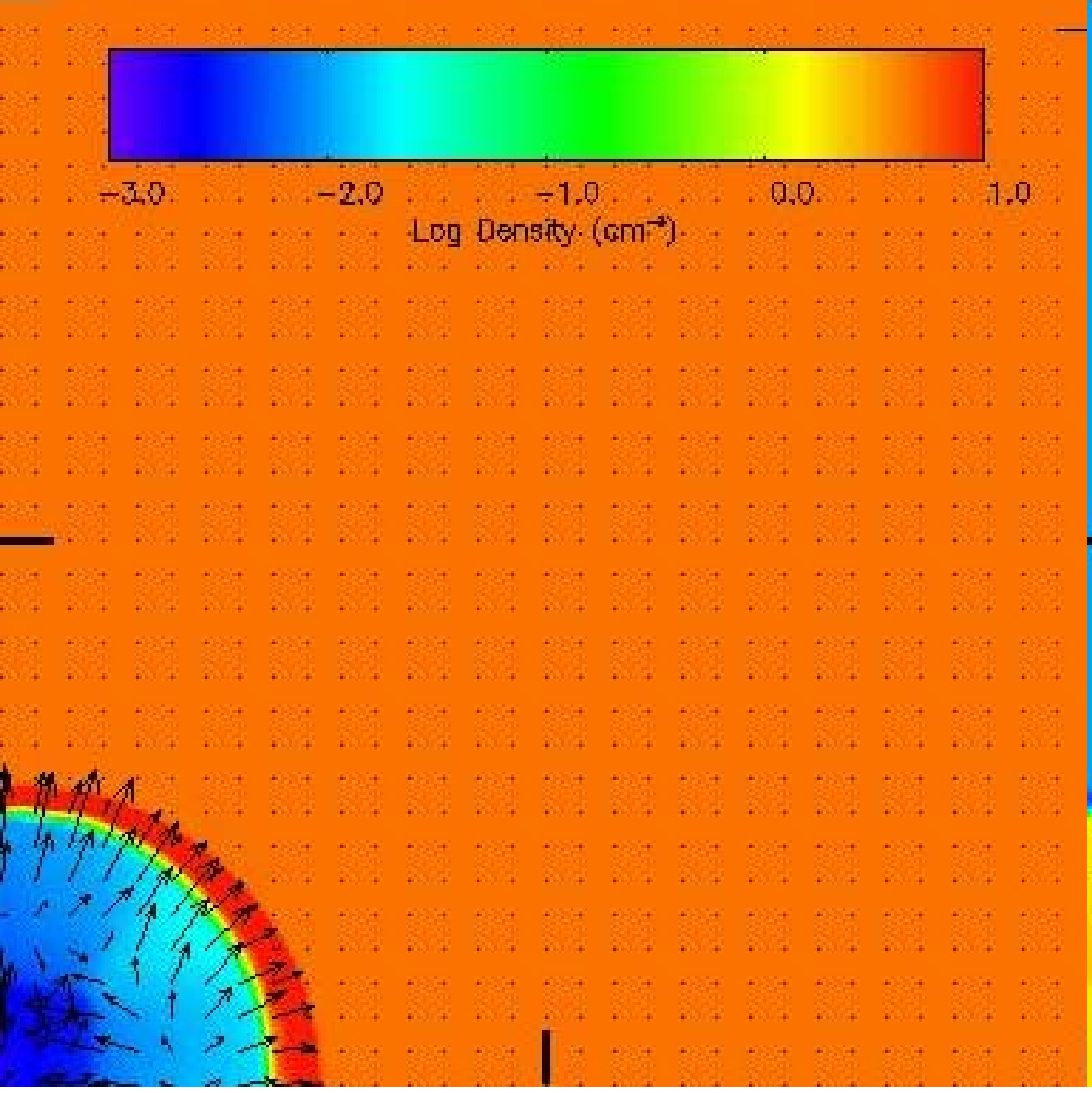

Fig. 9a

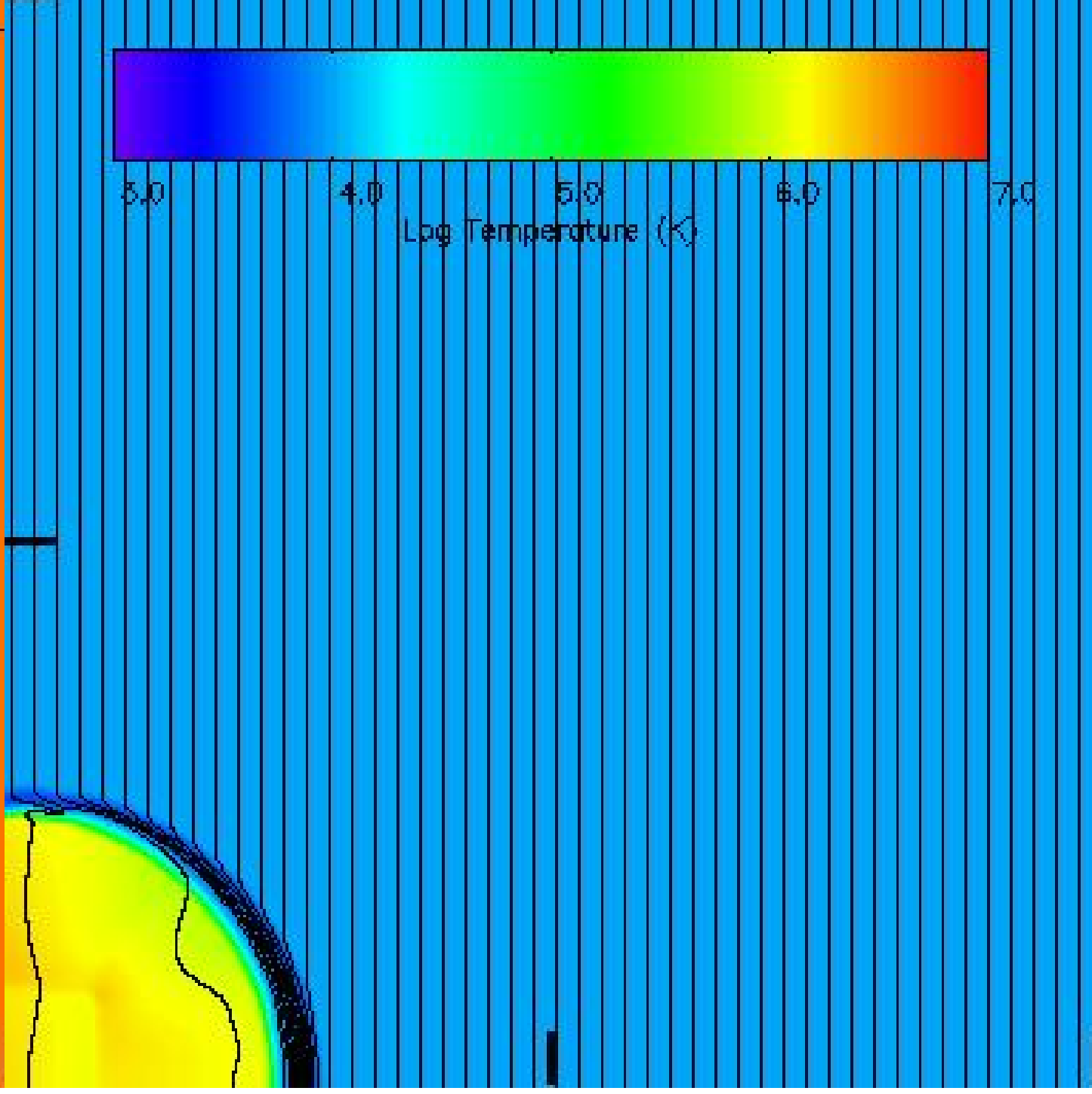

Fig. 9b

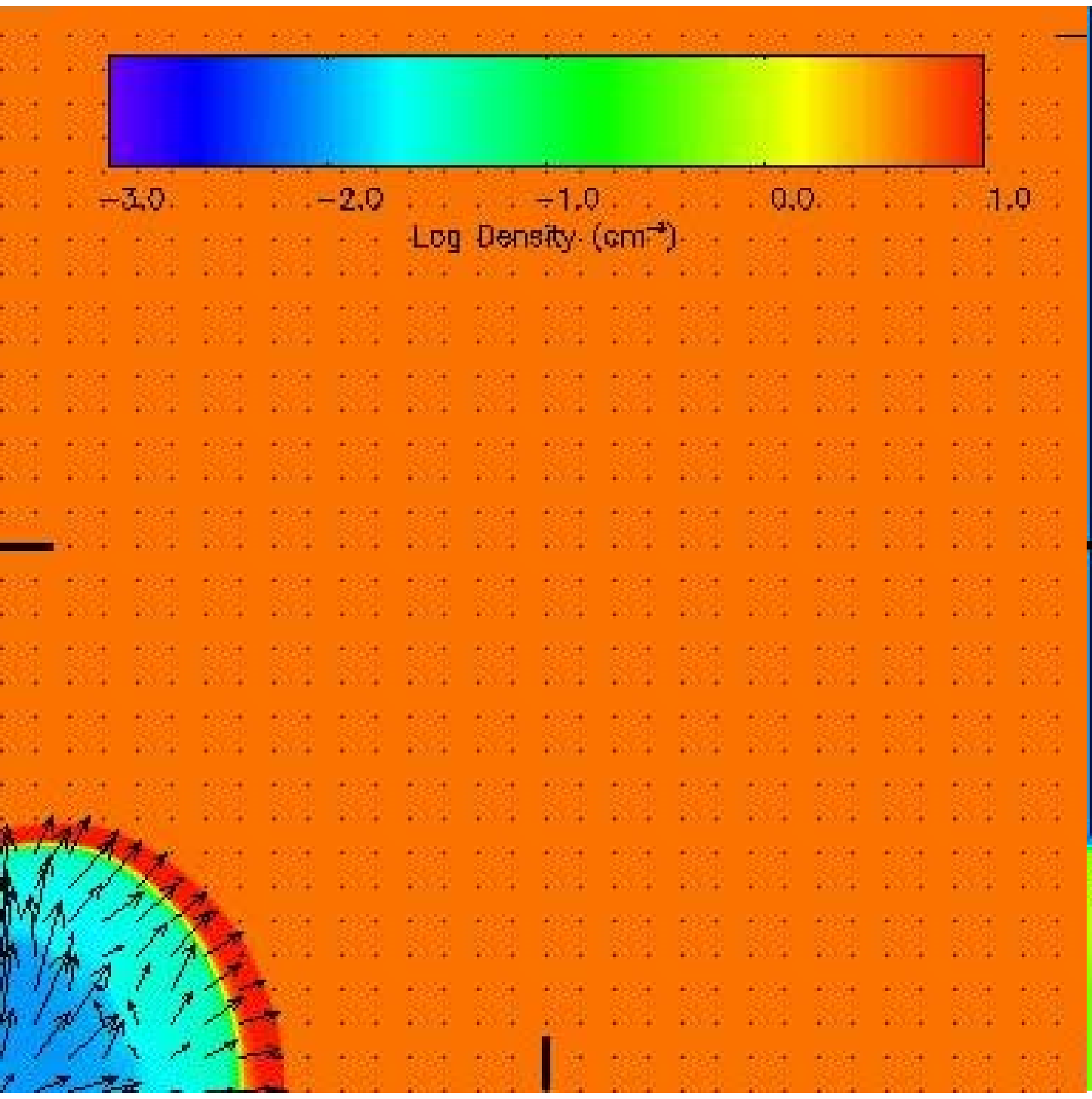

Fig. 9c

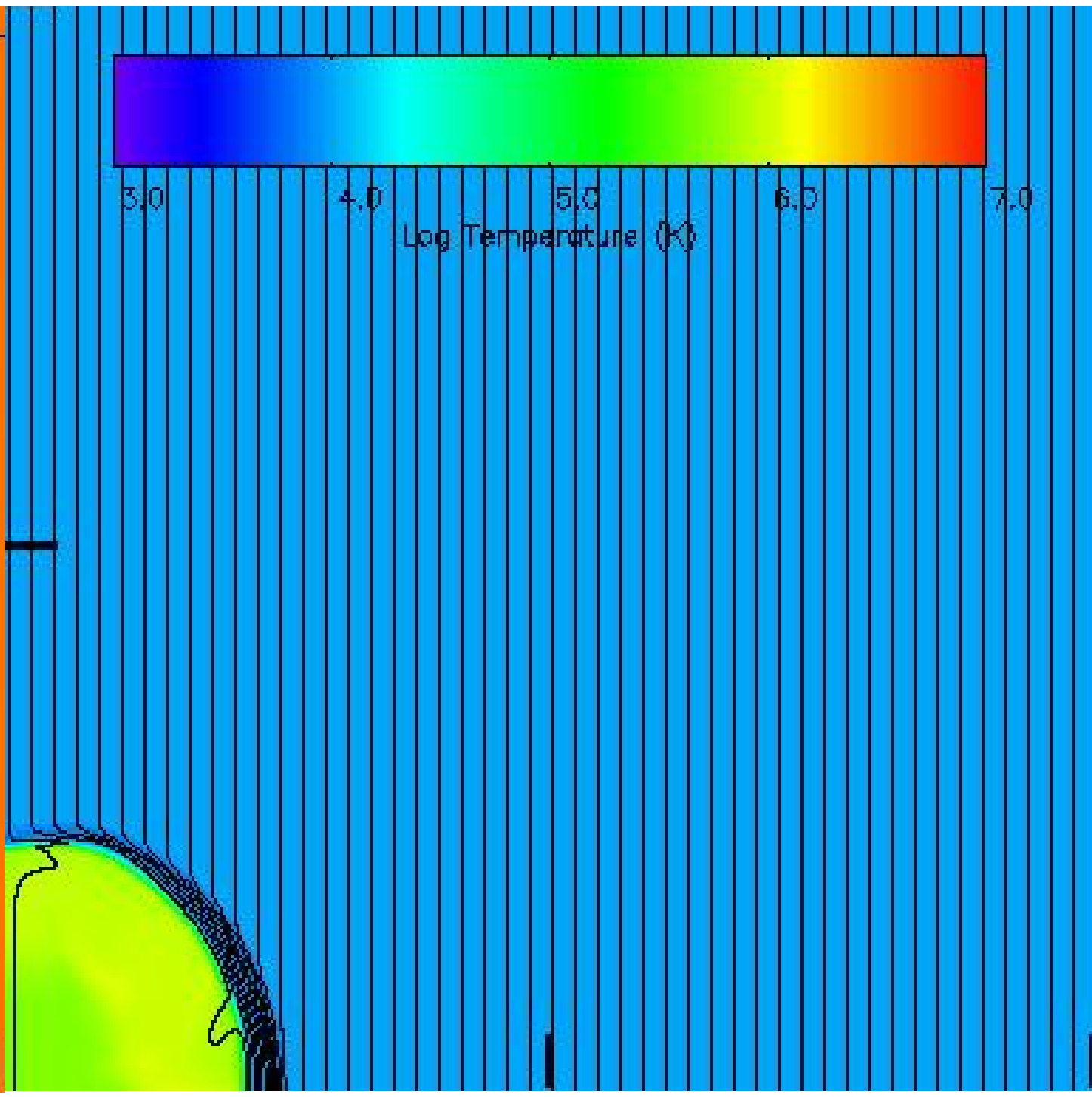

Fig. 9d

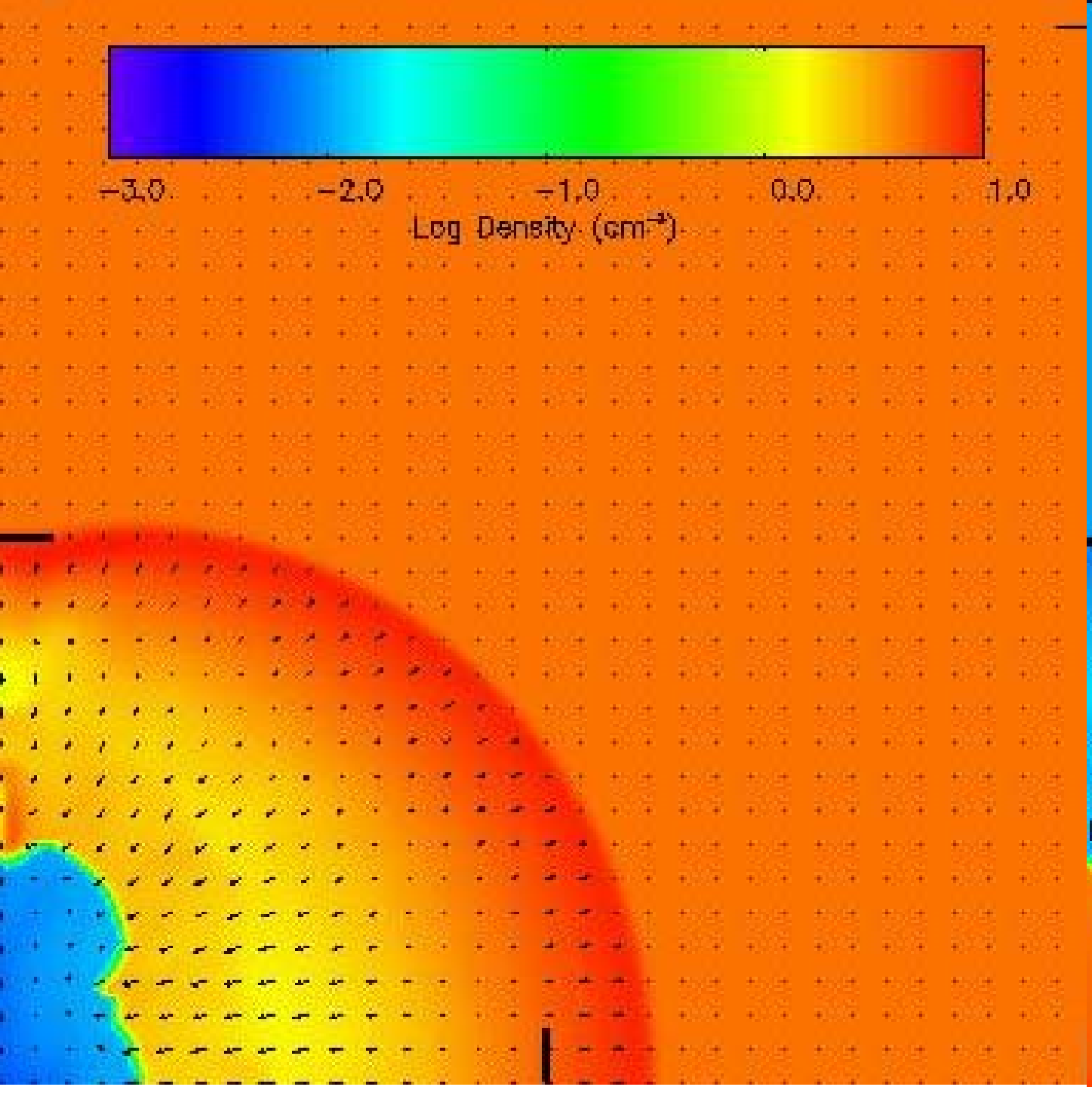

Fig. 9e

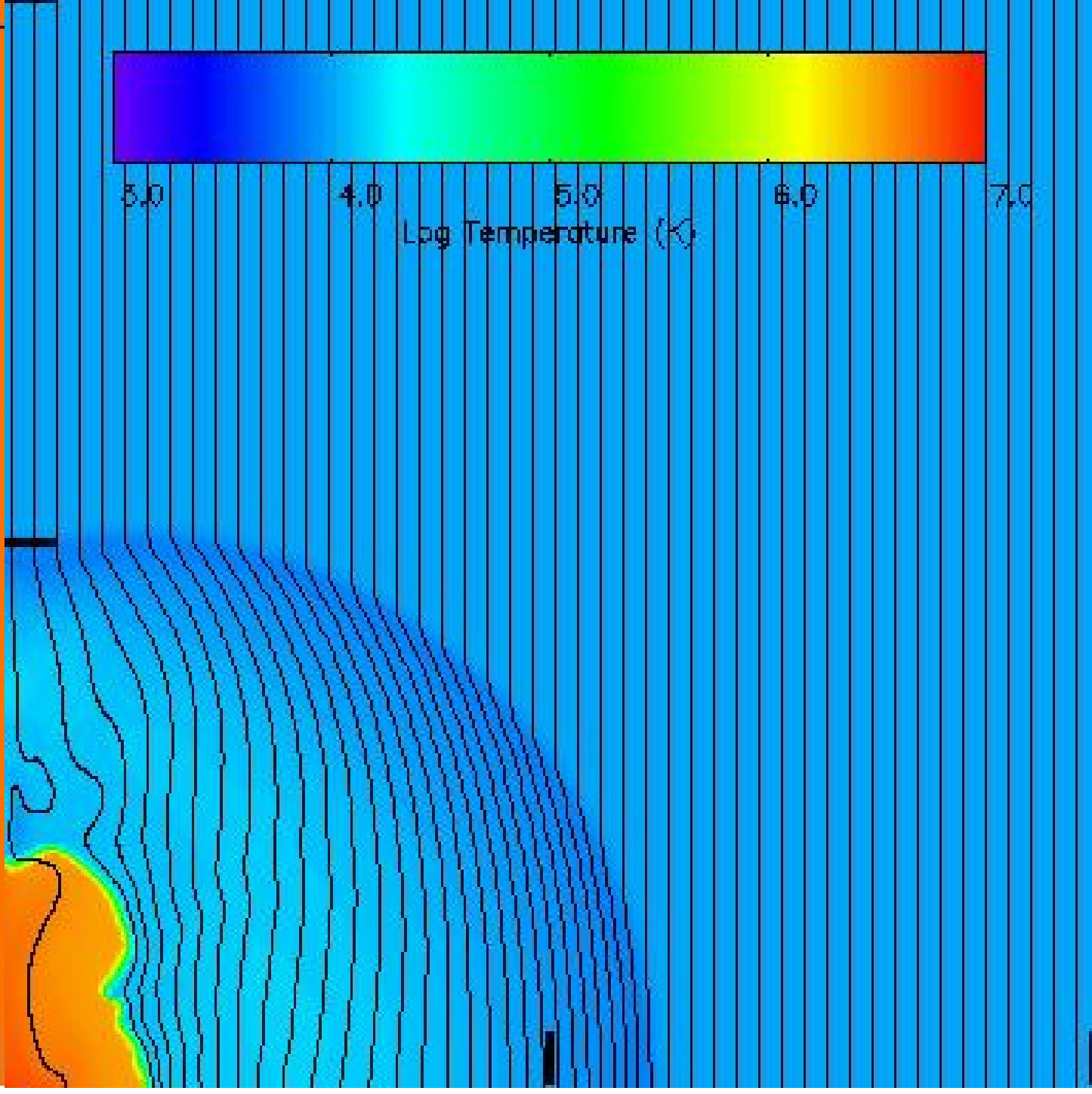

Fig. 9f

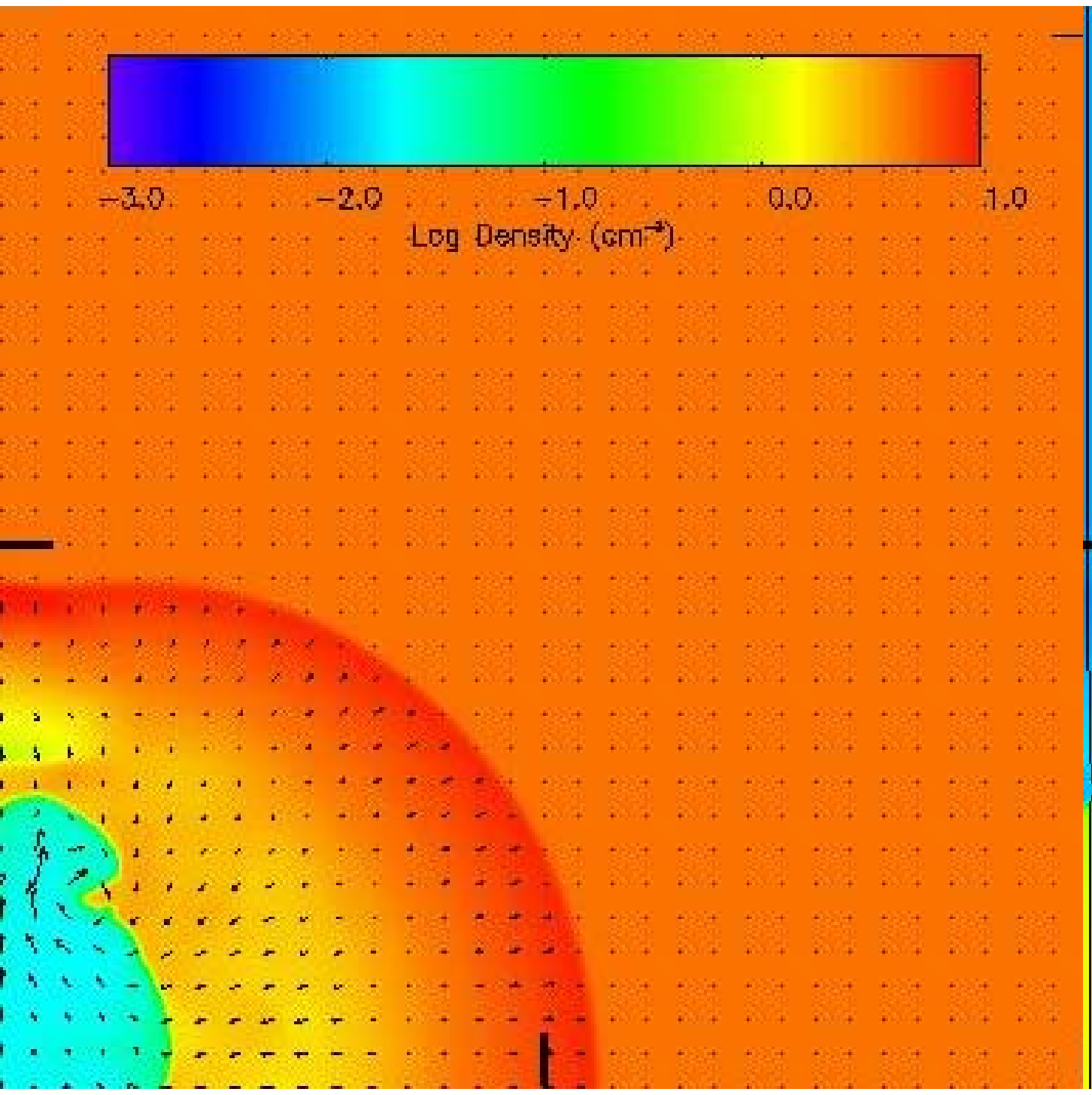

Fig. 9g

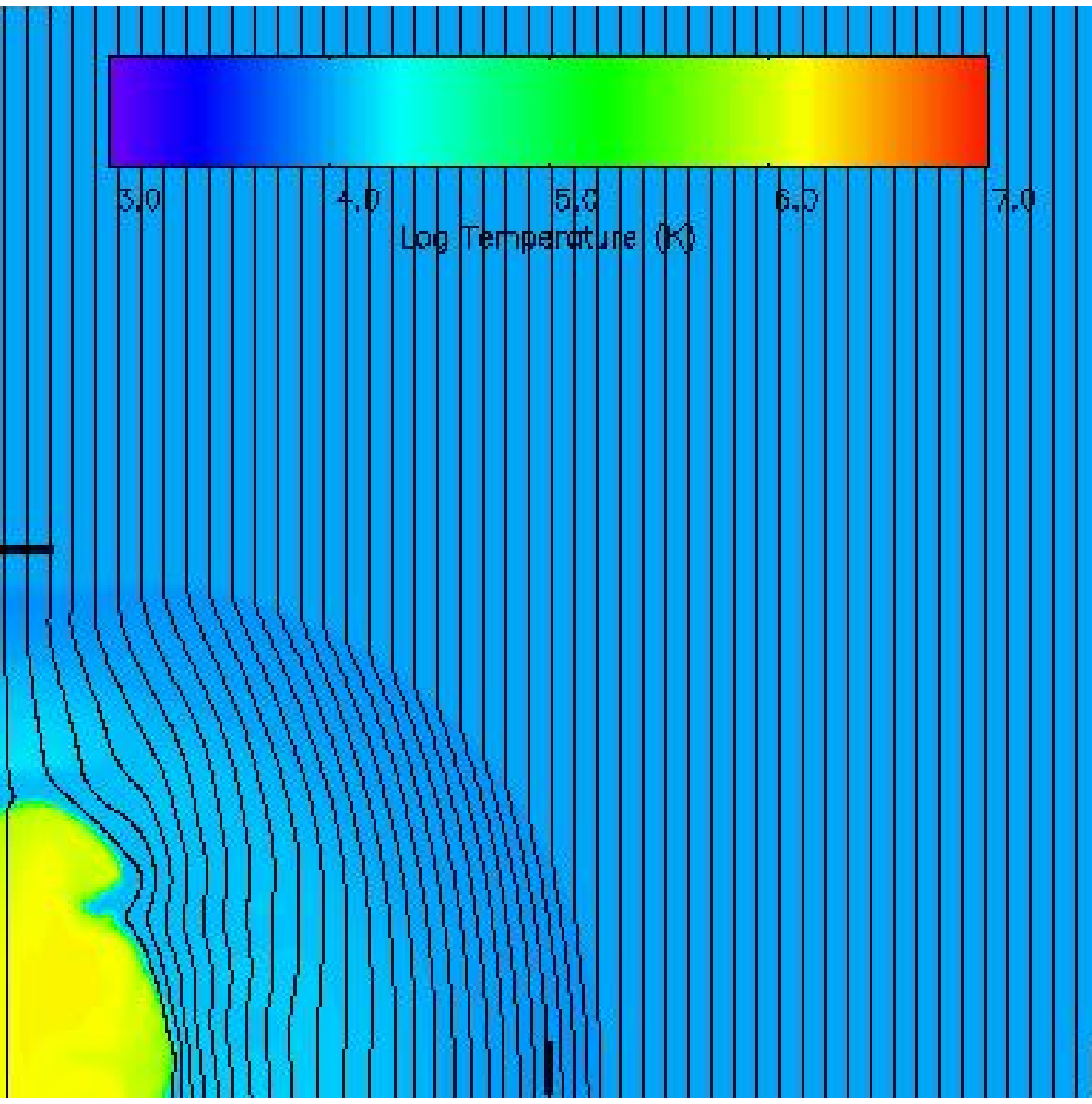

Fig. 9h

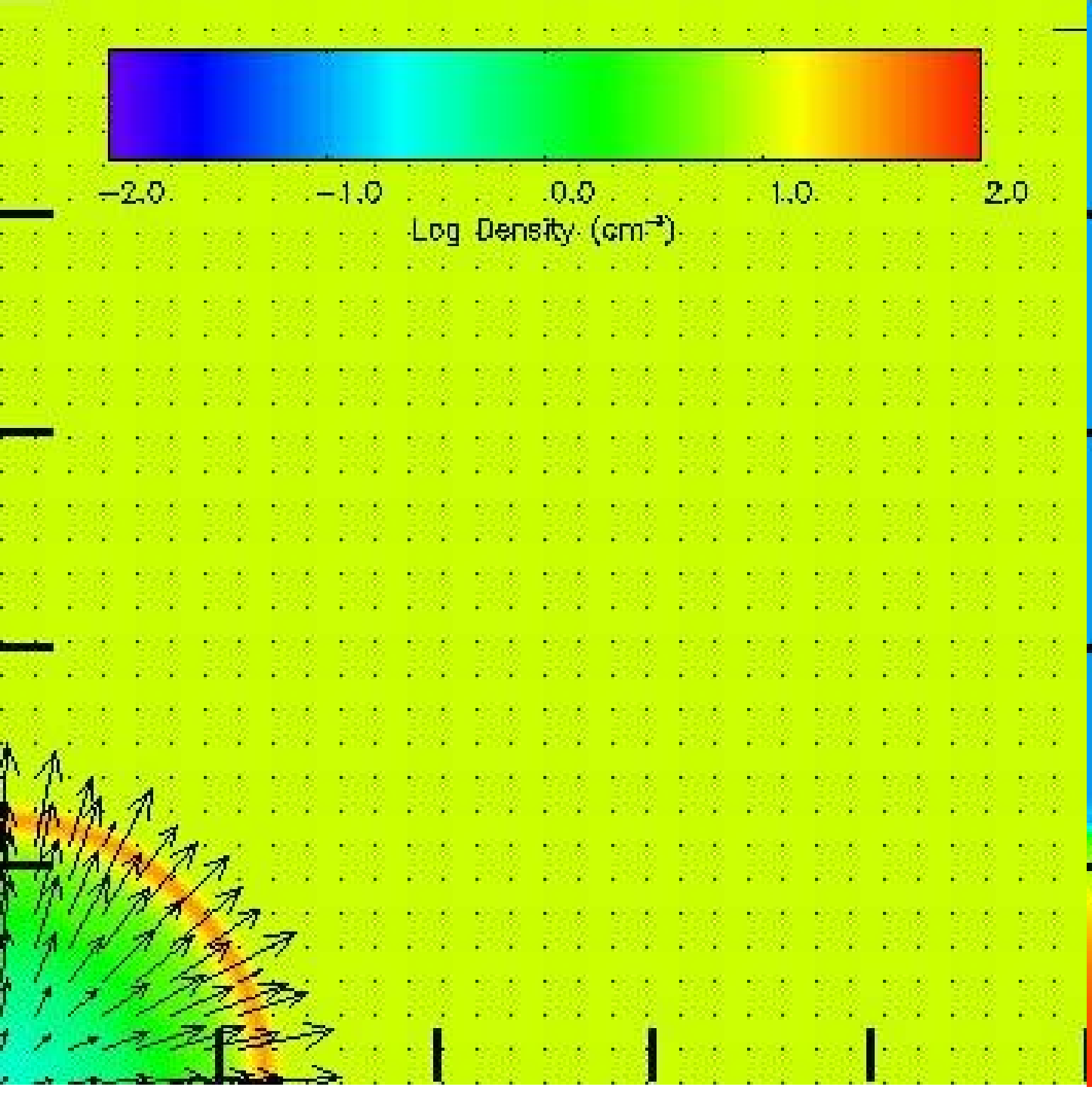

Fig. 10a

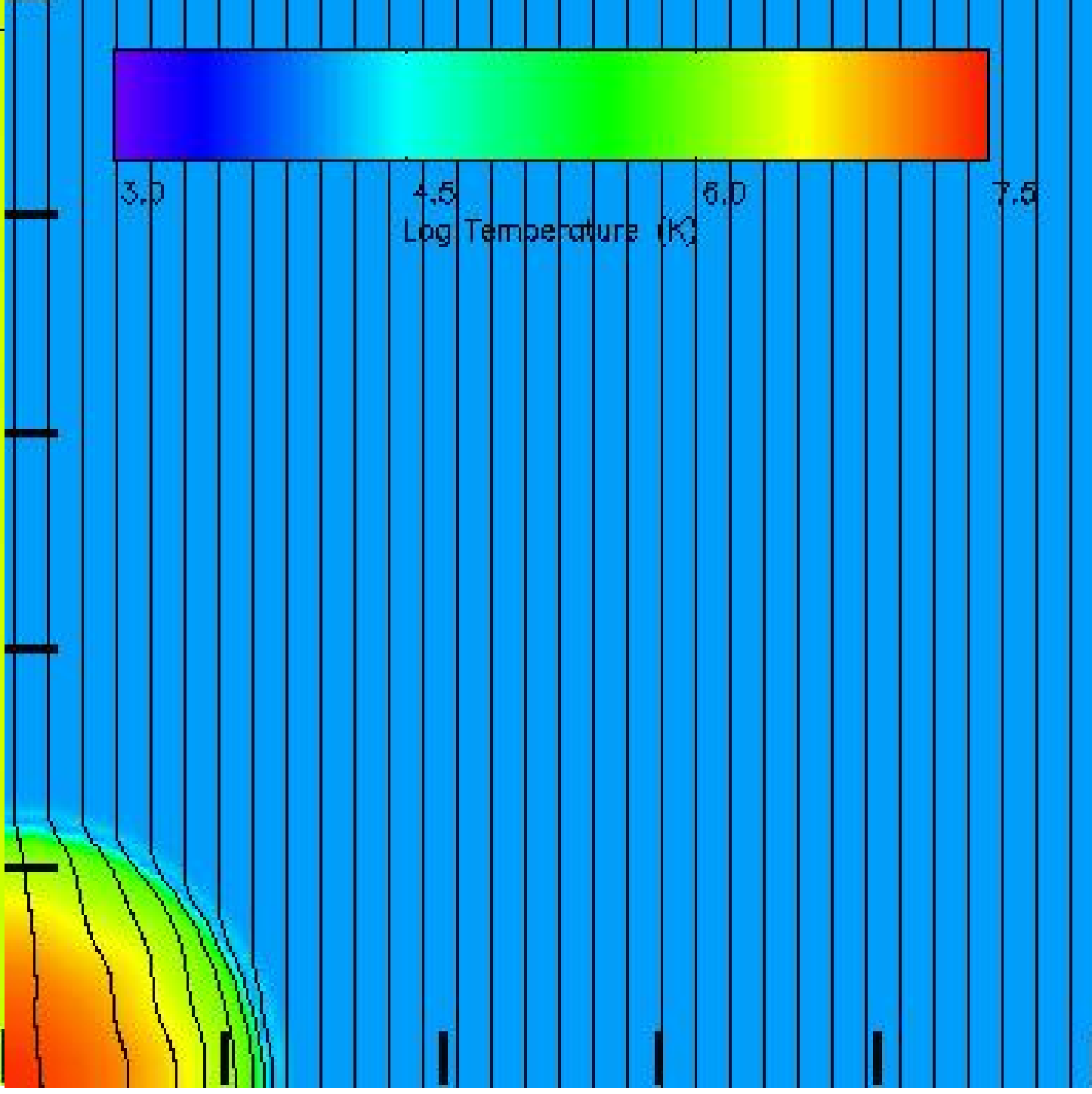

Fig. 10b

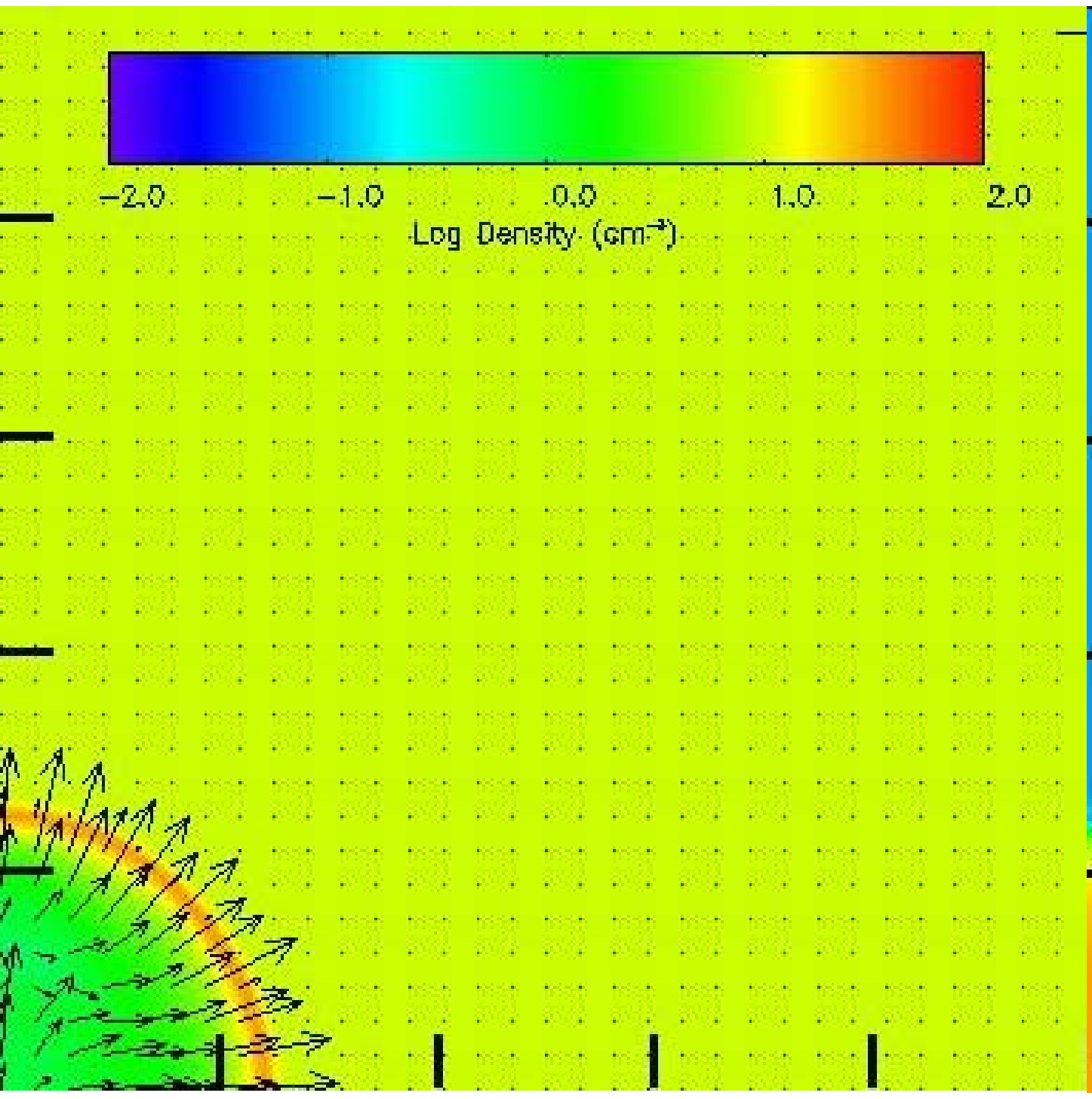

Fig. 10c

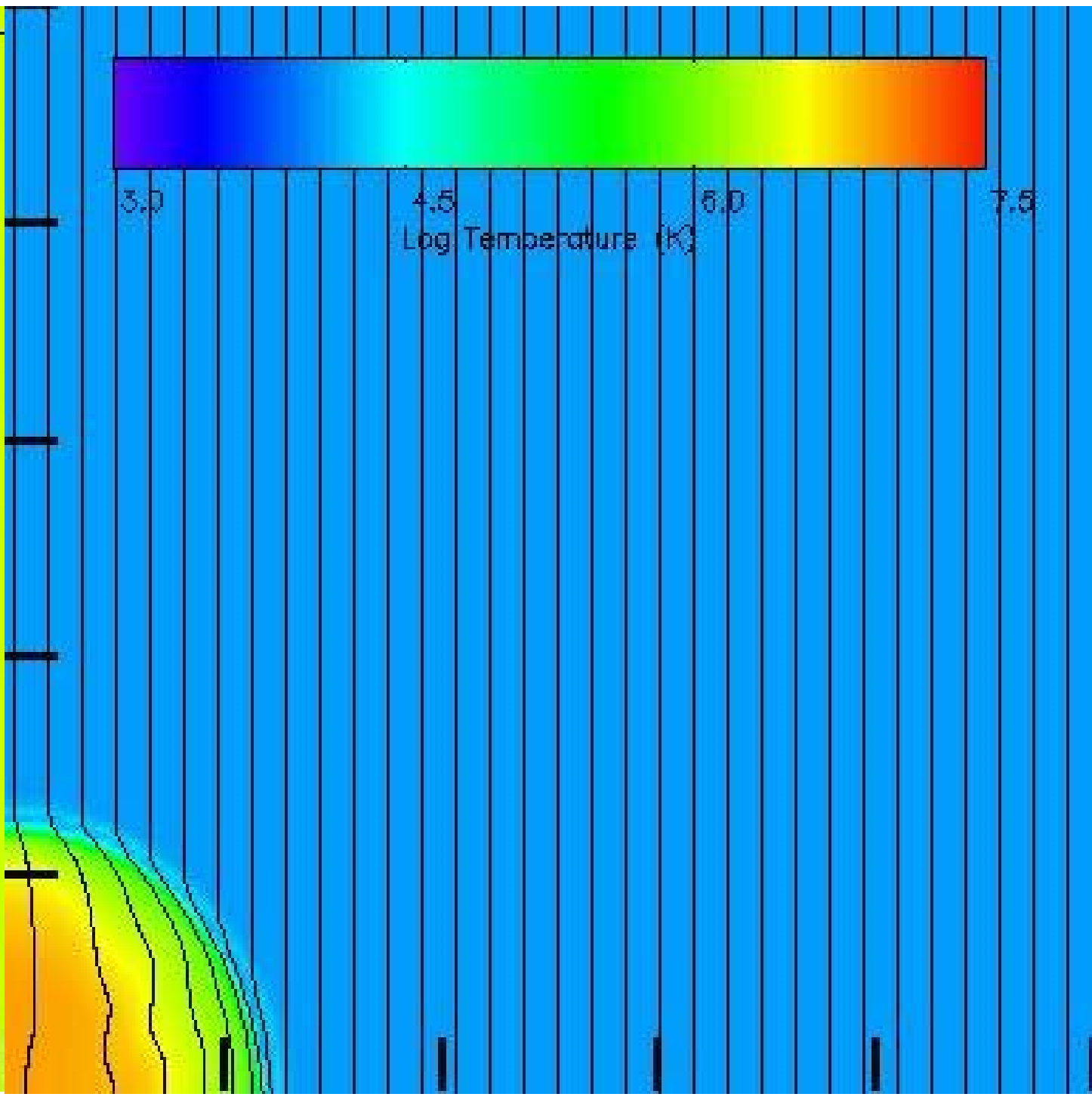

Fig. 10d

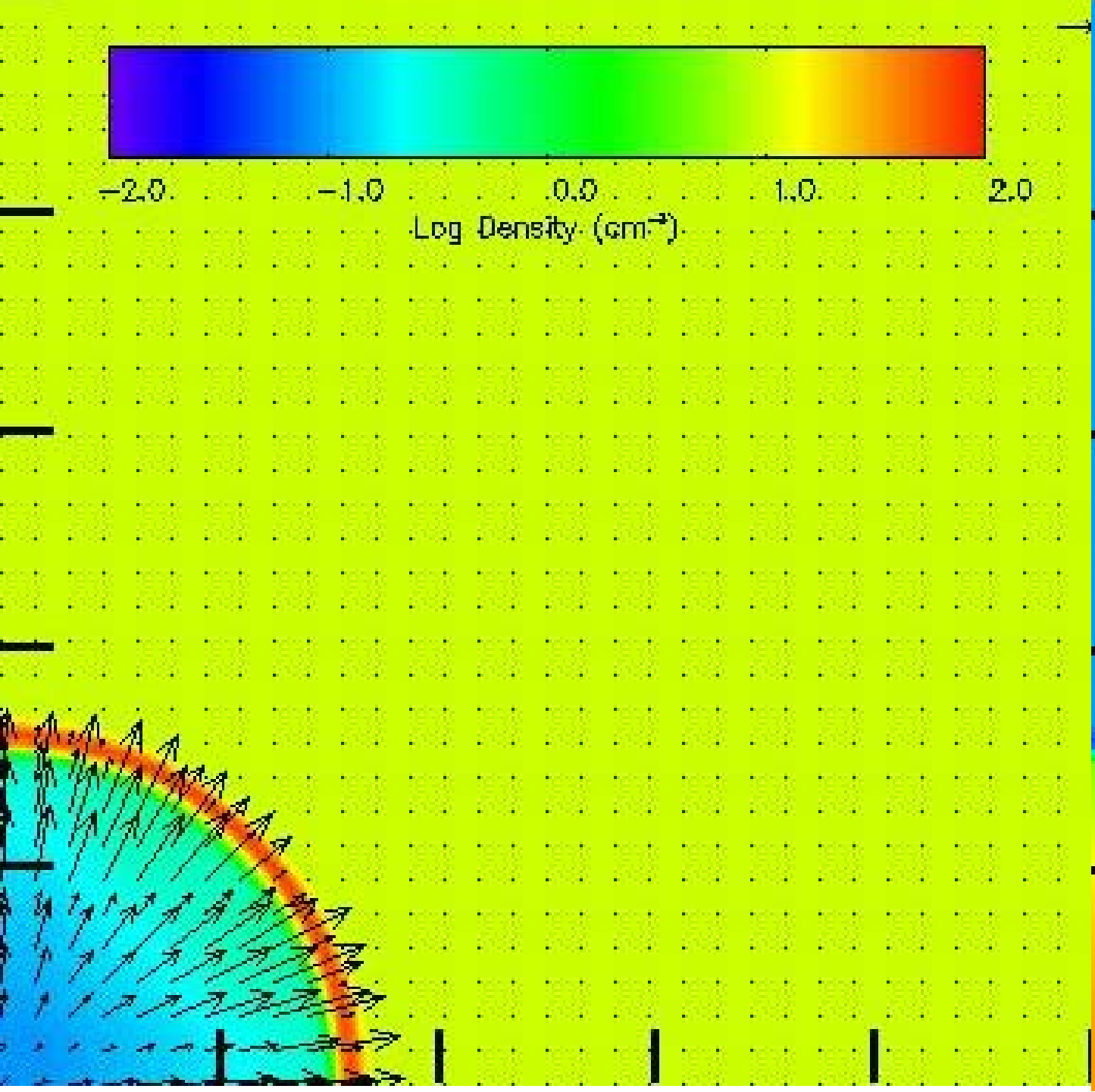

Fig. 10e

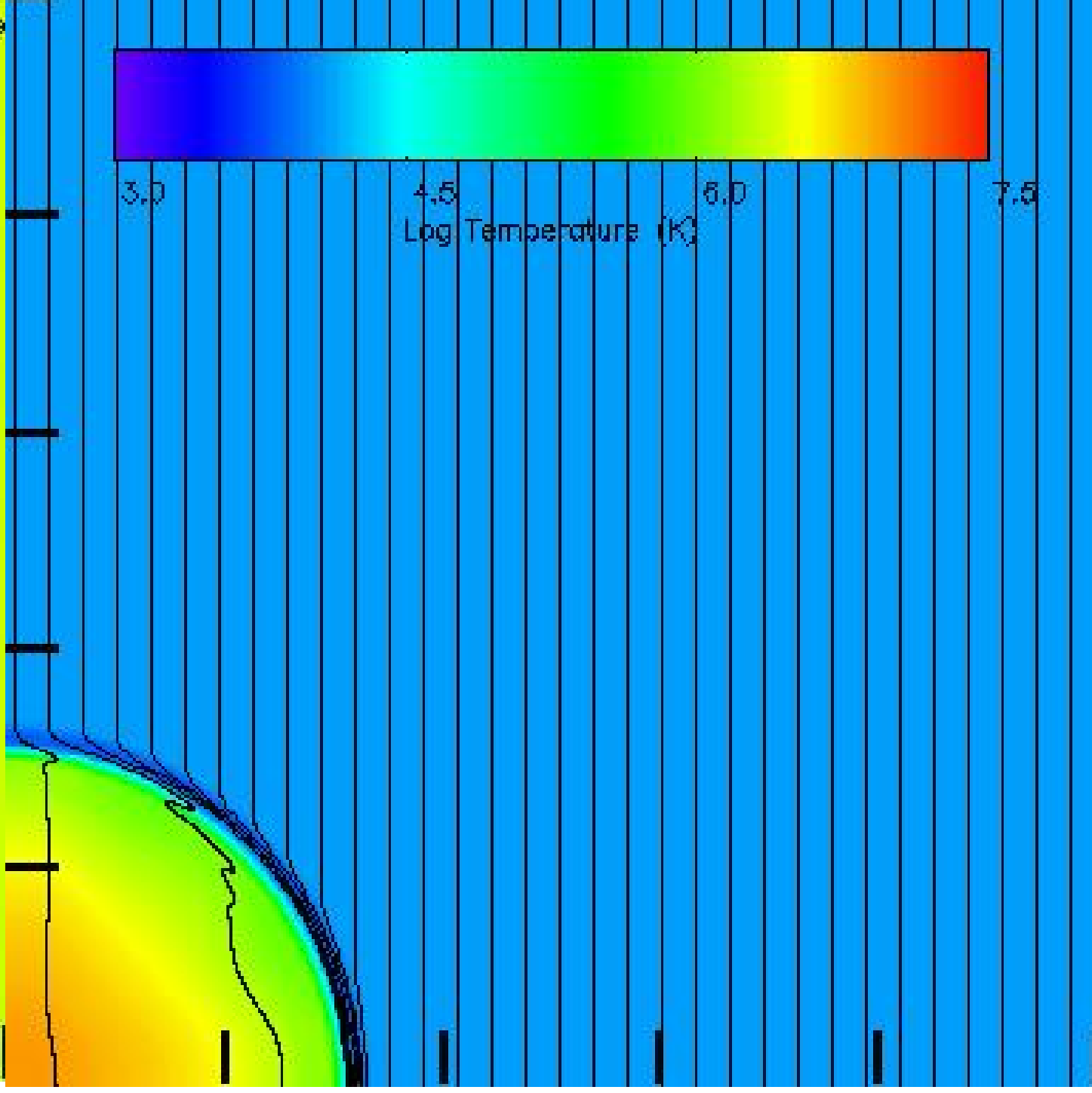

Fig. 10f

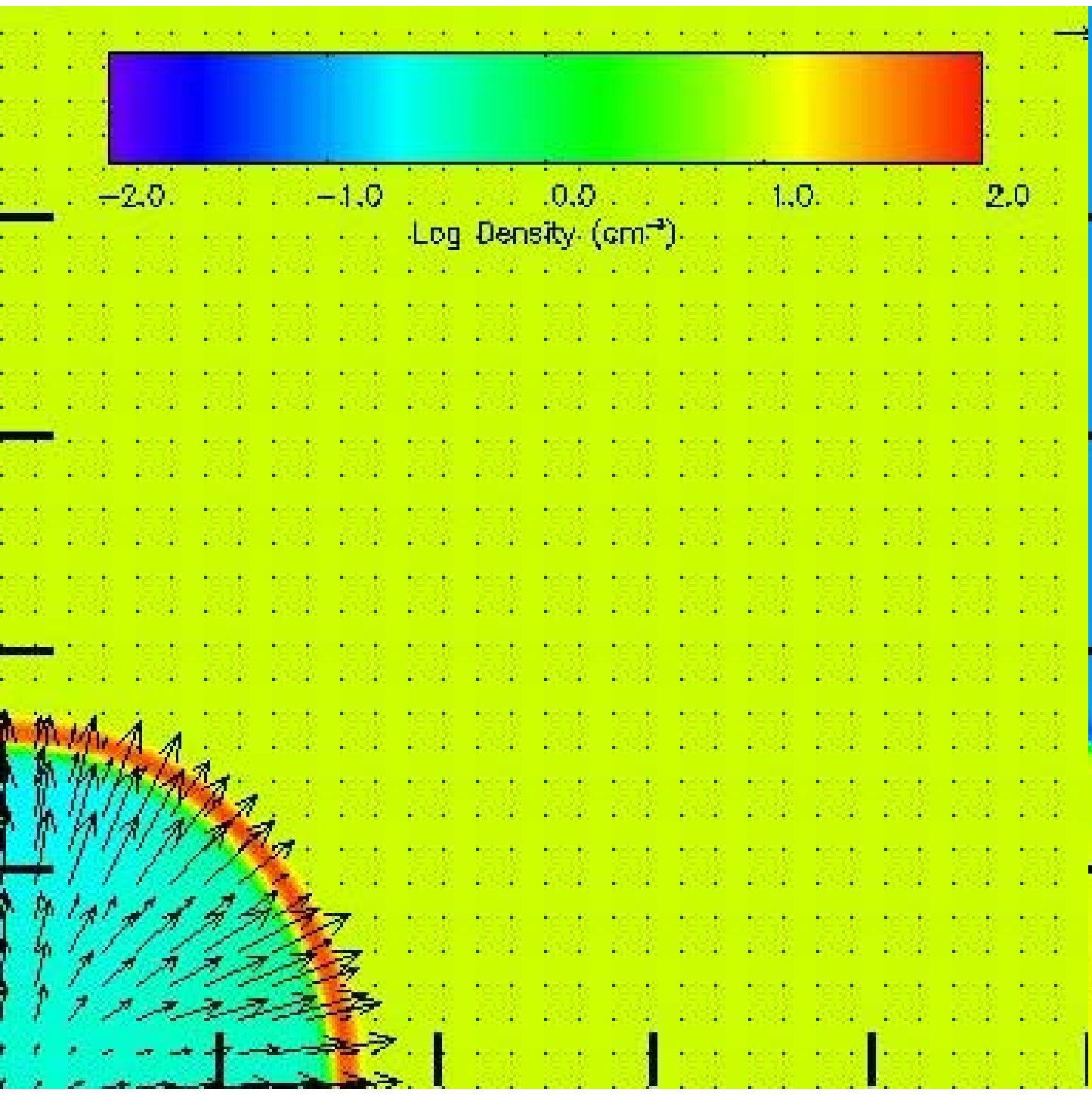

Fig. 10g

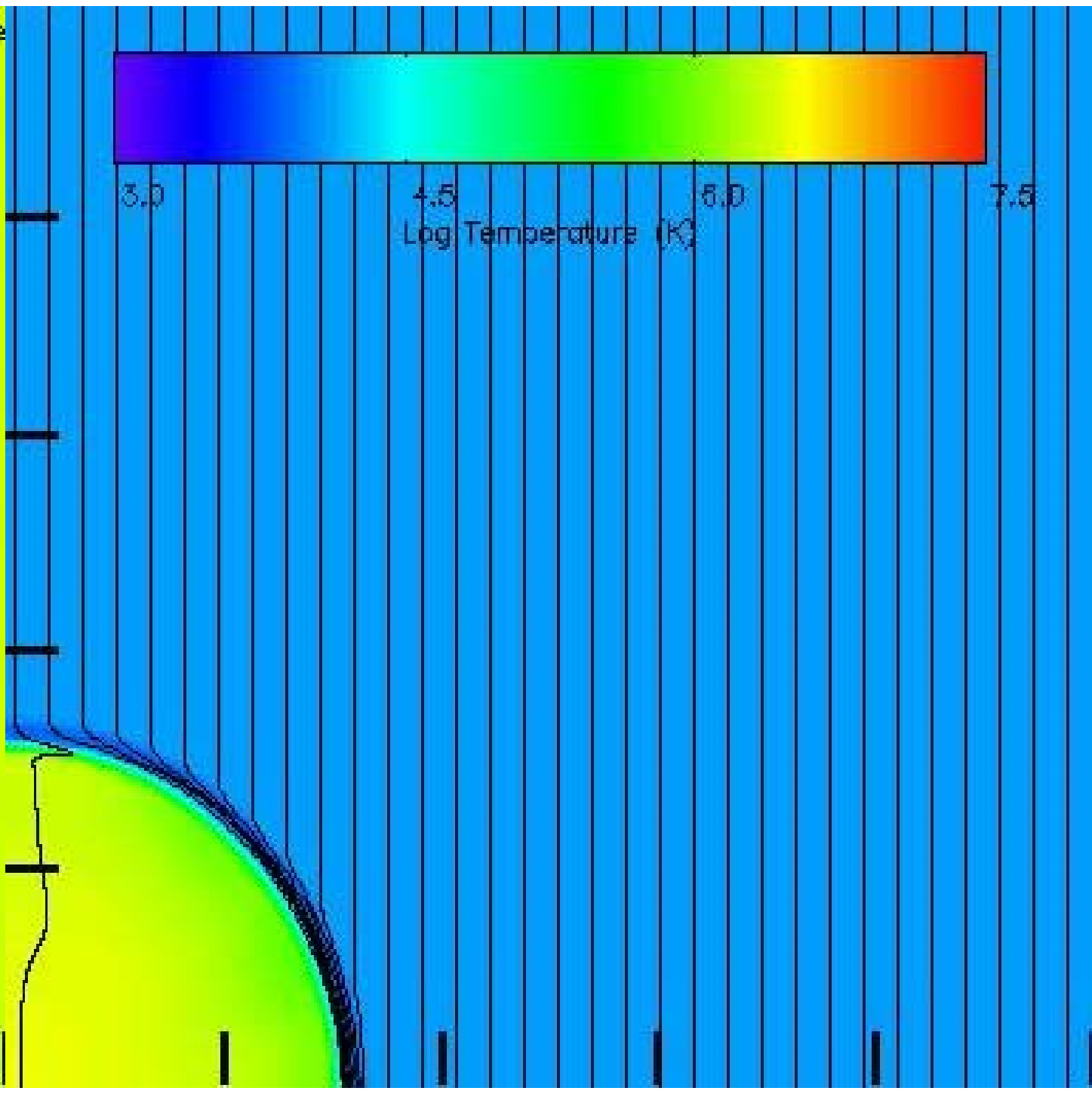

Fig. 10h

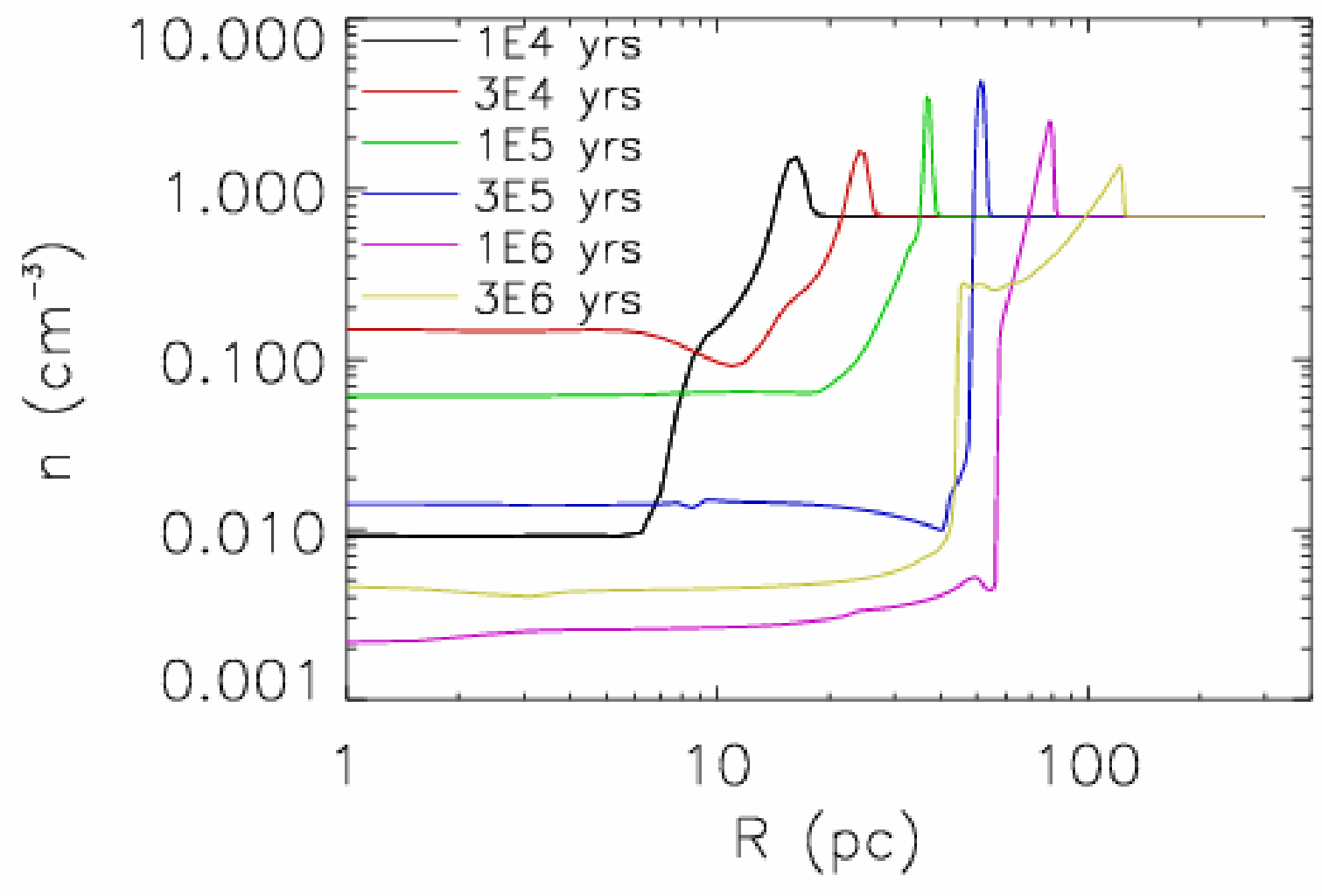

Fig. 11a

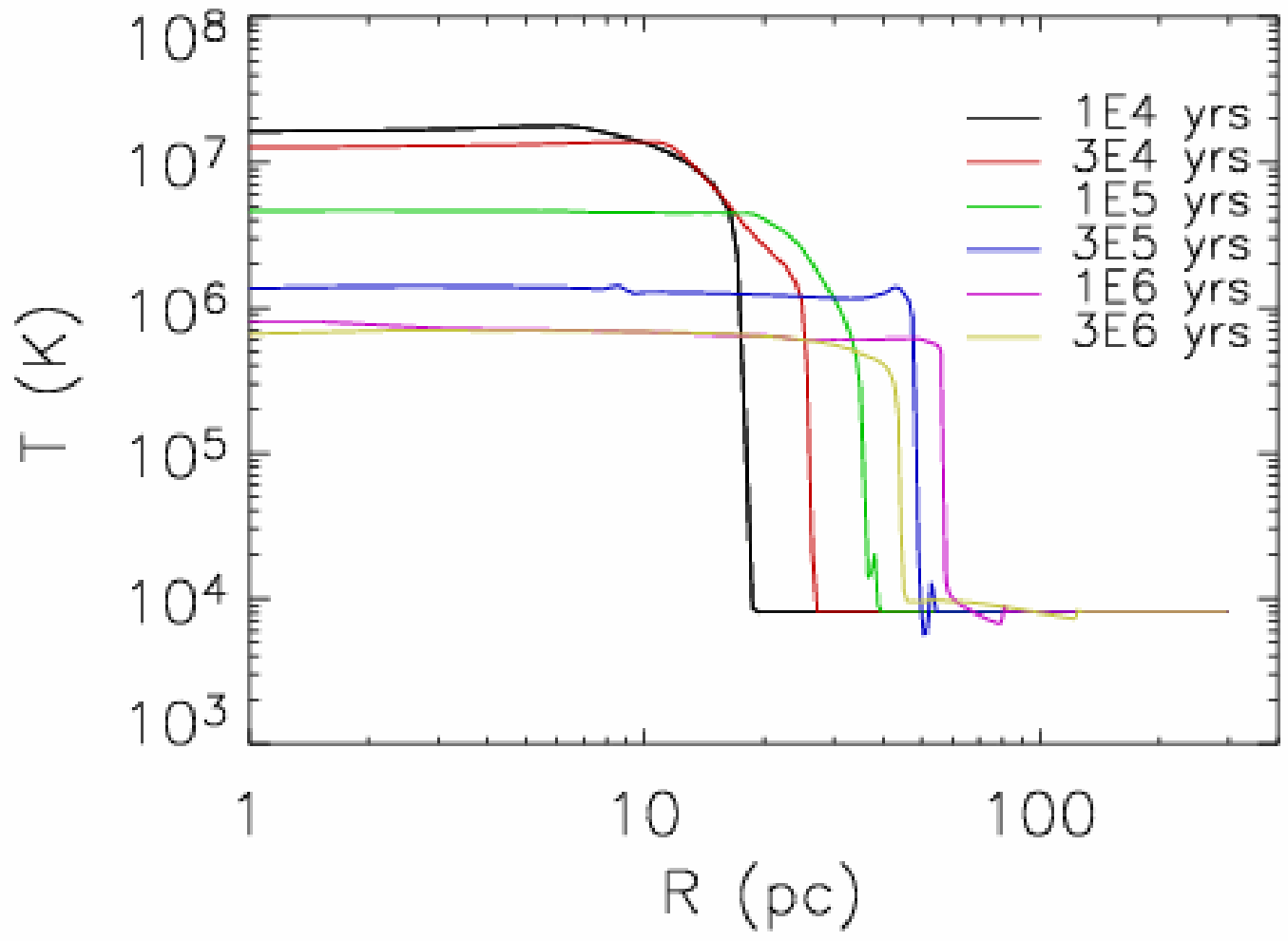

Fig. 11b

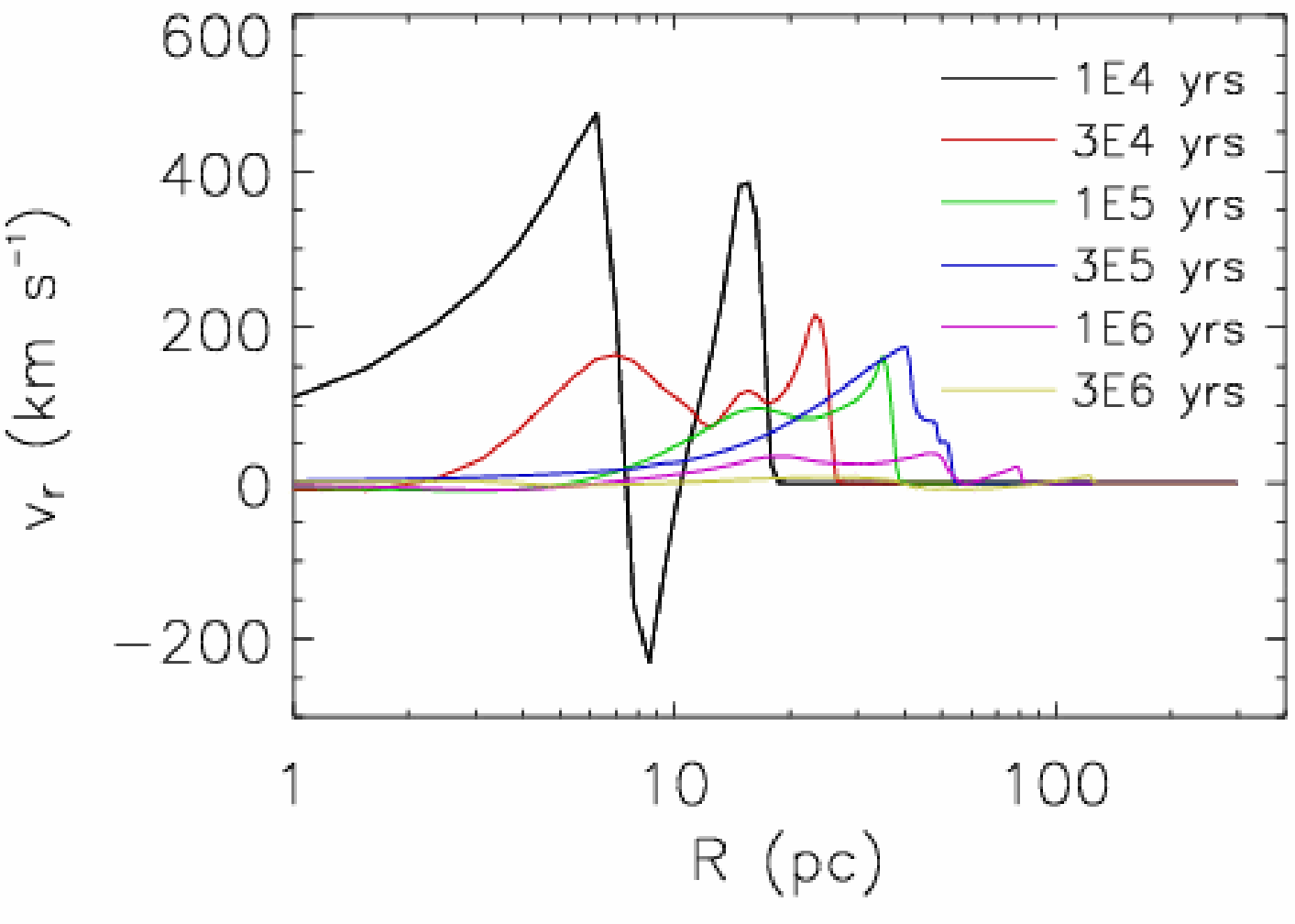

Fig. 11c

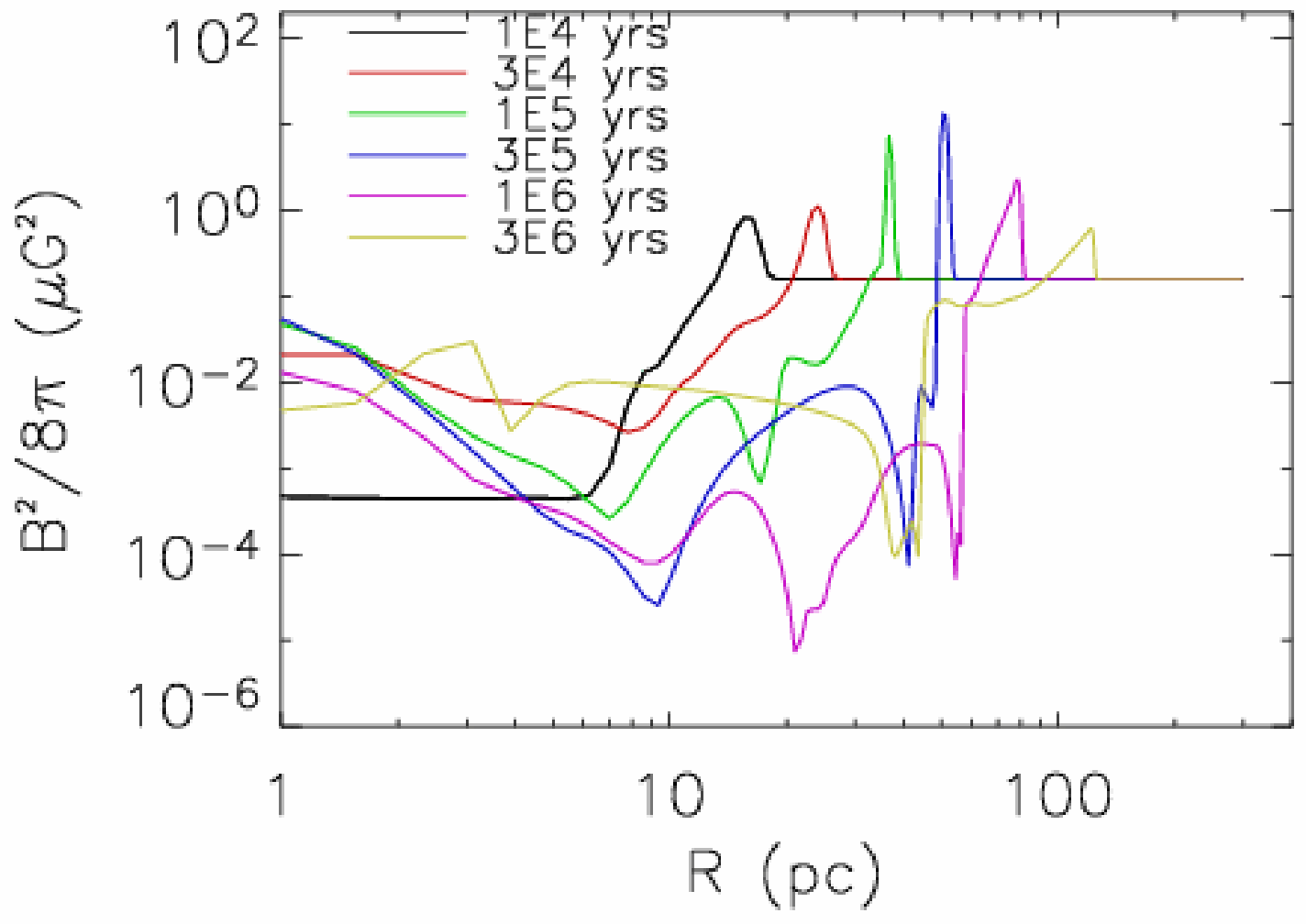

Fig. 11d

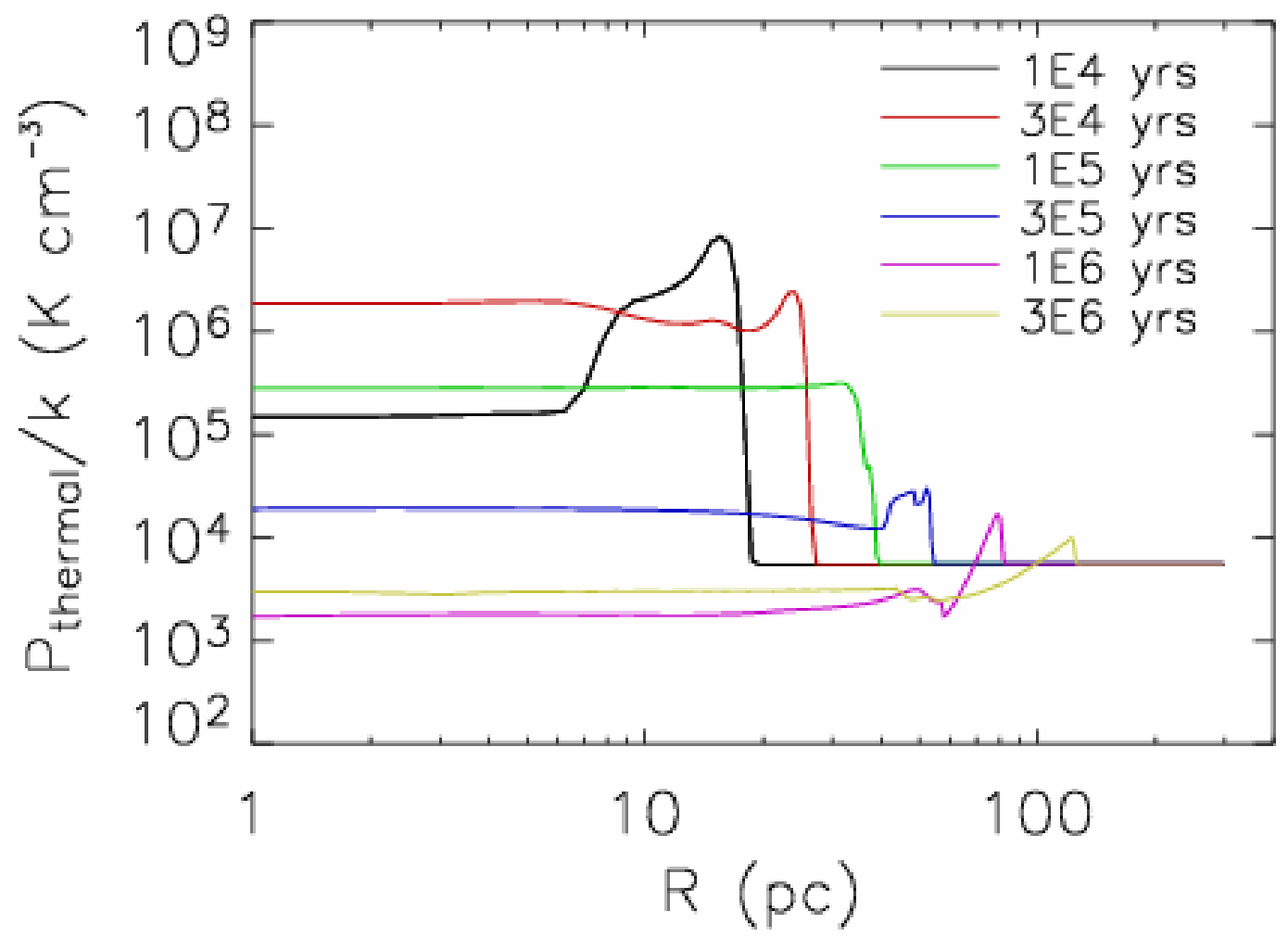

Fig. 11e

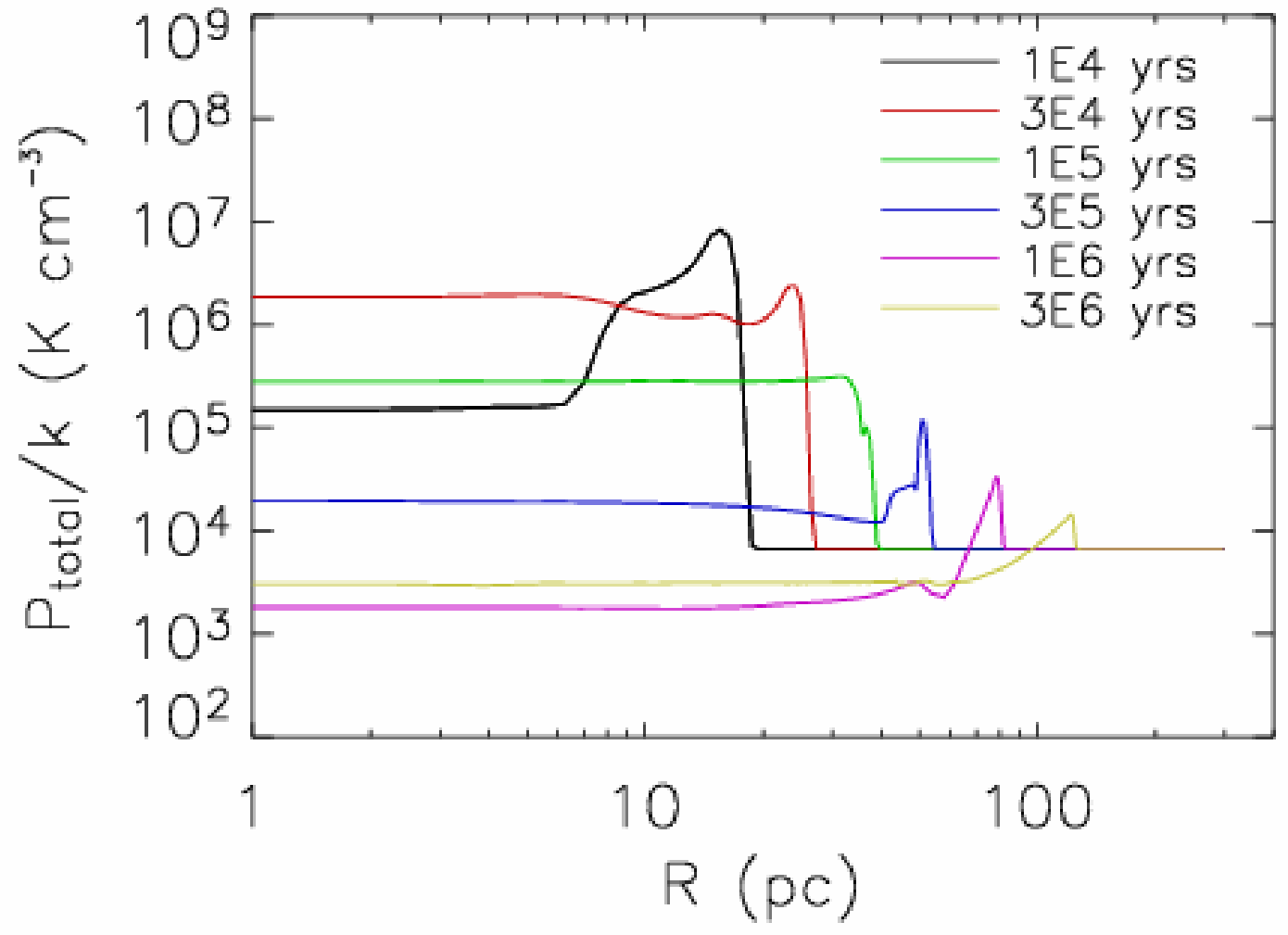

Fig. 11f

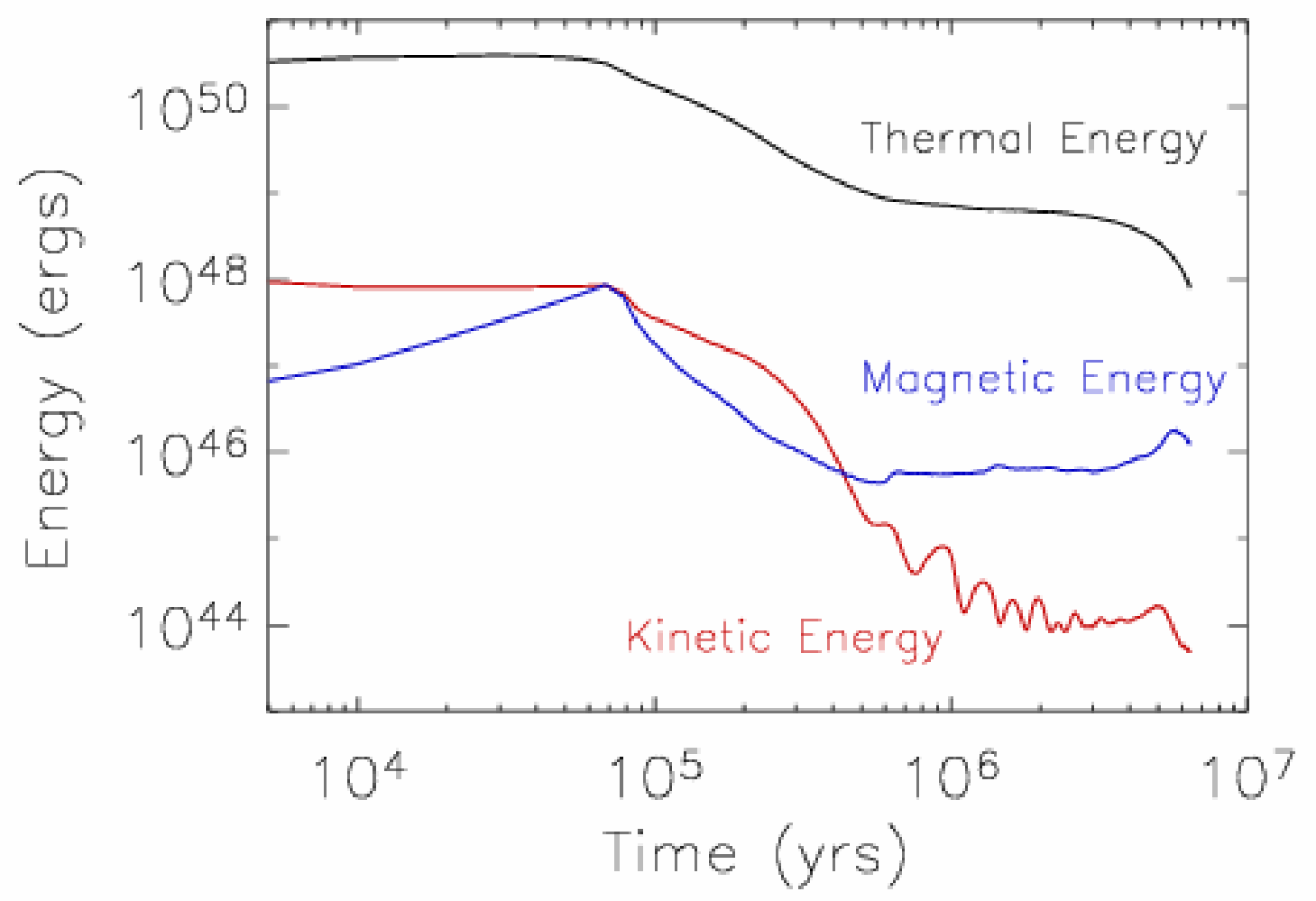

Fig. 11g

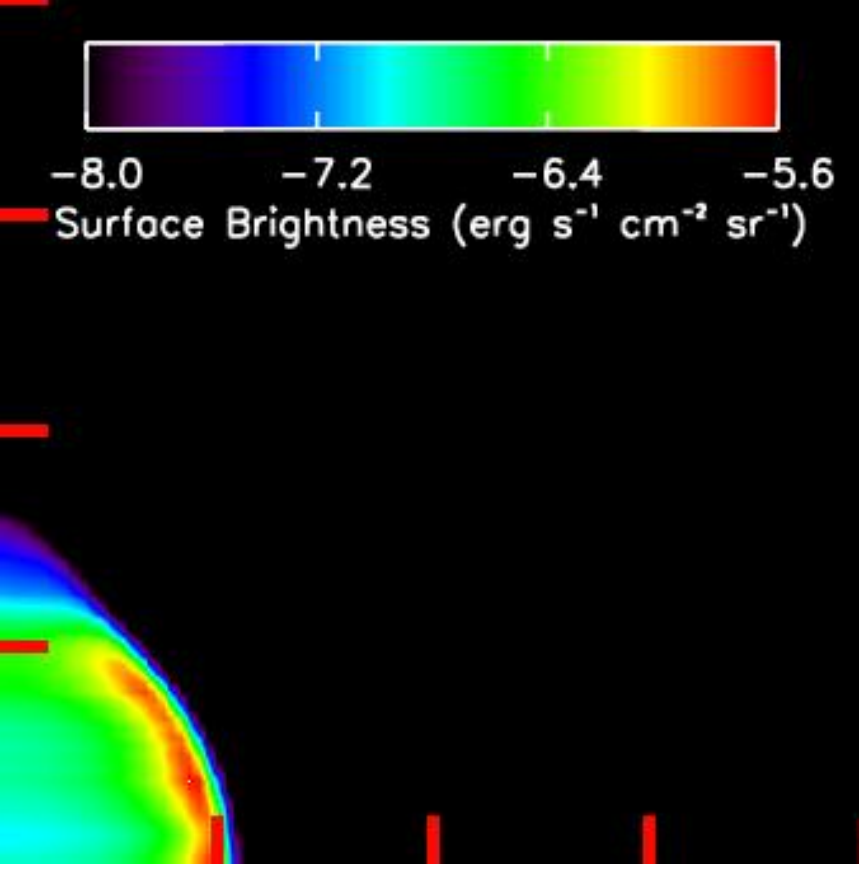

Fig. 12a

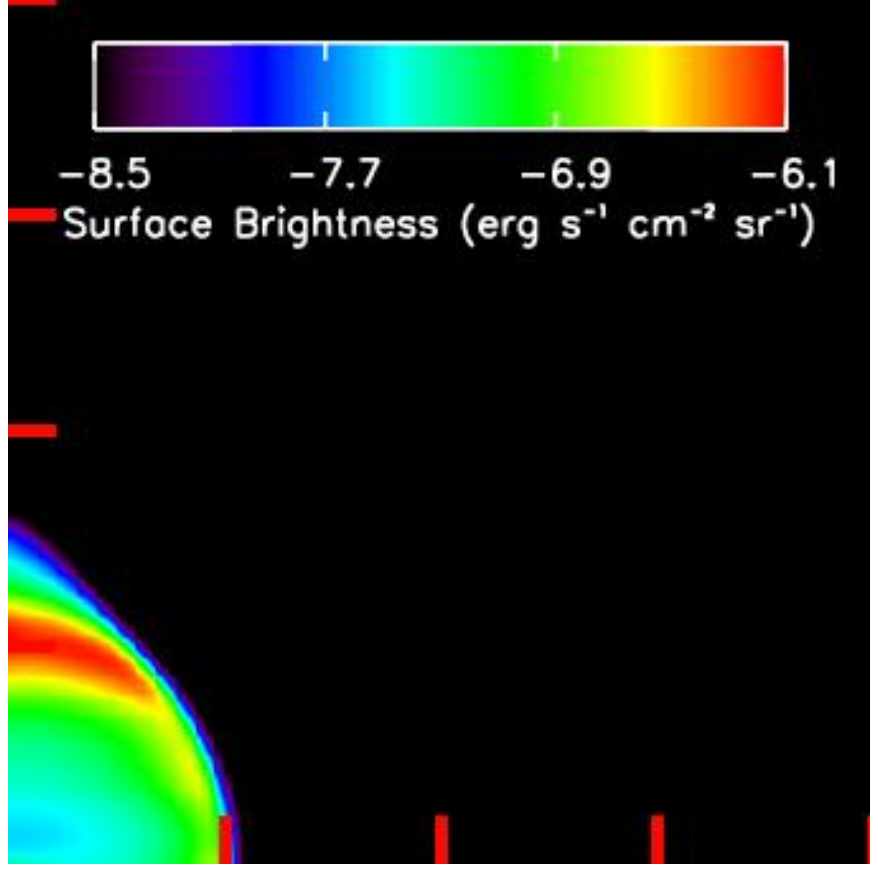

Fig. 12b

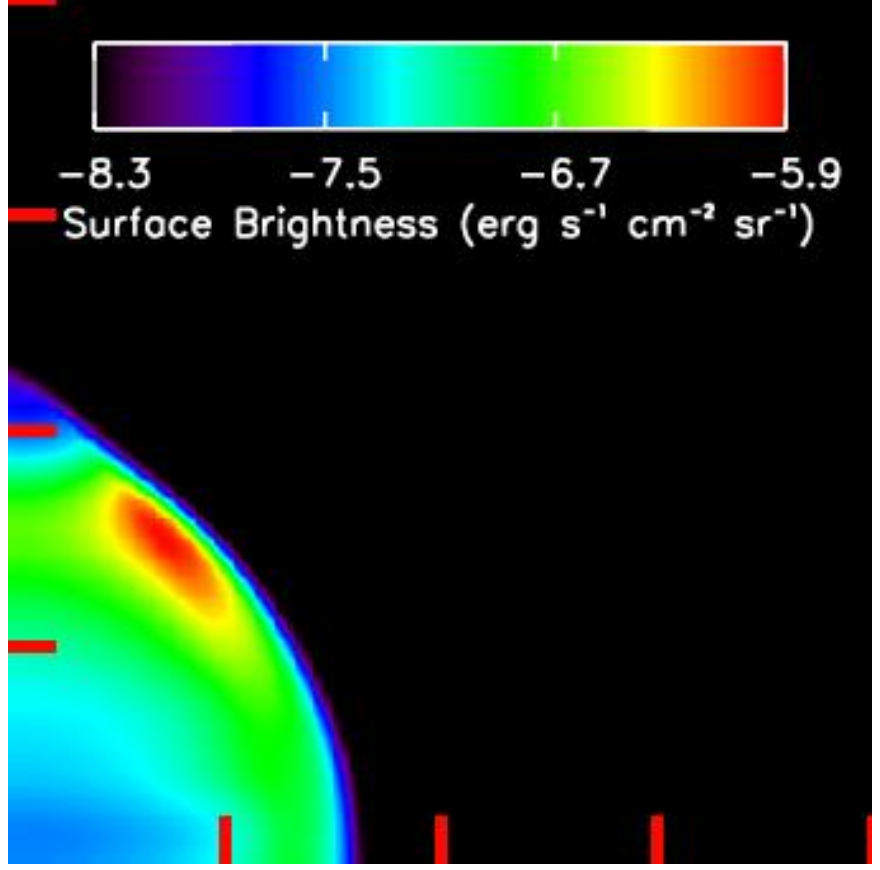

Fig. 12c

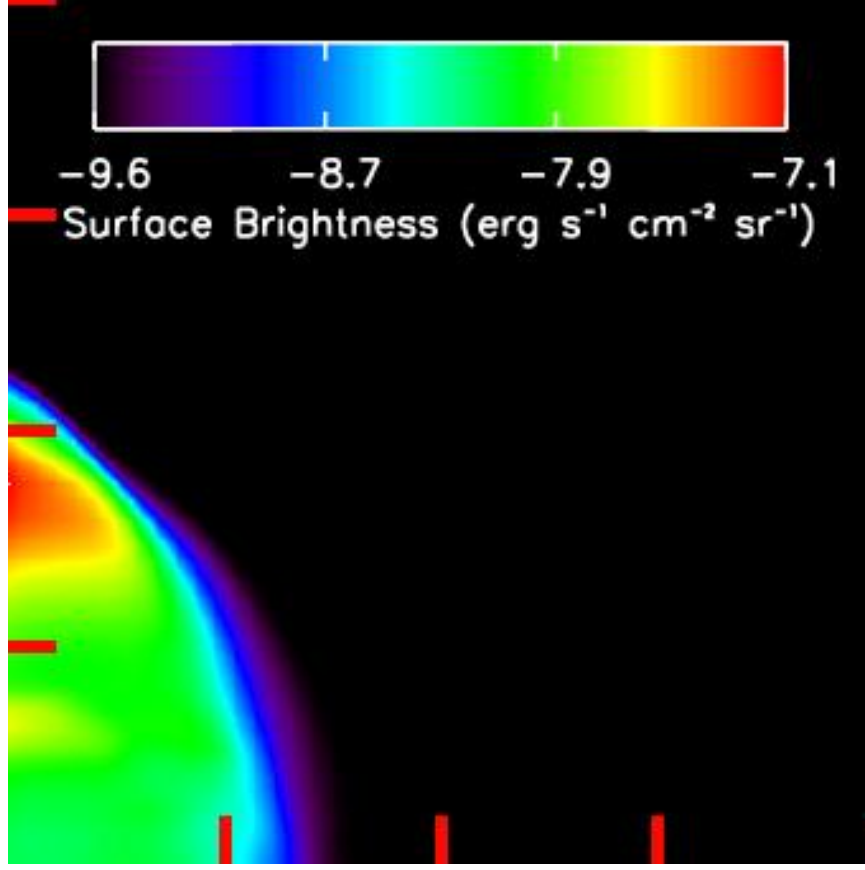

Fig. 12d

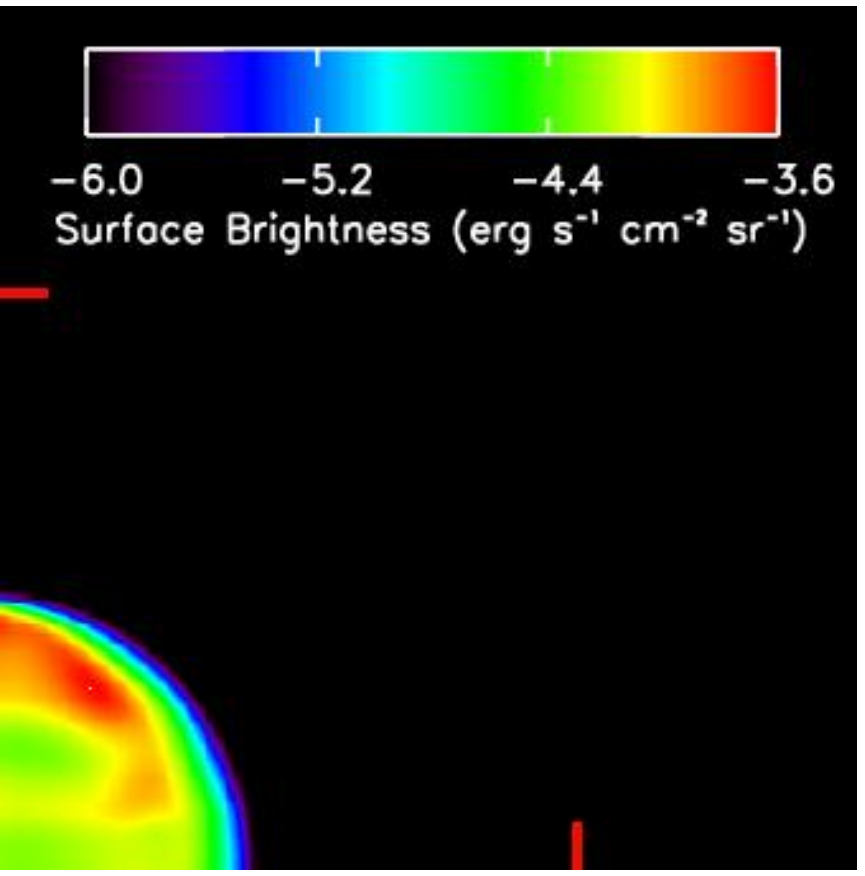

Fig. 12e

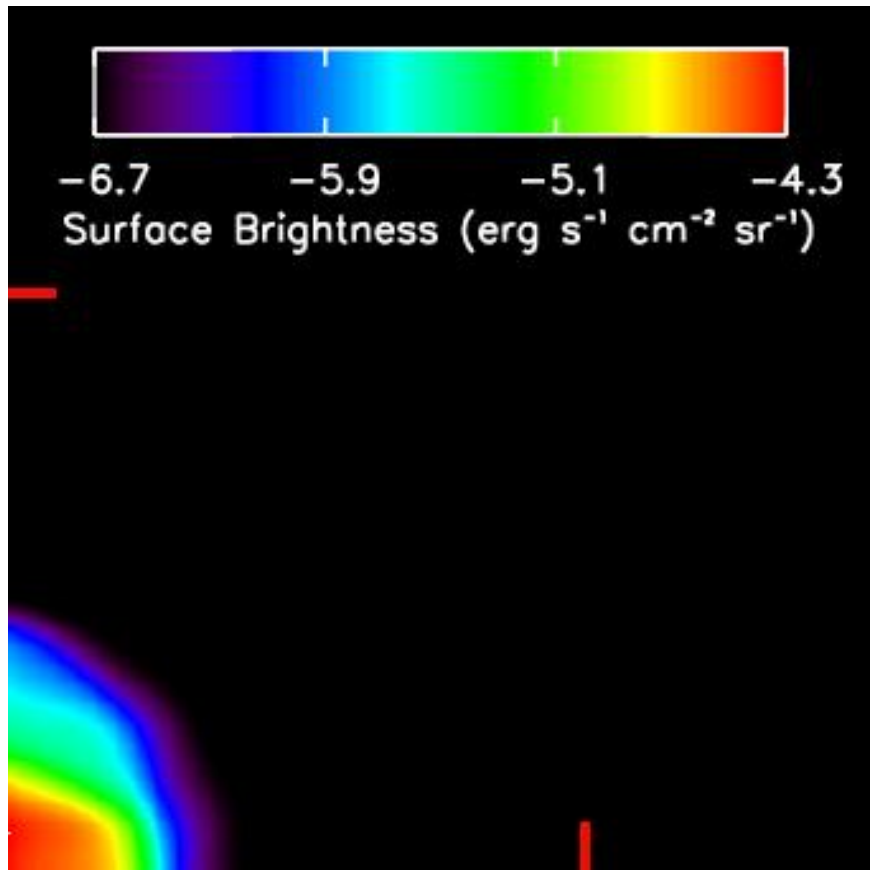

Fig. 12f

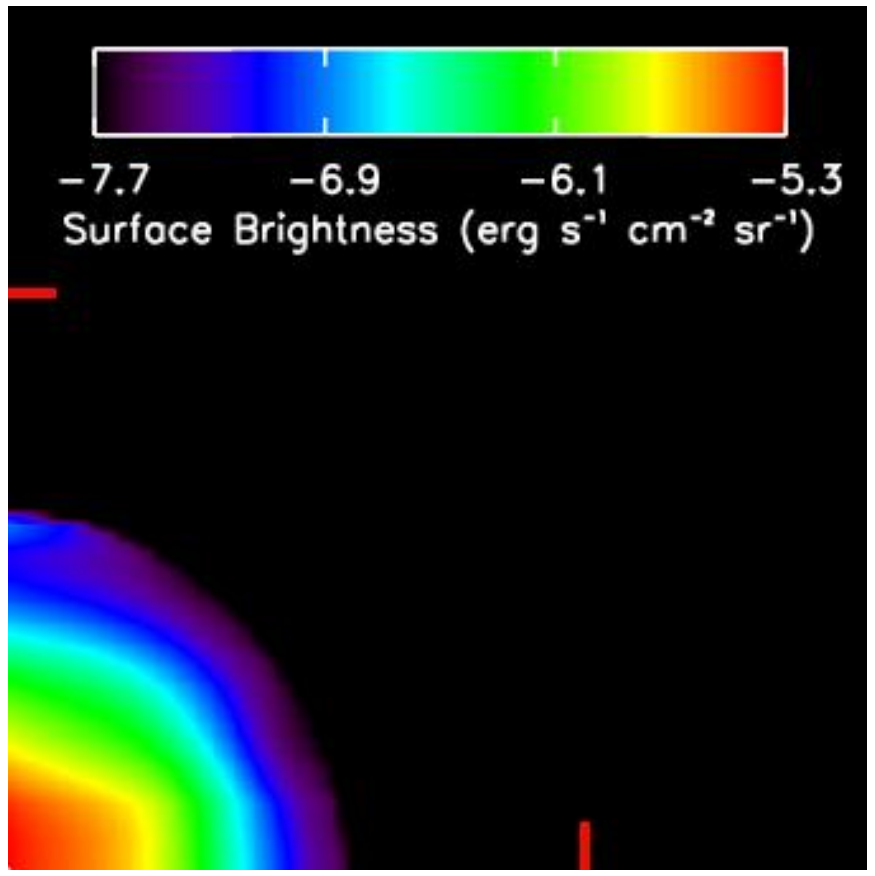

Fig. 12g

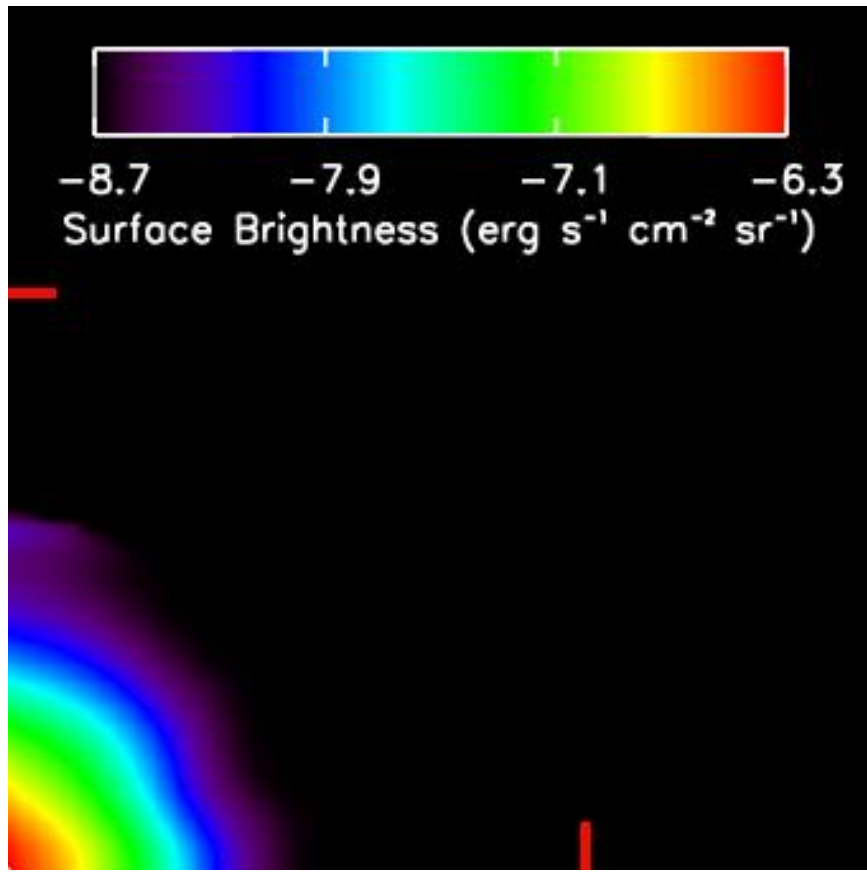

Fig. 12h